\documentclass[aps,twocolumn]{revtex4}

\usepackage{amssymb}
\usepackage{amsmath}
\usepackage{bm}
\usepackage{graphicx}
\usepackage{textcomp}

\begin{document}

\title{The complexity classes of angular diagrams of the metal conductivity
in strong magnetic fields.}

\author{A.Ya. Maltsev.}

\affiliation{
\centerline{\it L.D. Landau Institute for Theoretical Physics}
\centerline{\it 142432 Chernogolovka, pr. Ak. Semenova 1A,
maltsev@itp.ac.ru}}

\begin{abstract}
We consider angular diagrams describing the dependence of the magnetic 
conductivity of metals on the direction of the magnetic field in rather
strong fields. As it can be shown, all angular conductivity diagrams can be 
divided into a finite number of classes with different complexity.
The greatest interest among such diagrams is represented by diagrams with 
the maximal complexity, which can occur for metals with rather complicated 
Fermi surfaces. In describing the structure of complex diagrams, in addition 
to the description of the conductivity itself, the description of the Hall 
conductivity for different directions of the magnetic field plays very 
important role. For the evaluation of the complexity of angular diagrams 
of the conductivity of metals, it is convenient also to compare such diagrams 
with the full mathematical diagrams that are defined (formally) for the 
entire dispersion relation.
\end{abstract}

\maketitle

\vspace{5mm}

\section{Introduction}

 In this paper, we consider the angular diagrams of the electrical 
conductivity of metals with arbitrary Fermi surfaces in strong magnetic 
fields. It is well known that the specific features of the conductivity 
of metals in this limit are primarily due to specific features of 
the semiclassical electron trajectories arising on the Fermi surface in 
the presence of an external magnetic field. More specifically, many 
nontrivial features in the behavior of conductivity in this case are 
mainly associated with the appearance of open electron trajectories 
that appear on Fermi surfaces of a rather complex shape. As a consequence, 
the complexity of the angular diagrams of conductivity is determined directly 
by the presence and type of open trajectories appearing on the Fermi surface 
for different directions of $\, {\bf B} \, $. The angular diagram can then 
be referred to the ``quite complicated'' type if it contains regions 
corresponding to the appearance of open trajectories on the Fermi surface, 
i.e. the appearance of stable open quasiclassical electron trajectories.
This paper will focus mainly on such angular diagrams of conductivity.
As it turns out, diagrams of this class can be naturally divided into two 
types - simpler (type A) and diagrams that can be called ultra-complex 
(type B) in accordance with their really complicated structure. In this 
paper, we will describe and compare the complexity of both types of diagrams 
with the most complex diagrams, namely, diagrams defined for the entire 
dispersion law. Diagrams of the latter type are somewhat abstract from 
the point of view of the theory of normal metals; nevertheless, consideration 
of such diagrams provides a convenient tool for constructing a general 
approach to the description of diagrams determining the conductivity of 
normal metals. In particular, consideration of such diagrams allows one 
to describe the ``areas of occurrence'' of type A or B diagrams and to 
estimate the ``probability'' of the appearance of both types of diagrams 
for metals with a dispersion law of a given type. Consideration of diagrams 
defined for the entire dispersion law also makes it possible to naturally 
include simpler diagrams in the classification, corresponding to the appearance 
of only closed or unstable open trajectories on the Fermi surface.

 Below we give a brief description of the relationship between the 
behavior of magnetic conductivity and the behavior of semiclassical 
electron trajectories on the Fermi surface of a metal, which will allow us, 
in fact, to treat angular diagrams describing the behavior of trajectories 
on the Fermi surface as conductivity diagrams in strong magnetic fields.
In subsequent chapters, we will consider in more detail the structure of 
the Fermi surface when stable open trajectories appear on it, the features 
of angular diagrams of conductivity for the entire dispersion law and for 
a fixed Fermi surface, and compare the structures of the diagrams of both 
types. As a result of our consideration, we will suggest the general scheme 
of the natural separation of angular diagrams of conductivity in metals into 
different classes.

\vspace{1mm}

 As is well known, electron states in a crystal are parametrized by the 
values of the quasi-momentum $\, {\bf p} \, = \, (p_{1}, p_{2}, p_{3}) \, $,
and any two values of $\, {\bf p} \, $, which differ by a vector of the 
reciprocal lattice
\begin{equation}
\label{VekObrResh}
{\bf a} \,\,\, = \,\,\, n_{1} {\bf a}_{1} \,\, + \,\,
n_{2} {\bf a}_{2} \,\, + \,\, n_{3} {\bf a}_{3} \,\,\, ,
\quad n_{1}, n_{2}, n_{3} \, \in \, \mathbb{Z} \,\,\, , 
\end{equation}
specify the same electron state. As is also well known, the basis vectors 
$\, ({\bf a}_{1}, {\bf a}_{2}, {\bf a}_{3}) \, $ of the reciprocal lattice 
can be chosen in the form
$${\bf a}_{1} \,\,\, = \,\,\, 2 \pi \hbar \,\,
{{\bf l}_{2} \, \times \, {\bf l}_{3} \over
({\bf l}_{1}, \, {\bf l}_{2}, \, {\bf l}_{3} )} \,\,\, , \quad \quad
{\bf a}_{2} \,\,\, = \,\,\, 2 \pi \hbar \,\,
{{\bf l}_{3} \, \times \, {\bf l}_{1} \over
({\bf l}_{1}, \, {\bf l}_{2}, \, {\bf l}_{3} )} \,\,\, ,  $$
$${\bf a}_{3} \,\,\, = \,\,\, 2 \pi \hbar \,\,
{{\bf l}_{1} \, \times \, {\bf l}_{2} \over
({\bf l}_{1}, \, {\bf l}_{2}, \, {\bf l}_{3} )} \,\,\, , $$
where $\, ({\bf l}_{1}, {\bf l}_{2}, {\bf l}_{3}) \, $
represent the basis of a direct crystal lattice.

 The change in the values of the quasimomentum in the presence of an external 
constant magnetic field is described by the system
(see for example \cite{Kittel,Ziman,Abrikosov})
\begin{equation}
\label{MFSyst}
{\dot {\bf p}} \,\,\,\, = \,\,\,\, {e \over c} \,\,
\left[ {\bf v}_{\rm gr} ({\bf p}) \, \times \, {\bf B} \right]
\,\,\,\, \equiv \,\,\,\, {e \over c} \,\, \left[ \nabla \epsilon ({\bf p})
\, \times \, {\bf B} \right] \,\,\, , 
\end{equation}
where $\, \epsilon ({\bf p}) \, $ determines the dependence of the electron 
state energy on the quasi-momentum (dispersion relation) for a fixed conduction 
band. Due to the fact that any two values of the quasi-momentum that 
differ by a reciprocal lattice vector specify the same electron state, the 
function $\, \epsilon ({\bf p}) \, $ is a 3-periodic function in the
$\, {\bf p}$ - space with periods
$\, {\bf a}_{1}, {\bf a}_{2}, {\bf a}_{3} \, $.

\begin{figure}[t]
\begin{center}
\includegraphics[width=0.9\linewidth]{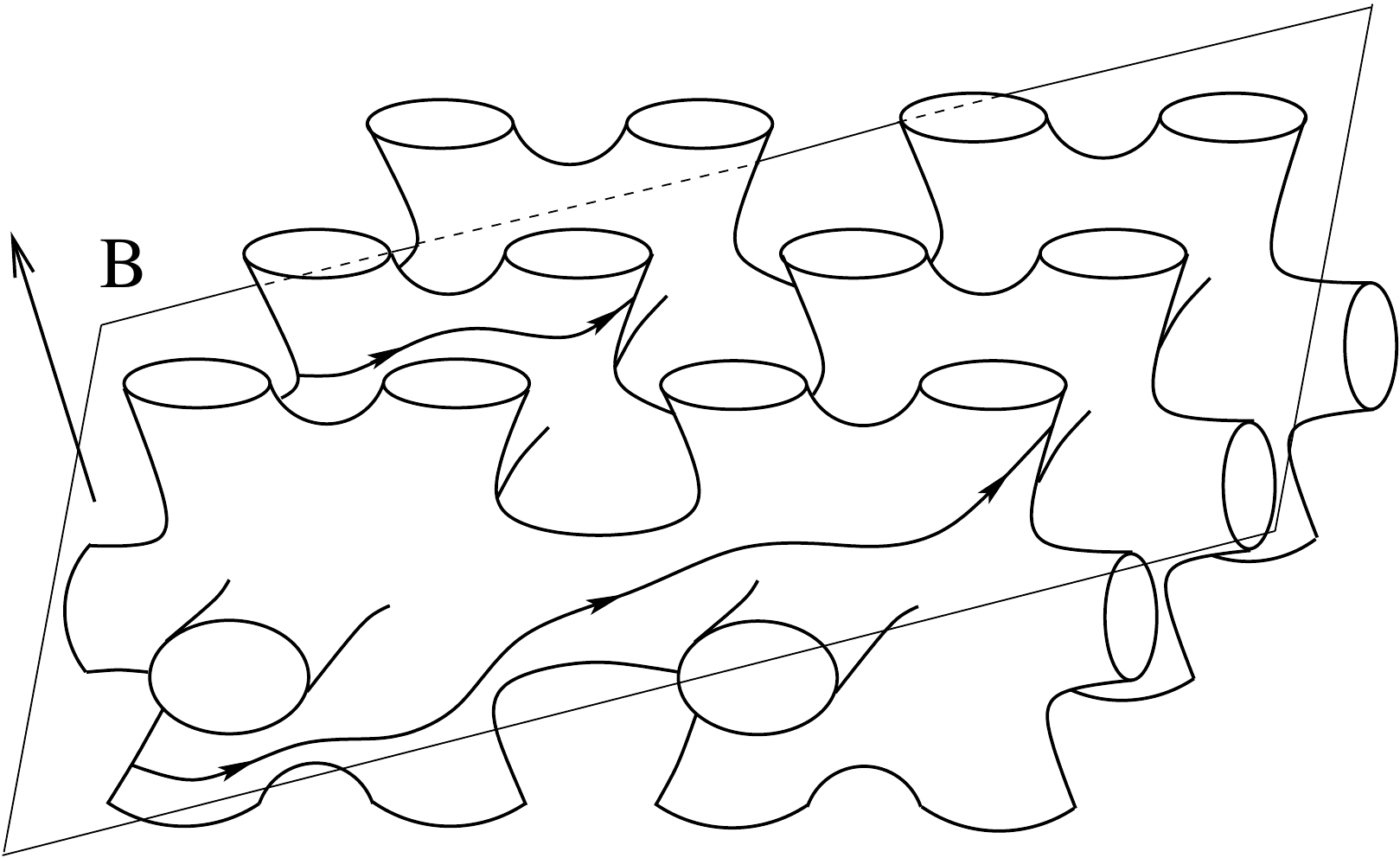}
\end{center}
\caption{The trajectories of the system (\ref{MFSyst}) on a constant energy 
surface of a rather complicated form.}
\label{ComplSurf}
\end{figure}

 The system (\ref{MFSyst}) is analytically integrable, in particular, 
its trajectories in the $\, {\bf p}$ - space are given by the intersections 
of the surfaces of constant energy
$\, \epsilon ({\bf p}) \, = \, const \, $ with planes orthogonal to 
$\, {\bf B} \, $. It should be noted here that the description of the 
geometry of such trajectories for a really complex 3-periodic surface in
$\, {\bf p}$ - space is a highly nontrivial problem (see Fig. \ref{ComplSurf}). 
It is also easy to see that the most complicated is the description of the 
geometry of open (non-closed) trajectories of (\ref{MFSyst}) in the 
$\, {\bf p}$ - space.

 It is also well known that the trajectories of the system (\ref{MFSyst})
in the $\, {\bf p}$ - space correspond to quasiclassical electron trajectories 
in the coordinate space, which have a somewhat more complicated form 
(in particular, they are not plane). At the same time, however, the shape 
of the electron trajectories in the $\, {\bf x}$ - space is in fact closely 
related to their shape in the $\, {\bf p}$ - space, for example, the projections 
of the trajectories in the $\, {\bf x}$ - space on the plane orthogonal to
$\, {\bf B} \, $ are similar to trajectories in the $\, {\bf p}$ - space, 
rotated by $90^{\circ}$. The latter property makes the geometry of the 
trajectories of system (\ref{MFSyst}) extremely important for the description of 
transport phenomena in normal metals in the presence of strong magnetic fields.

 The extremely important role of the geometry of the trajectories of the system 
(\ref{MFSyst}) in the description of galvanomagnetic phenomena in normal metals 
was indicated in the works of the school of I.M. Lifshits in the 1950s
(\cite{lifazkag,lifpes1,lifpes2}). During this period, many issues related 
to the features of electronic phenomena due to the nontrivial geometry of 
the Fermi surface, in particular, issues related to the geometry 
of the trajectories of system (\ref{MFSyst}), were studied, and many important 
examples of different types of trajectories on different Fermi surfaces
were considered (see for example 
\cite{lifazkag,lifpes1,lifpes2,lifkag1,lifkag2,lifkag3,
etm,ElPr,KaganovPeschansky,Kittel,Ziman,Abrikosov}). 
We, of course, will not be able to give here even a brief overview of these 
issues or provide any full bibliography on this topic. In this paper, we will, 
in fact, be interested in the picture of a complete mathematical description 
of the structure of the system (\ref{MFSyst}), based on the results of the study 
of this system in a later period. We note here that the shape of the trajectories
of (\ref{MFSyst}) starts to play a really significant role in the  considered 
phenomena under the condition
$\, \omega_{B} \tau \, \gg \, 1 \, $, where
$\, \omega_{B} \, = \, e B / m^{*} c \, $ has the meaning of some effective 
cyclotron frequency in a crystal and $\, \tau \, $ represents the electron 
mean free time (\cite{lifazkag}).

 The problem of complete classification of various types of open trajectories 
of the system (\ref{MFSyst}) with arbitrary (periodic) dispersion laws 
$\, \epsilon ({\bf p}) \, $ was set by S.P. Novikov in the 1980s
(\cite{MultValAnMorseTheory}) and was actively investigated in his topological 
school in the following decades. It can be said that at present a quite
complete description of all types of trajectories of (\ref{MFSyst}), 
based on rather deep mathematical theorems, has been obtained. It should 
be noted that the most serious breakthroughs in solving this problem were 
made in \cite{zorich1,dynn1992,dynn1}, where a description of open trajectories 
of (\ref{MFSyst}), which in a certain sense can be called stable open
trajectories, was obtained. 

\begin{figure}[t]
\begin{center}
\includegraphics[width=0.9\linewidth]{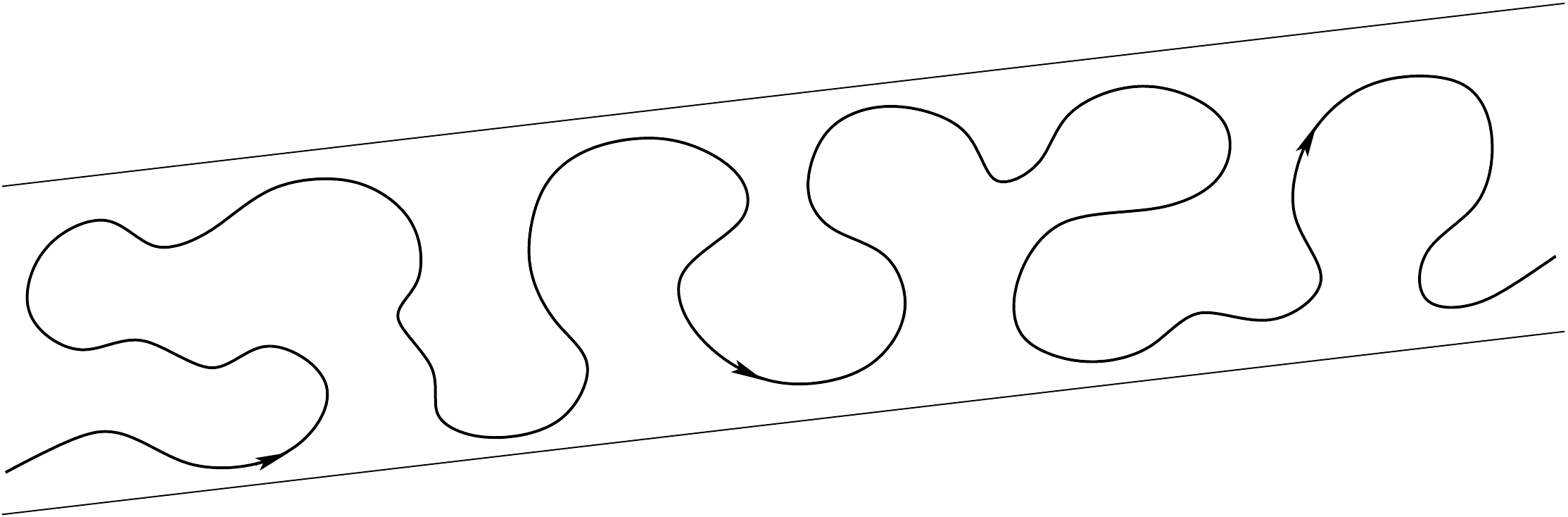}
\end{center}
\caption{The form of a stable open trajectory of the system (\ref{MFSyst}) 
in the plane orthogonal to $\, {\bf B} \, $.}
\label{StableTr}
\end{figure}

 We specify here what we shall call open trajectories of the system
(\ref{MFSyst}) stable if they do not disappear and retain their global 
geometry under small rotations of the direction of $\, {\bf B} \, $,
as well as variations of the energy level $\, \epsilon \, $. 
As follows from the results of the works \cite{zorich1,dynn1992,dynn1}, 
the stable open trajectories of the systems (\ref{MFSyst}) have 
the following remarkable properties:

\vspace{1mm}

\noindent
1) Each stable open trajectory of the system (\ref{MFSyst})
lies in a straight strip of finite width in the plane orthogonal to 
$\, {\bf B} \, $, passing it through (Fig. \ref{StableTr});

\vspace{1mm}

\noindent
2) The mean direction of all stable open trajectories in
$\, {\bf p}$ - space is the same for a given direction of
$\, {\bf B} \, $ and is given by the intersection of the plane 
orthogonal to $\, {\bf B} \, $ and some integral plane $\, \Gamma \, $,
unchanged for small rotations of $\, {\bf B} \, $ and variations of the 
level $\, \epsilon \, $.

\vspace{1mm}

 Properties (1) and (2) of stable open trajectories have direct manifestation 
in the behavior of galvanomagnetic phenomena in metals in the presence of strong 
magnetic fields. Namely, the strong anisotropy of the electron trajectories
(in $\, {\bf x} \, - \, $ and $\, {\bf p}$ - space) leads to a sharp anisotropy 
of conductivity in the plane orthogonal to $\, {\bf B} \, $ in the limit
$\, \omega_{B} \tau \, \gg \, 1 \, $. The limiting values of the conductivity 
tensor in the $\, {\bf x}$ - space can be represented in this situation as
\begin{equation}
\label{LimitValue}
\sigma^{kl}_{\infty} \,\,\,\,  =  \,\,\,\,
{n e^{2} \tau \over m^{*}} \,\, \left(
\begin{array}{ccc}
 0  &  0  &  0  \cr
 0  &  *  &  *  \cr
 0  &  *  &  *
\end{array}  \right) 
\end{equation}
provided that the $\, x \, $ axis coincides with the mean direction of the 
stable open trajectories in the $\, {\bf p}$ - space.

 It is easy to see that the contribution (\ref{LimitValue}) differs 
significantly from the contribution of closed trajectories in the same 
limit (see \cite{lifazkag})
$$\sigma^{kl}_{\infty} \,\,\,\,  =  \,\,\,\,
{n e^{2} \tau \over m^{*}} \,\, \left(
\begin{array}{ccc}
 0  &  0  &  0  \cr
 0  &  0  &  0  \cr
 0  &  0  &  *
\end{array}  \right) $$
(in both formulas, the notation $\, * \, $ represents some dimensionless 
constant of order 1).

 The properties of the tensor (\ref{LimitValue}) served as the basis for the 
introduction in \cite{PismaZhETF} (see also \cite{UFN}) of new topological 
characteristics (topological quantum numbers) observable in the conductivity 
of normal metals with complex Fermi surfaces.  

  Indeed, as is easy to see, the mean direction of stable open trajectories 
in the $\, {\bf p}$ - space coincides with the direction of the greatest 
suppression of conductivity in the plane orthogonal to $\, {\bf B} \, $,
and, thus, is observable experimentally. Due to the stability of such 
trajectories, the experimentally observable is actually also the integral 
plane $\, \Gamma \, $, swept out by the directions of the greatest suppression 
of conductivity at small rotations of the direction of $\, {\bf B} \, $. 
The integer parameters of the plane $\, \Gamma \, $ represent in this case 
the topological quantum numbers observable in conductivity in strong magnetic 
fields.

 We note here that the integerness of the plane $\, \Gamma \, $ means that 
it is generated by two reciprocal lattice vectors (\ref{VekObrResh}).
In particular, it does not have to coincide in the general case with any 
of the crystallographic planes; instead, it is orthogonal to some 
crystallographic direction in the $\, {\bf x}$ - space. Thus, the integral 
plane $\, \Gamma \, $ observable in the measurements of the magnetic 
conductivity can be represented by some triple of integers 
$\, (M^{1}, M^{2}, M^{3}) \, $, specifying some integer direction in the 
crystal lattice.

\begin{figure}[t]
\begin{center}
\includegraphics[width=0.9\linewidth]{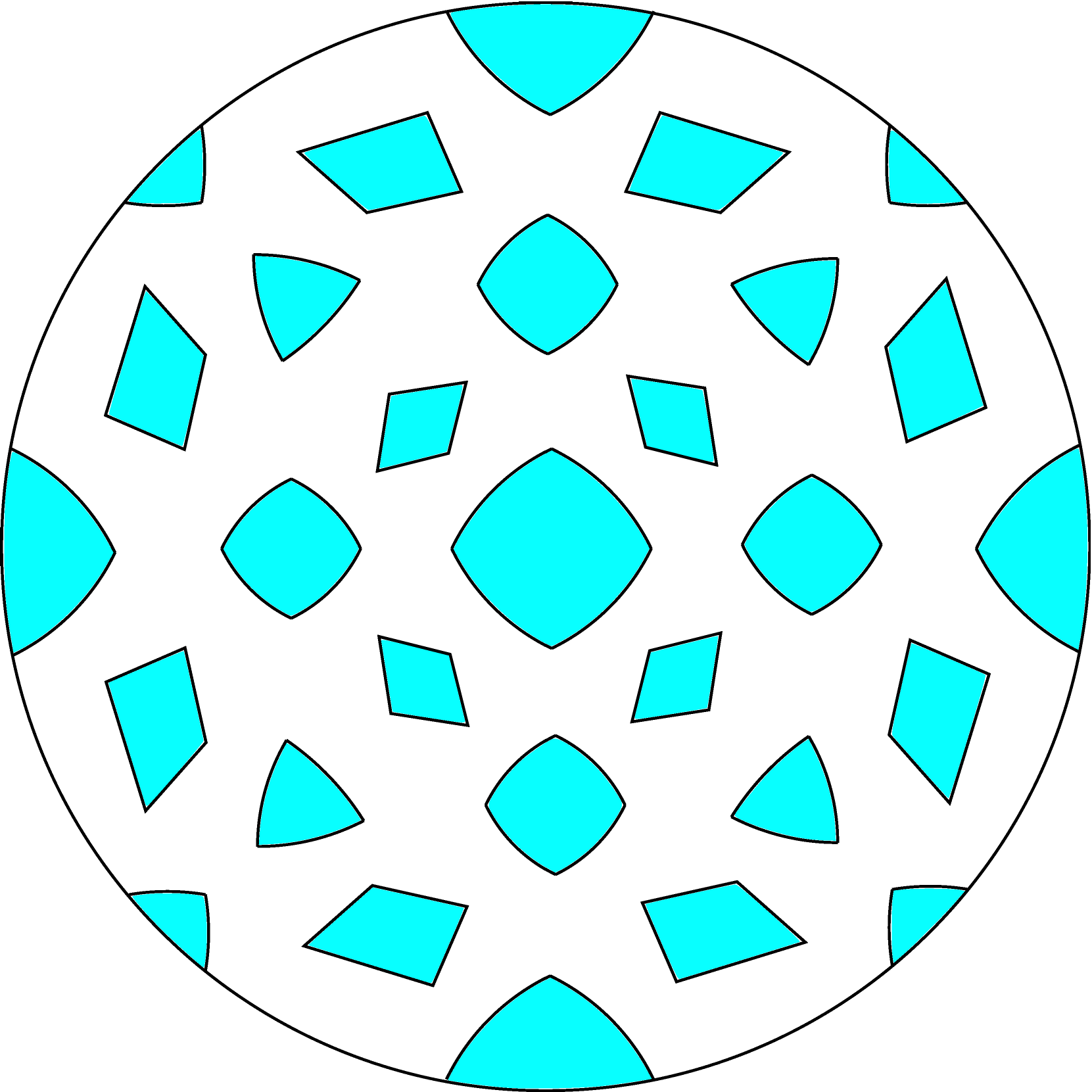}
\end{center}
\caption{Stability Zones $\, \Omega_{\alpha} \, $ in the space of directions 
of $\, {\bf B} \, $ corresponding to the appearance of families of stable open 
trajectories of the system (\ref{MFSyst}) with different numbers
$\, (M_{\alpha}^{1}, M_{\alpha}^{2}, M_{\alpha}^{3}) \, $
on the Fermi surface.}
\label{AngleDiag}
\end{figure}

 Each of the observable topological triples 
$\, (M_{\alpha}^{1}, M_{\alpha}^{2}, M_{\alpha}^{3}) \, $ is connected
with a certain family of stable open trajectories of the system
(\ref{MFSyst}) corresponding to a certain Stability Zone 
$\, \Omega_{\alpha} \, $ in the space of directions of $\, {\bf B} \, $
(Fig. \ref{AngleDiag}). As we have said, in the theory of normal metals 
it is natural to consider the trajectories of the system (\ref{MFSyst}) 
only at one energy level $\, \epsilon ({\bf p}) = \epsilon_{F} \, $
corresponding to the Fermi energy of a metal. Thus, it is natural to define 
the boundaries of the Stability Zones $\, \Omega_{\alpha} \, $ for a normal 
metal by the condition of the existence of open trajectories of the 
corresponding family on the Fermi surface
$$S_{F} \, :  \quad \epsilon ({\bf p}) \, = \, \epsilon_{F} $$

 Note also that the relation (\ref{LimitValue}) is true, in particular, also 
for periodic open trajectories (see \cite{lifazkag}), which have rational 
directions in the $\, {\bf p}$  - space. The latter, however, can both belong 
to a stable family (if $\, {\bf B}/B \, \in \, \Omega_{\alpha} $), and also be 
unstable (if $\, {\bf B}/B \, \notin \, \Omega_{\alpha} $).

 It must now be said that in addition to stable open trajectories and periodic 
trajectories of the system (\ref{MFSyst}), having a relatively simple form, 
trajectories with more complicated ``chaotic'' behavior can also appear on 
complex enough Fermi surfaces.

 The first example of a trajectory of this type was constructed by S.P. Tsarev 
in 1992 (\cite{Tsarev}) for a direction of $\, {\bf B} \, $ of irrationality 2 
(the plane orthogonal to $\, {\bf B} \, $ contains a reciprocal lattice vector).
Tsarev trajectories can arise on Fermi surfaces of the genus $\, g \geq 3 \, $
and have an obvious chaotic behavior on these surfaces. At the same time, in 
planes orthogonal to $\, {\bf B} \, $ Tsarev-type trajectories have an asymptotic 
direction and their contribution to conductivity is in fact also strongly 
anisotropic. In particular, we should also have here the relation 
(\ref{LimitValue}) in the limit 
$\, \omega_{B} \tau \, \rightarrow \, \infty \, $.
On closer examination, however, the contribution of Tsarev's trajectories 
to electrical conductivity actually differs from the contributions of stable 
open or periodic trajectories, which can be revealed in accurate enough 
experiments. Tsarev-type trajectories are unstable both with respect to small 
rotations of the direction of $\, {\bf B} \, $,  and with respect to variations 
of the Fermi level.

\begin{figure}[t]
\begin{center}
\includegraphics[width=0.9\linewidth]{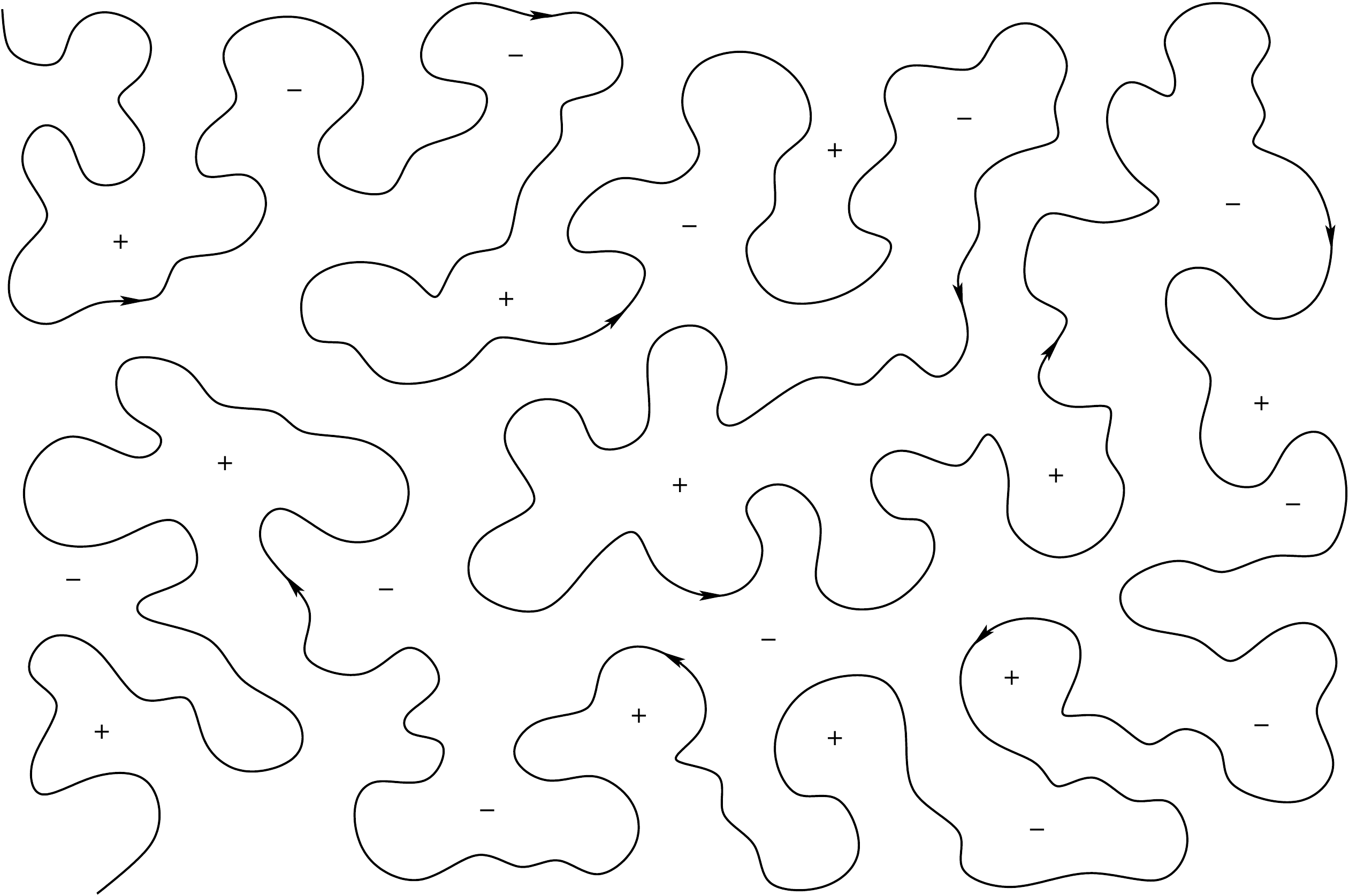}
\end{center}
\caption{The form of chaotic trajectories of Dynnikov type in planes orthogonal 
to $\, {\bf B} \, $ in the $\, {\bf p}$ - space.}
\label{DynnTr}
\end{figure}

 More complex examples of chaotic trajectories on complex Fermi surfaces are 
given by trajectories of Dynnikov type (\cite{dynn2}). Dynnikov trajectories 
can arise only for directions of $\, {\bf B} \, $ of maximal irrationality and 
possess obvious chaotic behavior both on the Fermi surface and in the covering
$\, {\bf p}$ - space (Fig. \ref{DynnTr}). The main feature of the contribution 
of such trajectories to the electrical conductivity in this case is a sharp 
suppression of conductivity along the direction of $\, {\bf B} \, $  in the 
limit $\, \omega_{B} \tau \, \rightarrow \, \infty \, $ (\cite{ZhETF1997}).
In general, the contribution of Dynnikov-type trajectories to the conductivity 
tensor is characterized by a decrease of conductivity in all directions in the 
limit $\, \omega_{B} \tau \, \rightarrow \, \infty \, $ and the appearance of 
fractional powers of the parameter $\, \omega_{B} \tau \, $  in the asymptotics 
of the tensor components (\cite{ZhETF1997,TrMian}). Like Tsarev-type trajectories, 
Dynnikov-type trajectories are unstable both with respect to small rotations 
of the direction of $\, {\bf B} \, $ and with respect to variations 
of the Fermi level. As we will see below, the appearance of chaotic trajectories
(of both types) will be connected mainly with the most complex angular diagrams 
(of type B), where the corresponding directions of $\, {\bf B} \, $  will play 
(together with the Zones $\, \Omega_{\alpha} $) a very important role in the 
structure of such diagrams.

\begin{figure}[t]
\begin{center}
\includegraphics[width=0.9\linewidth]{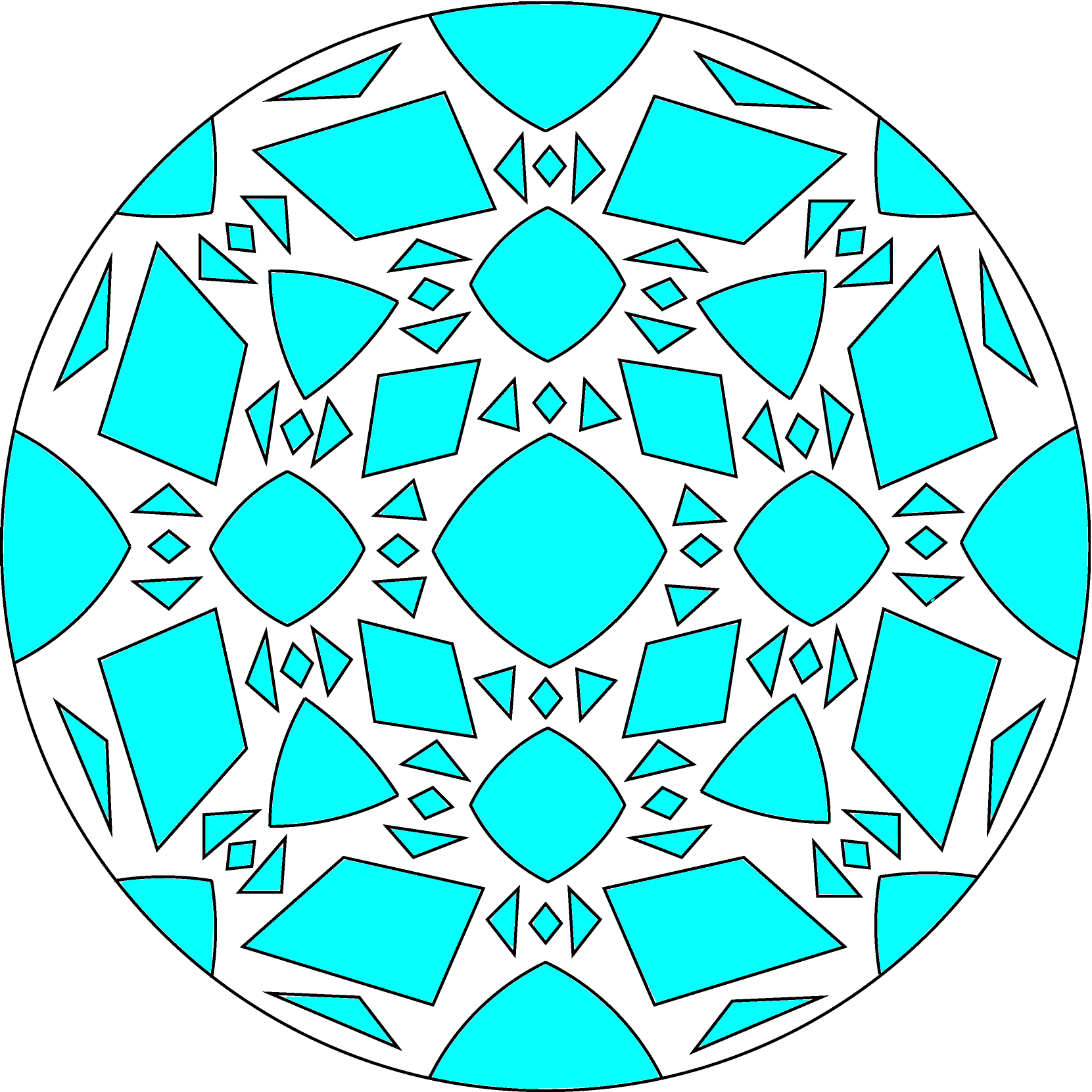}
\end{center}
\caption{Stability Zones $\, \Omega^{*}_{\alpha} \, $ corresponding to the 
appearance of families of stable open trajectories of the system (\ref{MFSyst})
with different numbers $\, (M_{\alpha}^{1}, M_{\alpha}^{2}, M_{\alpha}^{3}) \, $
at least at one of the energy levels $\, \epsilon ({\bf p}) \, = \, const \, $ 
(very schematically, only a finite number of Zones is shown).}
\label{DispRel}
\end{figure}

 As we have said above, in addition to the Stability Zones defined by some fixed 
Fermi surface, one can in fact also introduce Stability Zones
$\, \Omega^{*}_{\alpha} \, $ corresponding to the entire dispersion relation
$\, \epsilon ({\bf p}) \, $ (\cite{dynn3}). The Zones $\, \Omega^{*}_{\alpha} \, $ 
form, as a rule, a more complex set in the space of directions of $\, {\bf B} \, $ 
(Fig. \ref{DispRel}), in particular, each of the Zones $\, \Omega_{\alpha} \, $ 
represents a subset of some bigger Zone $\, \Omega^{*}_{\alpha} \, $. In this paper 
we will discuss the complexity of angular diagrams on $\, \mathbb{S}^{2} \, $
representing the Stability Zones for a fixed Fermi surface, and, in particular, 
we will make some comparison with the diagrams defined for the whole dispersion law.

 We also note here that a rather wide range of issues relating to the description 
of stable trajectories of the system (\ref{MFSyst}), both from the point of view 
of their topological description, and from the point of view of the theory of 
transport phenomena in metals, was presented in the works
\cite{DynnBuDA,JETP1,BullBrazMathSoc,JournStatPhys,DynSyst,DeLeoPhysB,
AnProp,CyclRes,SecBound}.

\section{Topological representation of the Fermi surface and the Stability Zone
in the case of presence of stable open trajectories}
\setcounter{equation}{0}

\begin{figure}[t]
\begin{center}
\includegraphics[width=\linewidth]{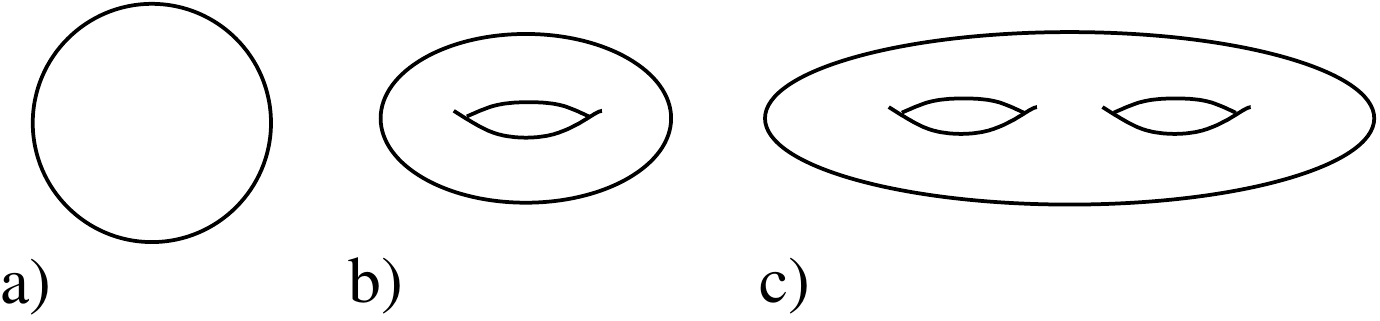}
\end{center}
\caption{Topology of closed surfaces of genus 0, 1, 2 (etc.).}
\label{Genus}
\end{figure}

 In this paper we will focus on Fermi surfaces, which have a rather complex 
shape. Let us determine at once what we mean by this. First of all, we note 
the well-known fact that the total space of physical states for a fixed conduction 
band represents a three-dimensional torus $\, \mathbb{T}^{3} \, $ from the 
topological point of view. Indeed, by virtue of the above equivalence of states 
with quasi-momenta differing by the reciprocal lattice vectors, the complete 
set of states for a given conduction band (Brillouin zone) is a factor space
$$\mathbb{T}^{3} \,\,\, = \,\,\, \mathbb{R}^{3} / L $$
(note also that we do not consider here spin variables that will not play 
a significant role in our reasoning). As a consequence, any (non-singular) 
energy surface $\, \epsilon ({\bf p}) = const \, $ can also be considered 
(after factorization) as a smooth compact surface embedded into
$\, \mathbb{T}^{3} \, $. As a compact smooth two-dimensional surface, each 
such surface (in particular, the Fermi surface) represents a surface of 
a certain genus $\, g \geq 0 \, $ (Fig. \ref{Genus}).

\begin{figure}[t]
\begin{center}
\includegraphics[width=\linewidth]{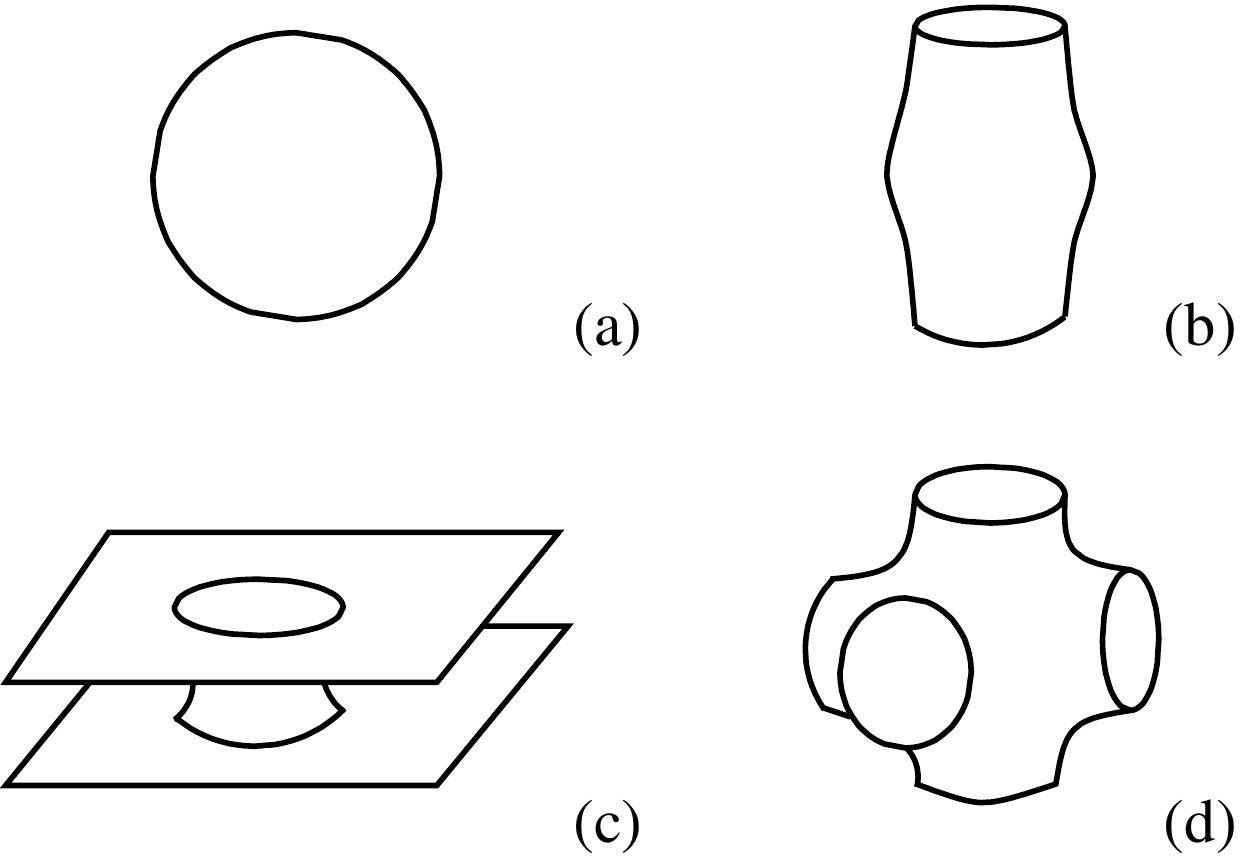}
\end{center}
\caption{Examples of the Fermi surfaces of rank 0, 1, 2, and 3, respectively.}
\label{Rang}
\end{figure}

 In addition to the topological genus, the Fermi surface $\, S_{F} \, $
is also characterized by the rank defined as the dimension of the image 
of the mapping
$$H_{1} (S_{F}) \,\,\, \rightarrow \,\,\, 
H_{1} (\mathbb{T}^{3}) $$
under the embedding $\, S_{F} \, \rightarrow \, \mathbb{T}^{3} \, $.
It is easy to see that the rank of the Fermi surface can take the values 
0, 1, 2 and 3 (Fig. \ref{Rang}). Let us say at once that we will consider 
here the Fermi surface rather complicated if its rank is equal to 3.
It can be shown that for the genus of the Fermi surface we must in this 
case also have the relation $\, g \geq 3 \, $.

 We now give a brief description of the structure of a complex Fermi surface 
in the presence of stable open trajectories of the system (\ref{MFSyst}).
Note that the presence of such a structure is a common property of the system
(\ref{MFSyst}) and follows from the rigorous topological results presented 
in the papers \cite{zorich1,dynn1,dynn3}. It will be most convenient for us 
to use the description presented in the paper \cite{dynn3} where all open 
trajectories of the system (\ref{MFSyst}), arising for a given dispersion 
relation $\, \epsilon ({\bf p}) \, $, were considered. For simplicity, we will 
assume here that the Fermi surface is represented by a single connected component 
having rank 3. Here we will be interested, first of all, in the structure of 
the Fermi surface for a direction of $\, {\bf B} \, $, lying in one of the 
Stability Zones $\, \Omega_{\alpha} \, $, for simplicity, we can assume also
that $\, {\bf B} \, $ has a generic direction (in particular, the plane 
orthogonal to $\, {\bf B} \, $ does not contain reciprocal lattice vectors).

\begin{figure}[t]
\begin{center}
\includegraphics[width=\linewidth]{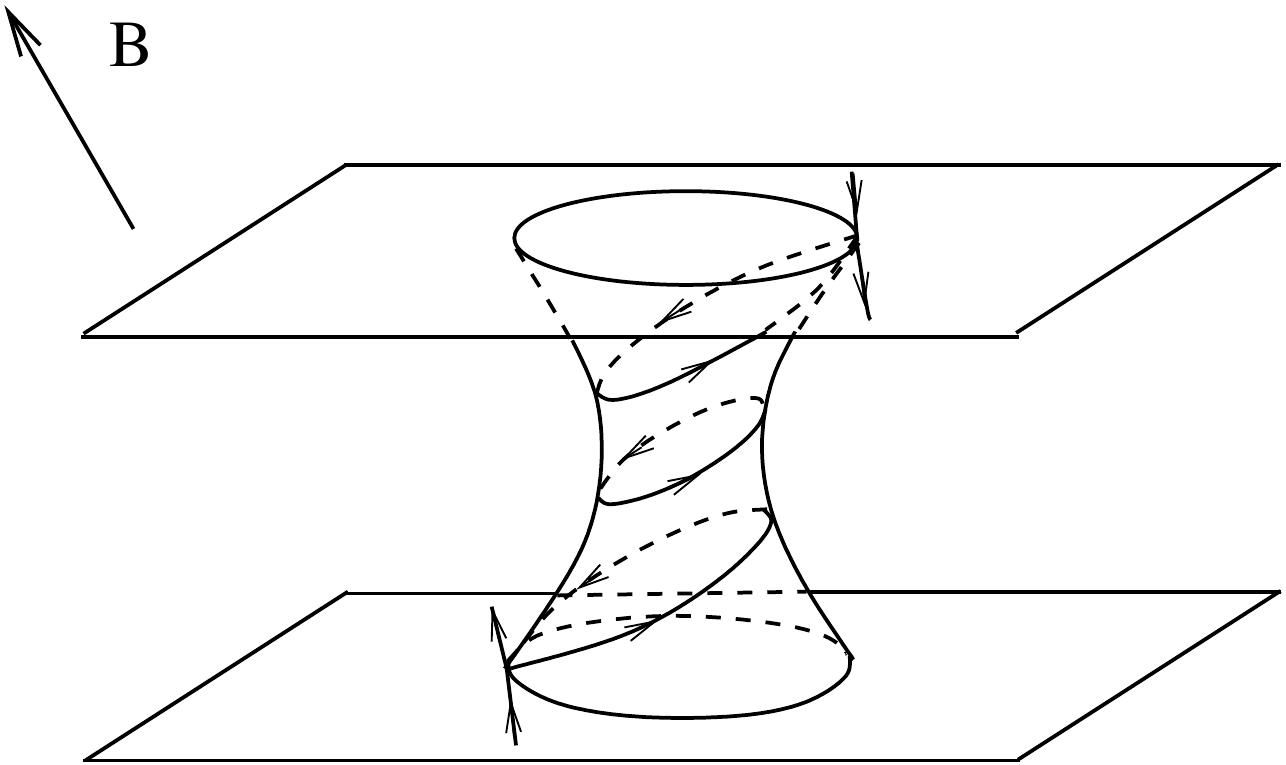}
\end{center}
\caption{Cylinder of closed trajectories of (\ref{MFSyst}),
bounded by singular closed trajectories.}
\label{CylindClosed}
\end{figure}

 Consider, following \cite{dynn3}, the set of nonsingular closed trajectories 
of the system (\ref{MFSyst}) with $\, {\bf B} / B \, \in \, \Omega_{\alpha} \, $. 
It is not difficult to see that this set is a finite set of (nonequivalent) 
cylinders bounded by singular closed trajectories of the system (\ref{MFSyst}) 
(Fig. \ref{CylindClosed}). We now remove all such cylinders of closed 
trajectories from the Fermi surface. The remaining part of the Fermi surface 
is thus a part filled with open trajectories of the system (\ref{MFSyst}). 
It is not difficult to see that the new surface thus obtained should consist 
of a finite number (nonequivalent) of connected components, which can be 
called ``carriers'' of open trajectories. Note also that the holes formed 
after removing the cylinders of closed trajectories can be ``filled'' with 
flat disks orthogonal to $\, {\bf B} \, $, so that a new (reduced) surface 
can also be viewed as a closed periodic surface in $\, {\bf p}$ - space.
The resulting surface can also be factorized by the reciprocal lattice vectors 
and be considered as a closed two-dimensional surface embedded in
$\, \mathbb{T}^{3} \, $.

 The most important property of the Fermi surface thus reduced in the presence 
of stable open trajectories of the system (\ref{MFSyst}) on it can be formulated 
as follows (\cite{zorich1,dynn1}):

 Each of the carriers of open trajectories represents a two-dimensional 
torus $\, \mathbb{T}^{2} \, $ embedded in $\, \mathbb{T}^{3} \, $ with 
a fixed mapping in homology uniquely defined for the entire Stability Zone
$\, \Omega_{\alpha} \, $.

\begin{figure}[t]
\begin{center}
\includegraphics[width=\linewidth]{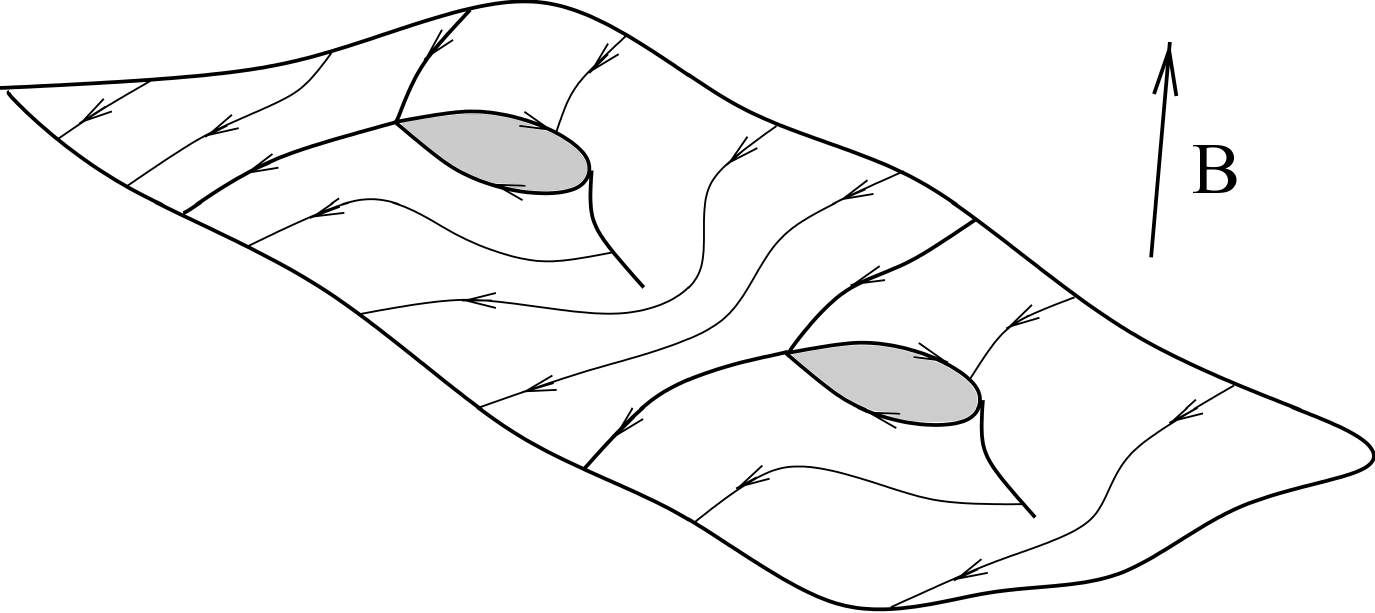}
\end{center}
\caption{The carrier of stable open trajectories of the system (\ref{MFSyst})
in the extended $\, {\bf p}$ - space.}
\label{IntPlane}
\end{figure}

 Coming back to the extended $\, {\bf p}$ - space, we can also formulate 
the above statement in the form:

 Each of the carriers of open trajectories in $\, {\bf p}$ - space represents 
an integral periodically deformed plane of a given direction, unchanged for the 
entire Stability Zone $\, \Omega_{\alpha} \, $ (Fig. \ref{IntPlane}).

 The total reduced Fermi surface thus represents an infinite set of integral
periodically deformed parallel planes of a fixed direction in the 
$\, {\bf p}$ - space. The number of non-equivalent integral planes (that is, 
representing different electron states) in the reduced Fermi surface is given 
by a finite even number equal to the number of two-dimensional tori 
$\, \mathbb{T}^{2} \, $ in the Brillouin zone ($\, \mathbb{T}^{3} \, $).

\begin{figure}[t]
\begin{center}
\includegraphics[width=\linewidth]{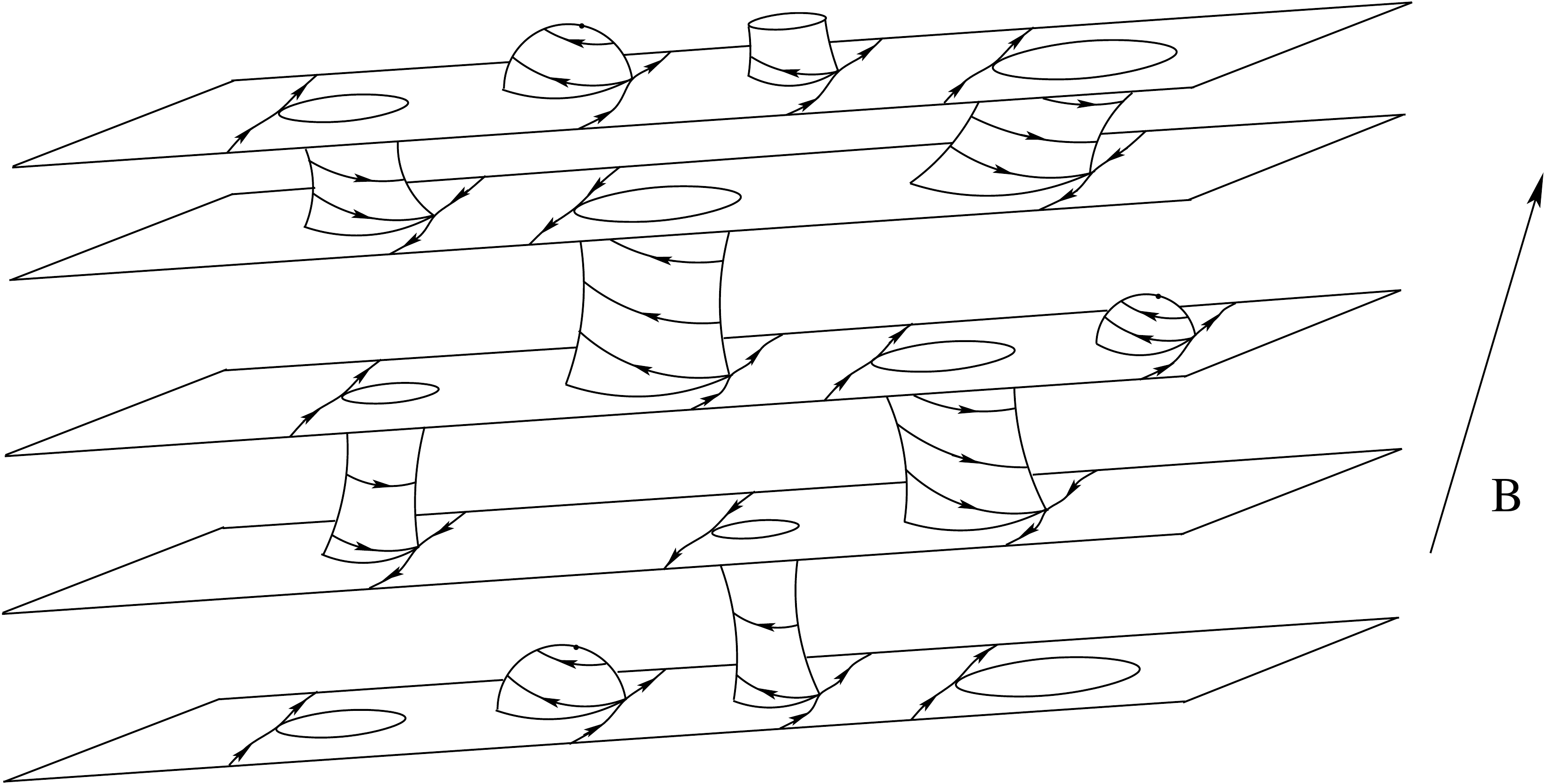}
\end{center}
\caption{Representation of the Fermi surface carrying stable open 
trajectories of the system (\ref{MFSyst}).}
\label{FullFermSurf}
\end{figure}

 It can be seen then that for generic directions of $\, {\bf B} \, $
lying in one of the Stability Zones $\, \Omega_{\alpha} \, $ the full 
Fermi surface is always representable as a set of parallel integral planes 
(with holes) carrying open trajectories of system (\ref{MFSyst}),
connected by components consisting only of closed trajectories of  
(\ref{MFSyst}). In our case, we will always assume that such components 
represent separate cylinders of closed trajectories, bounded by singular
closed trajectories (Fig. \ref{FullFermSurf}). In fact, more complex structure
of these components (consisting of several connected cylinders of closed
trajectories with additional singular points) can bring some additional 
features to the general picture (for example, non-simply-connectedness
of Stability Zones). However, such structure is extremely unlikely  
in a real situation, so we will not consider it here.

 The representation of the Fermi surface in the form of 
Fig. \ref{FullFermSurf} is generally not unique and, in particular, 
we have different such representations of the same surface for different 
Stability Zones $\, \Omega_{\alpha} \, $, $\, \Omega_{\beta} \, $. 
It should also be noted that the representation at Fig. \ref{FullFermSurf} 
reflects the main features of the topological structure of the system
(\ref{MFSyst}) on the Fermi surface and can be significantly more complex 
from a visual point of view.

 It is easy to see that the presented structure of the Fermi surface 
is stable with respect to small rotations of the direction of the magnetic 
field and explains the remarkable properties of the stable open 
trajectories of the system (\ref{MFSyst}) presented above. It can also 
be seen that all carriers of open trajectories can actually be divided into 
two classes according to the direction of motion along the trajectories 
(forward or backward) in the extended $\, {\bf p}$ - space. In addition, 
the cylinders of closed trajectories presented at Fig. \ref{FullFermSurf}
can also be divided into two classes according to the type of the trajectories 
(electron or hole) on them. 

\begin{figure}[t]
\begin{center}
\includegraphics[width=\linewidth]{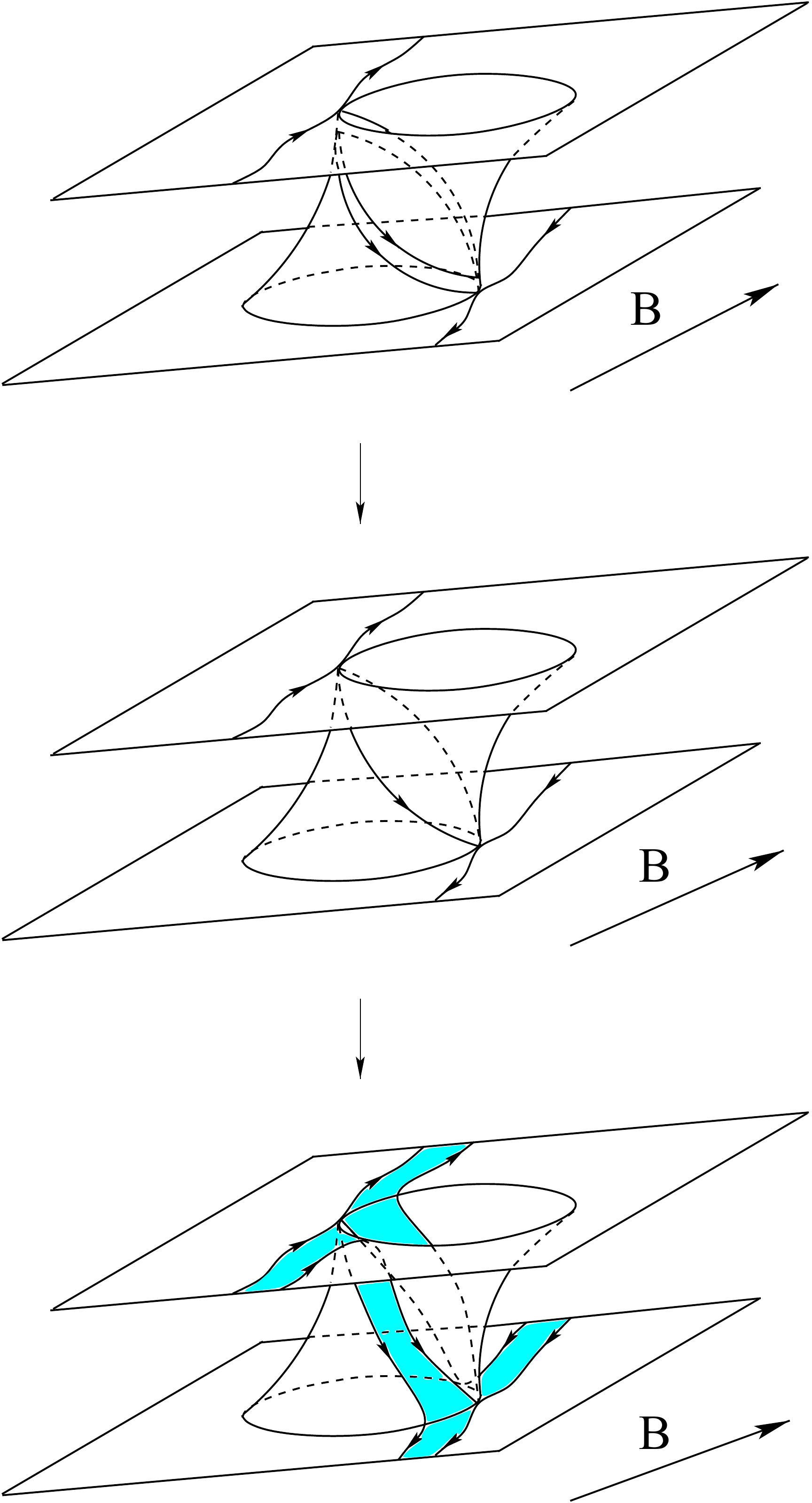}
\end{center}
\caption{The vanishing of the height of a cylinder of closed trajectories 
and the subsequent jump of open trajectories from one carrier to another at 
the boundary of a Stability Zone.}
\label{BoundCylind}
\end{figure}

 In accordance with the picture at Fig \ref{FullFermSurf}, the boundary of 
the Stability Zone $\, \Omega_{\alpha} \, $ must necessarily correspond 
to the vanishing of the height of one of the cylinders of closed trajectories 
(and its subsequent disappearance) for the corresponding directions of 
$\, {\bf B} \, $ (Fig. \ref{BoundCylind}). As it is easy to see, after 
crossing the boundary of a Stability Zone, open trajectories are able 
to ``jump'' from one carrier of open trajectories to another, which changes 
the overall structure of the trajectories on the Fermi surface.

\begin{figure*}
\begin{tabular}{lc}
\includegraphics[width=110mm]{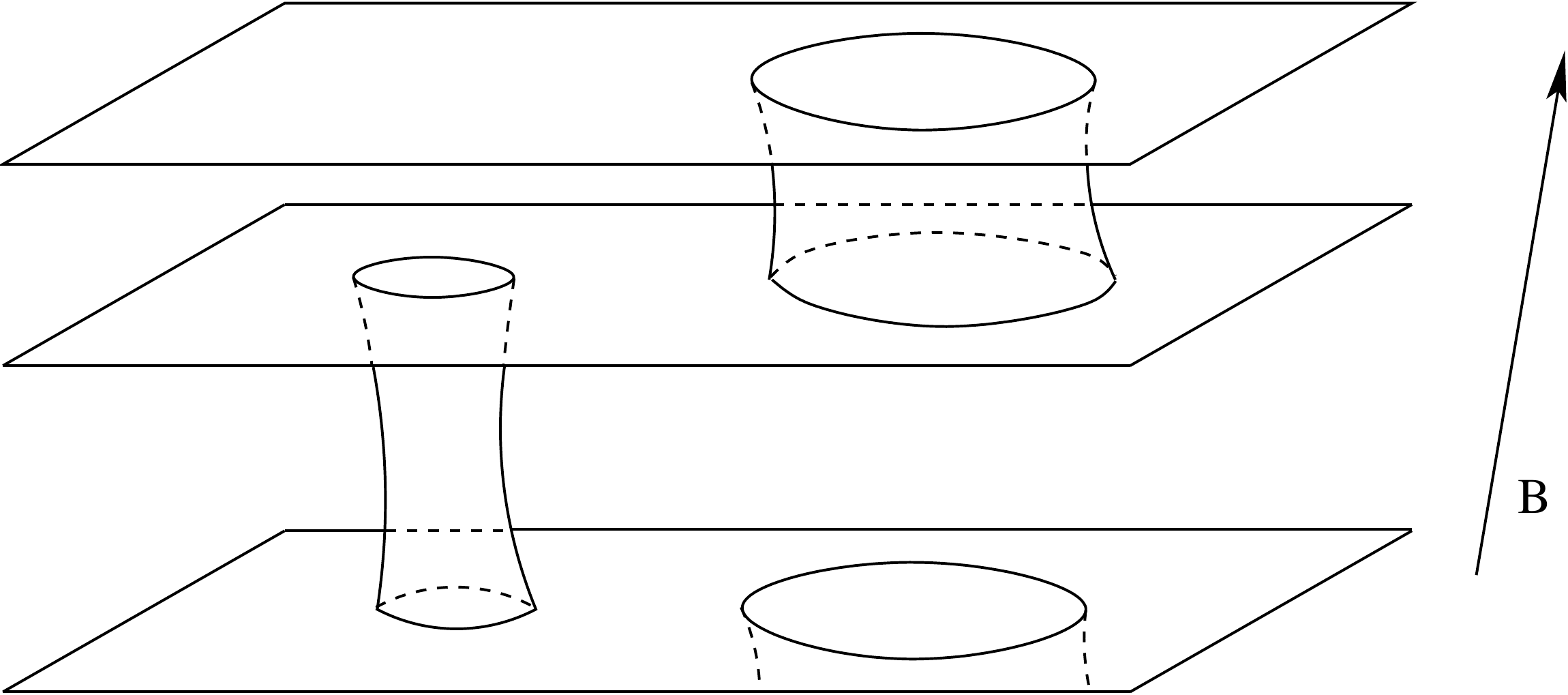}  &
\hspace{10mm}
\includegraphics[width=55mm]{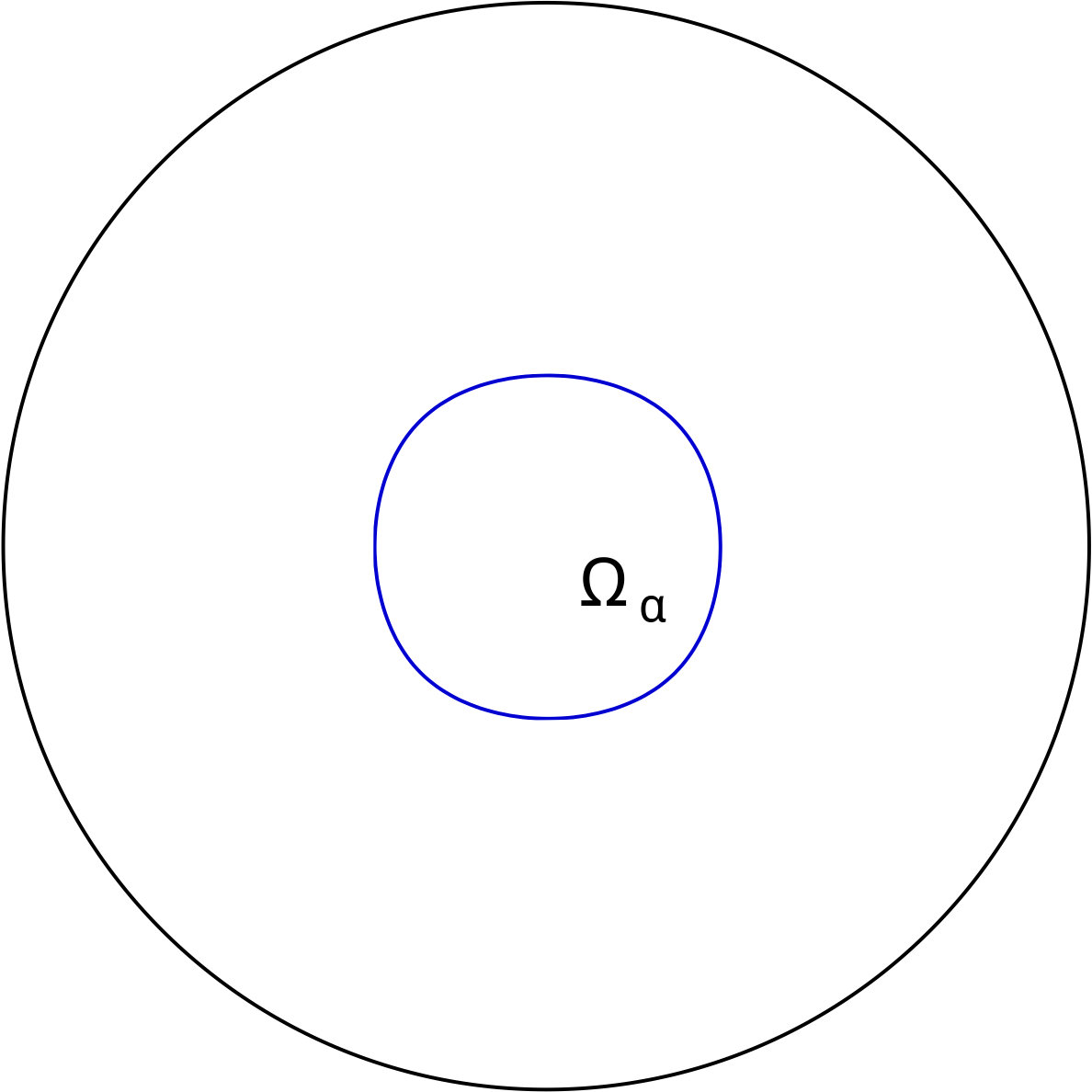}
\end{tabular}
\caption{The Fermi surface having a Stability Zone 
$\, \Omega_{\alpha} \, $ with the boundary defined by the disappearance 
of the same cylinder of closed trajectories at all its points.}
\label{SimpleFirstBound}
\end{figure*}

\begin{figure*}
\begin{tabular}{lc}
\includegraphics[width=110mm]{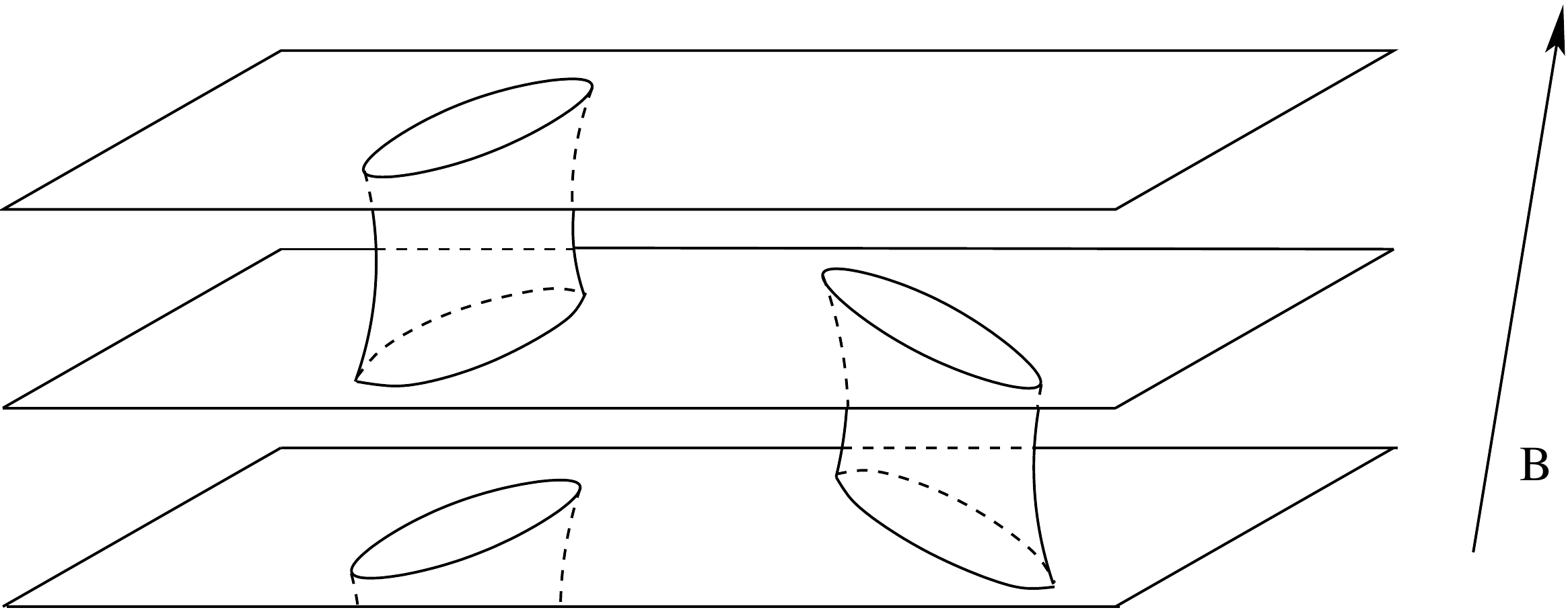}  &
\hspace{10mm}
\includegraphics[width=55mm]{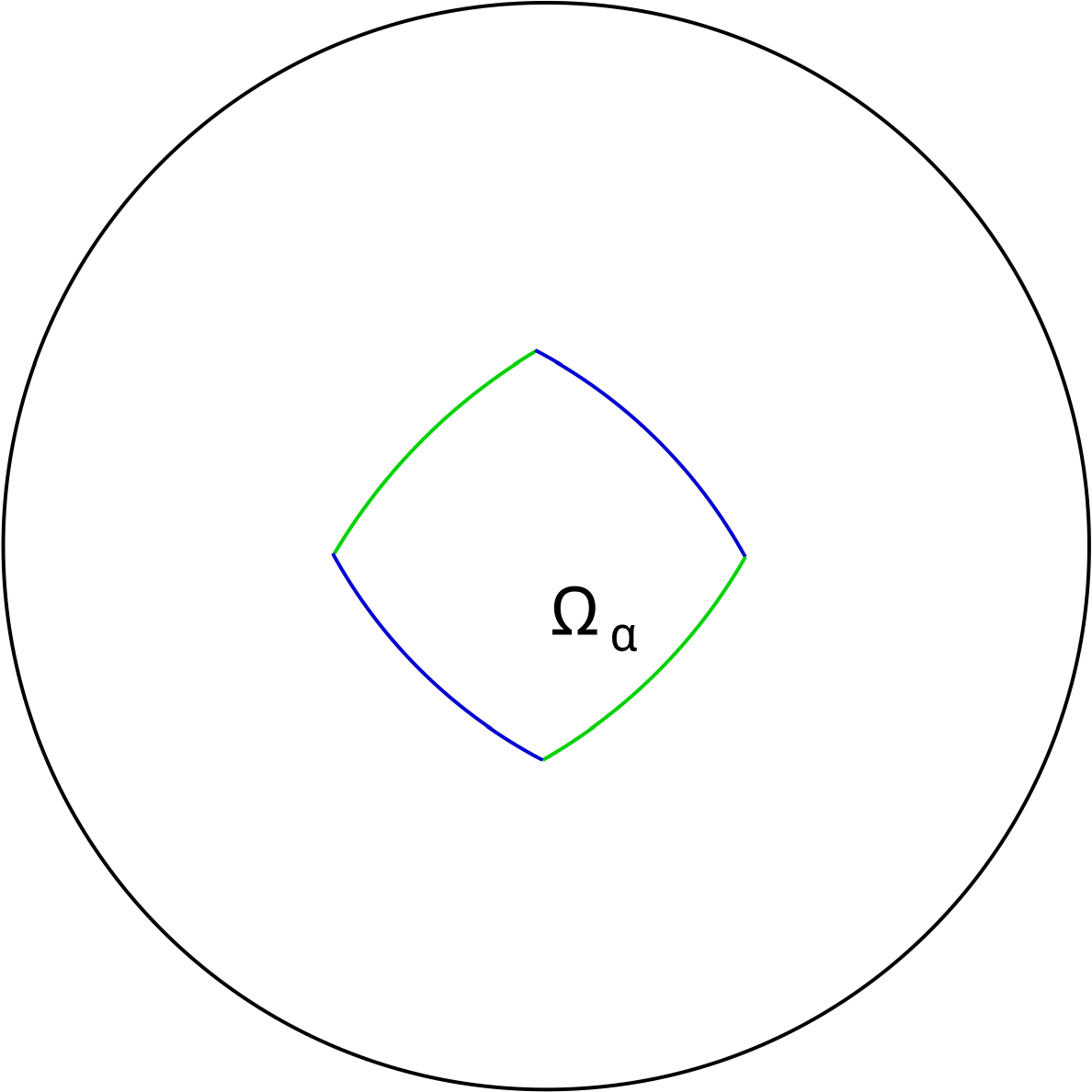}
\end{tabular}
\caption{The Fermi surface having a Stability Zone 
$\, \Omega_{\alpha} \, $ with the boundary defined by the disappearance
of different cylinders of different types at different parts of it.}
\label{CompoundFirstBound}
\end{figure*}

\begin{figure*}
\begin{center}
\includegraphics[width=175mm]{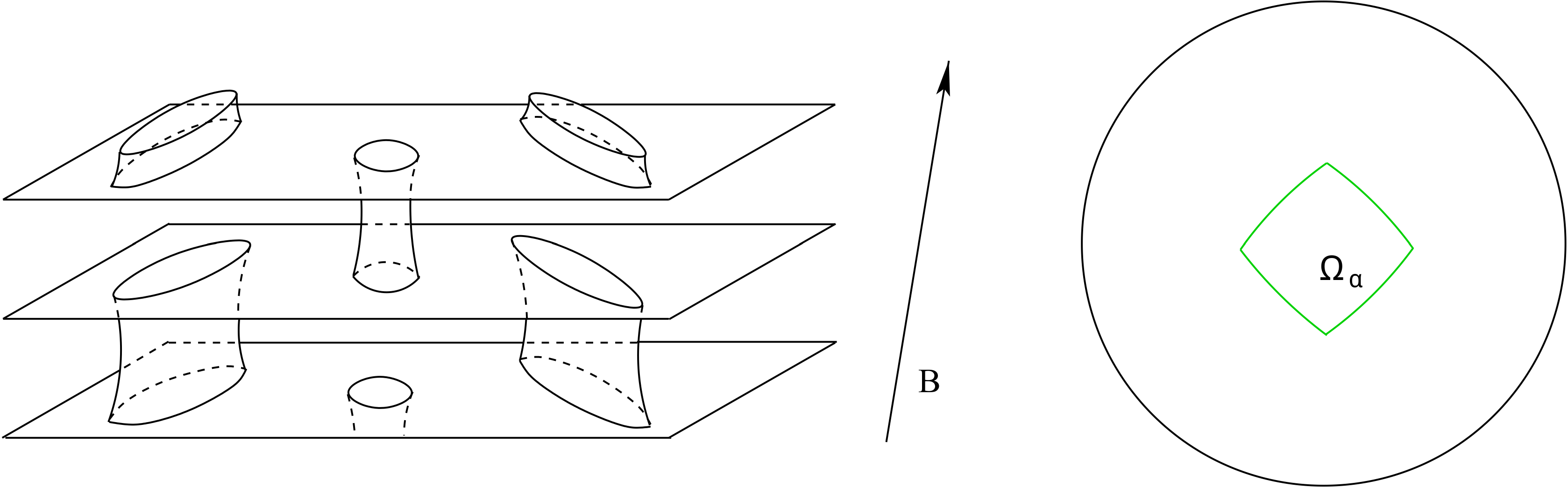}  
\end{center}
\caption{The Fermi surface having a Stability Zone 
$\, \Omega_{\alpha} \, $ with the boundary defined by the disappearance
of different cylinders of the same type at different parts of it.}
\label{CompoundOneType}
\end{figure*}

 It can also be seen that the full boundary of a Stability Zone can be 
either ``simple'', i.e. correspond to the disappearance of the same cylinder 
of closed trajectories at all its points (Fig. \ref{SimpleFirstBound}),
or  ``compound'', i.e. correspond to the disappearance of different 
cylinders of closed trajectories at its different parts  
(Fig. \ref{CompoundFirstBound}, \ref{CompoundOneType}).
In addition, it can be seen that in the case of a compound boundary
of a Stability Zone we can have both the situation when cylinders of closed 
trajectories of different types disappear at its different parts 
(Fig. \ref{CompoundFirstBound}), and also the situation when different 
cylinders of closed trajectories of the same type disappear at its different 
parts (Fig. \ref{CompoundOneType}).

 We note here that the type of vanishing cylinder of closed trajectories 
also plays an important role in this situation in the description of 
galvanomagnetic phenomena for the corresponding directions of 
$\, {\bf B} \, $. Thus, in particular, it determines the value of 
the Hall conductivity in a metal outside the Zone $\, \Omega_{\alpha} \, $ 
near the corresponding part of its boundary in the limit
$\, \omega_{B} \tau \, \rightarrow \, \infty \, $.
As was mentioned in \cite{SecBound}, the presence of at least one 
Stability Zone with a compound boundary, determined by the disappearance 
of cylinders of different types at its different parts, should lead to the 
presence of different values of Hall conductivity in different areas outside 
the Stability Zones and, as a result of this, to rather complex structure of 
the angular diagram as a whole.

\section{Features of angular diagrams defined for a dispersion law}
\setcounter{equation}{0}

 We now give a general description of the diagrams defined for the entire 
dispersion law $\, \epsilon ({\bf p}) \, $ (\cite{dynn3}). According to
\cite{dynn3}, the possibility of constructing such angular diagrams is based 
on the following important statements regarding the trajectories of the system
(\ref{MFSyst}):

\vspace{1mm}

\noindent
1) Consider an arbitrary periodic dispersion relation
$$\epsilon_{min} \,\,\, \leq \,\,\, \epsilon ({\bf p}) 
\,\,\, \leq \,\,\, \epsilon_{max} $$
and fix a direction of the magnetic field $\, {\bf B} \, $.
Then all energy levels $\, \epsilon ({\bf p}) = const \, $,
at which we have unclosed trajectories of the system (\ref{MFSyst}),
represent either a closed interval
$\, [ \epsilon_{1} ({\bf B}/B) \, , \, \epsilon_{2} ({\bf B}/B) ] \, $,
or a single point
$\, \epsilon_{0} \, = \, \epsilon_{1} ({\bf B}/B) \, = \,
\epsilon_{2} ({\bf B}/B) \, $.

\vspace{1mm}

\noindent

2) For generic directions of $\, {\bf B} \, $ (in particular, directions 
of maximal irrationality) the values $\, \epsilon_{1} ({\bf B}/B) \, $ and 
$\, \epsilon_{2} ({\bf B}/B) \, $ coincide with the values of some continuous 
functions $\, \tilde{\epsilon}_{1} ({\bf B}/B) \, $ and
$\, \tilde{\epsilon}_{2} ({\bf B}/B) \, $, defined everywhere on the unit 
sphere. For special directions of $\, {\bf B} \, $, corresponding to the 
appearance of strictly periodic open trajectories of system (\ref{MFSyst}),
the values $\, \epsilon_{1} ({\bf B}/B) \, $ and 
$\, \epsilon_{2} ({\bf B}/B) \, $ may differ from the values
$\, \tilde{\epsilon}_{1} ({\bf B}/B) \, $ and
$\, \tilde{\epsilon}_{2} ({\bf B}/B) \, $ and then we have the relations
$$\epsilon_{1} ({\bf B}/B) \,\,\, \leq \,\,\, 
\tilde{\epsilon}_{1} ({\bf B}/B) \,\,\, , \quad 
\epsilon_{2} ({\bf B}/B) \,\,\, \geq \,\,\, 
\tilde{\epsilon}_{2} ({\bf B}/B) $$

\vspace{1mm}

\noindent
3) In the case
$\, \tilde{\epsilon}_{2} ({\bf B}/B) \, > \, \tilde{\epsilon}_{1} ({\bf B}/B) \, $
for generic directions of $\, {\bf B} \, $ all non-singular open trajectories 
of system (\ref{MFSyst}) (at all energy levels) have the form shown at 
Fig. \ref{StableTr} with the same mean direction defined by the intersection 
of the plane orthogonal to $\, {\bf B} \, $ and some integral plane
$\, \Gamma \, $.

\vspace{1mm}

\noindent
4) The property
$\, \tilde{\epsilon}_{2} ({\bf B}/B) \, > \, \tilde{\epsilon}_{1} ({\bf B}/B) \, $,
as well as the integral plane $\, \Gamma \, $, are locally stable with respect to  
small rotations of $\, {\bf B} \, $ and thus define a certain Stability Zone
$\, \Omega^{*} \, $ on the angular diagram. The Zone $\, \Omega^{*} \, $ 
represents  an open domain with a piecewise smooth boundary on 
$\, \mathbb{S}^{2} \, $ and on the boundary $\, \partial \Omega^{*} \, $ 
we have the relation
$\, \tilde{\epsilon}_{2} ({\bf B}/B) \, = \, \tilde{\epsilon}_{1} ({\bf B}/B) \, $.

\vspace{1mm}

\noindent
5) The union of all the Stability Zones $\, \Omega^{*}_{\alpha} \, $ 
is everywhere dense on the unit sphere $\, \mathbb{S}^{2} \, $, and the sphere
$\, \mathbb{S}^{2} \, $ can contain either one Stability Zone 
$\, \Omega^{*} \, $, or infinitely many Zones $\, \Omega^{*}_{\alpha} \, $.

\vspace{1mm}

\begin{figure}[t]
\includegraphics[width=\linewidth]{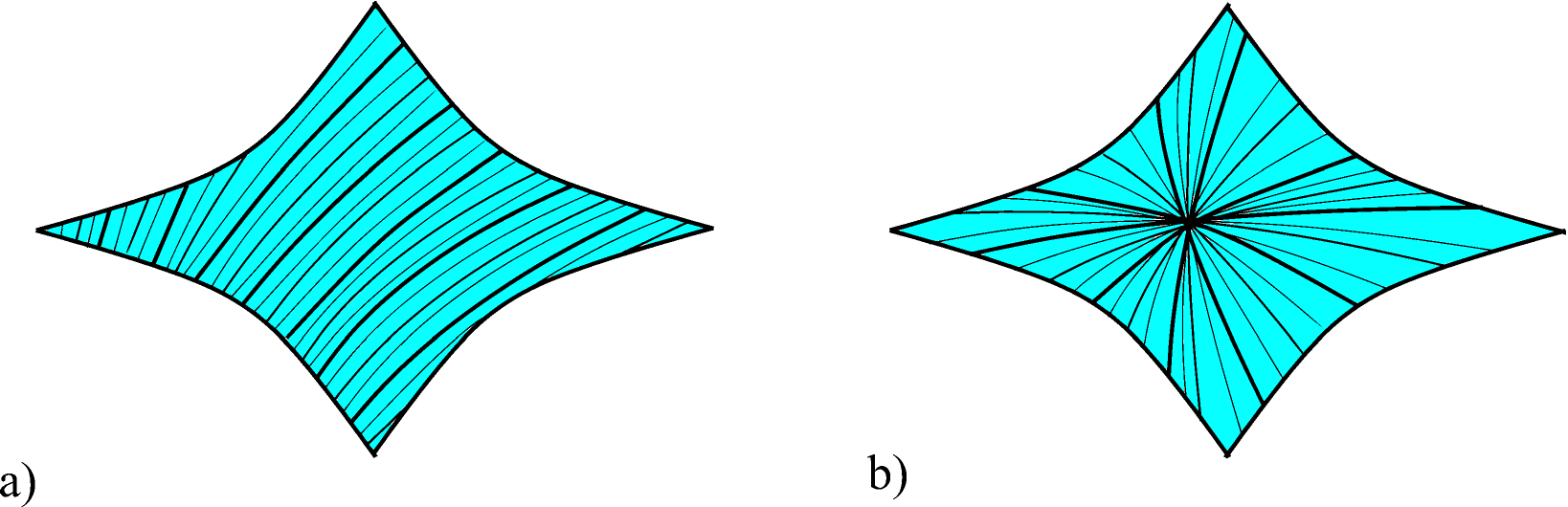} 
\caption{Everywhere dense set of directions of $\, {\bf B} \, $ in the Zone
$\, \Omega^{*}_{\alpha} \, $, corresponding to the appearance of periodic 
trajectories, in the cases when
$\, {\bf M}_{\alpha} / |{\bf M}_{\alpha}|
\, \notin \, \Omega^{*}_{\alpha} \, $ and 
$\, {\bf M}_{\beta} / |{\bf M}_{\beta}|
\, \in \, \Omega^{*}_{\alpha} \, $.}
\label{DenseNet}
\end{figure}

 It is easy to see that the angular diagrams for the dispersion law can 
naturally be divided into very simple (single Stability Zone) and very complex 
(infinitely many Stability Zones). Here we will be interested mainly in diagrams 
of the second type. The complexity of the angular diagrams of the second type is 
due, in particular, to the following features of their general structure
(see \cite{dynn3}):

 Namely, consider some Stability Zone 
$\, \Omega^{*}_{\alpha} \, $ corresponding to a certain integral plane
$\, \Gamma_{\alpha} \, $  (or topological numbers
$\, {\bf M}_{\alpha} \, = \, 
(M^{1}_{\alpha}, M^{2}_{\alpha}, M^{3}_{\alpha})$) and having a piecewise smooth 
boundary on $\, \mathbb{S}^{2} \, $. As already indicated, all non-singular open 
trajectories of the system (\ref{MFSyst}) have at 
$\, {\bf B}/B \, \in \, \Omega^{*}_{\alpha} \, $ the common mean direction given 
by the intersection of the plane $\, \Gamma_{\alpha} \, $ and the plane orthogonal 
to $\, {\bf B} \, $. It is not difficult to see that, whenever such an intersection 
has an integer direction, the corresponding trajectories are periodic in the
$\, {\bf p}$ - space. It can also be seen that the corresponding directions of
$\, {\bf B} \, $ form in $\, \Omega^{*}_{\alpha} \, $ an everywhere dense set 
consisting of segments of the big circles (Fig. \ref{DenseNet}). In addition, 
one can see that the corresponding set is also everywhere dense on the boundary 
of the domain $\, \Omega^{*}_{\alpha} \, $, forming on it a similarity of the set 
of rational numbers. We will call the points of the described set on the boundary
of $\, \Omega^{*}_{\alpha} \, $ ``rational'' boundary points.

\begin{figure}[t]
\includegraphics[width=\linewidth]{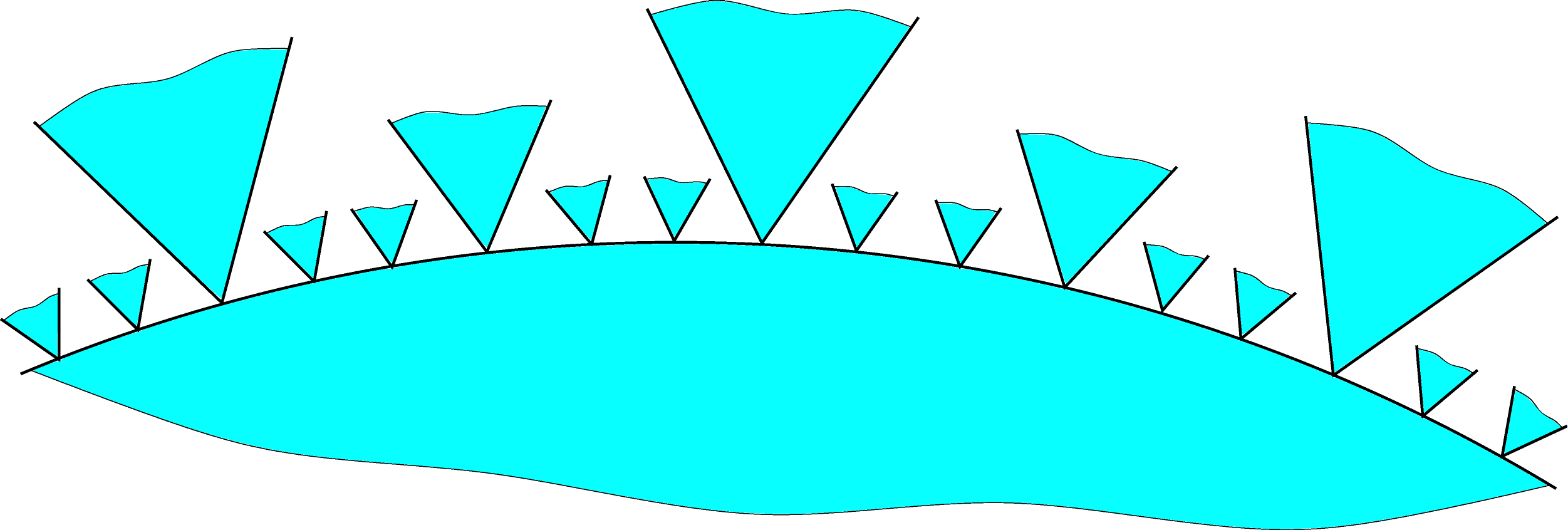} 
\caption{The set of Zones $\, \Omega^{*}_{\beta} \, $ adjoining the boundary
of a Zone $\, \Omega^{*}_{\alpha} \, $  at each of its ``rational'' points.}
\label{Adjacent}
\end{figure}

  An important feature of angular diagrams for the dispersion law is the fact 
that at each ``rational'' point of the boundary of any of the Stability Zones
$\, \Omega^{*}_{\alpha} \, $ another Stability Zone $\, \Omega^{*}_{\beta} \, $,
corresponding to another integral plane $\, \Gamma_{\beta} \, $,
is adjacent to this Zone (Fig. \ref{Adjacent}). The corresponding junction point 
represents a corner point of the boundary of $\, \Omega^{*}_{\beta} \, $,
and the topological numbers
$\, {\bf M}_{\alpha} \, = \, 
(M^{1}_{\alpha}, M^{2}_{\alpha}, M^{3}_{\alpha}) \, $ and
$\, {\bf M}_{\beta} \, = \, 
(M^{1}_{\beta}, M^{2}_{\beta}, M^{3}_{\beta}) \, $, 
as well as the corresponding direction of $\, {\bf B} \, $, 
are connected by the relation
$${\bf B} \,\,\, = \,\,\, 
\eta \, {\bf M}_{\alpha} \,\, + \,\, \theta \, {\bf M}_{\beta} \,\,\, ,$$
where $\, | \eta / \theta | \, < \, 1 \, $.

 It can be seen, therefore, that the structure of the angular diagrams for 
the dispersion relation, containing more than one Stability Zone, is extremely 
complex.

 To avoid ambiguity, we will always denote here by $\, \Omega^{*}_{\alpha} \, $ 
a complete (open) Stability Zone and by $\, {\overline{\Omega^{*}_{\alpha}}} \, $ 
- the Zone $\, \Omega^{*}_{\alpha} \, $ together with the boundary. The set
$${\cal D}^{*} \,\,\, = \,\,\, \mathbb{S}^{2}  \, \left\backslash  \, \cup \,
{\overline{\Omega^{*}_{\alpha}}} \right. $$
represents the set of directions of $\, {\bf B} \, $, for which open 
trajectories of the system (\ref{MFSyst}) exist on a single energy level 
and have generically a chaotic behavior in planes orthogonal to $\, {\bf B} \, $. 
According to the conjecture of S.P. Novikov (see \cite{DynSyst}),
the fractal dimension of this set for generic dispersion relations is 
strictly less than 2. We note here that the study of various features of 
both the set $\, {\cal D}^{*} \, $ and the corresponding chaotic trajectories
of (\ref{MFSyst}) is currently a rapidly developing area of the theory of 
dynamical systems
(see for example \cite{Tsarev,dynn2,dynn3,zorich2,Zorich1996, 
ZorichAMS1997, ZhETF1997,zorich3,DeLeo1,DeLeo2,DeLeo3,ZorichLesHouches,
DeLeoDynnikov1,DeLeoDynnikov2,Skripchenko1,Skripchenko2,DynnSkrip1,
DynnSkrip2,AvilaHubSkrip1,AvilaHubSkrip2,DeLeo2017,TrMian}).

 The above description of the angle diagrams for system (\ref{MFSyst}), 
built in \cite{dynn3}, is based on the results obtained in the works 
\cite{zorich1, dynn1} and leading to the above-described representation 
of the energy surfaces (Fig. \ref{FullFermSurf}) containing stable open 
trajectories of system (\ref{MFSyst}). Note here also that the boundary 
of a Stability Zone $\, \Omega^{*}_{\alpha} \, $, defined for the entire 
dispersion law, corresponds to the simultaneous disappearance of two cylinders 
of closed trajectories (of electron and hole type) connecting the carriers
of open trajectories. Here one can see the difference with the situation on 
the angular diagrams defined for a fixed value of energy, where each segment 
of the boundary of a Stability Zone corresponds to the disappearance of 
a cylinder of closed trajectories of a certain type. In addition, as was 
noted in \cite{dynn3}, the boundaries of different Stability Zones defined 
for a fixed energy value have no common points in the generic case, which 
indicates in general a somewhat less complex structure of such diagrams.

 In the next chapter, we will give a more detailed consideration of angular 
diagrams of conductivity corresponding to some fixed energy level (Fermi level) 
$\, \epsilon_{F} \, $, which corresponds to the situation of a normal metal.
As we will see, most of the consideration will need to be devoted to 
the study of the trajectories of system (\ref{MFSyst}) on the complex 
Fermi surfaces discussed in the previous chapter. This study will, 
in particular, contain a description of the most complex diagrams that arise 
in this situation and the investigation of how such diagrams can approach 
those described above in terms of complexity.

\section{The complexity of the angular diagrams defined for a fixed 
Fermi surface, and their relation to the diagrams for the dispersion law}
\setcounter{equation}{0}

 As we indicated above, each of the Stability Zones $\, \Omega_{\alpha} \, $, 
defined for a fixed Fermi surface, lies inside some (larger) Stability Zone
$\, \Omega^{*}_{\alpha} \, $, corresponding to the full dispersion relation
$\, \epsilon ({\bf p}) \, $. For directions 
$\, {\bf B}/B \, \in \, \Omega_{\alpha} \, $ we thus have stable open 
trajectories lying on the Fermi surface and corresponding to topological 
numbers $\, (M^{1}_{\alpha}, M^{2}_{\alpha}, M^{3}_{\alpha}) \, $.

\begin{figure}[t]
\includegraphics[width=\linewidth]{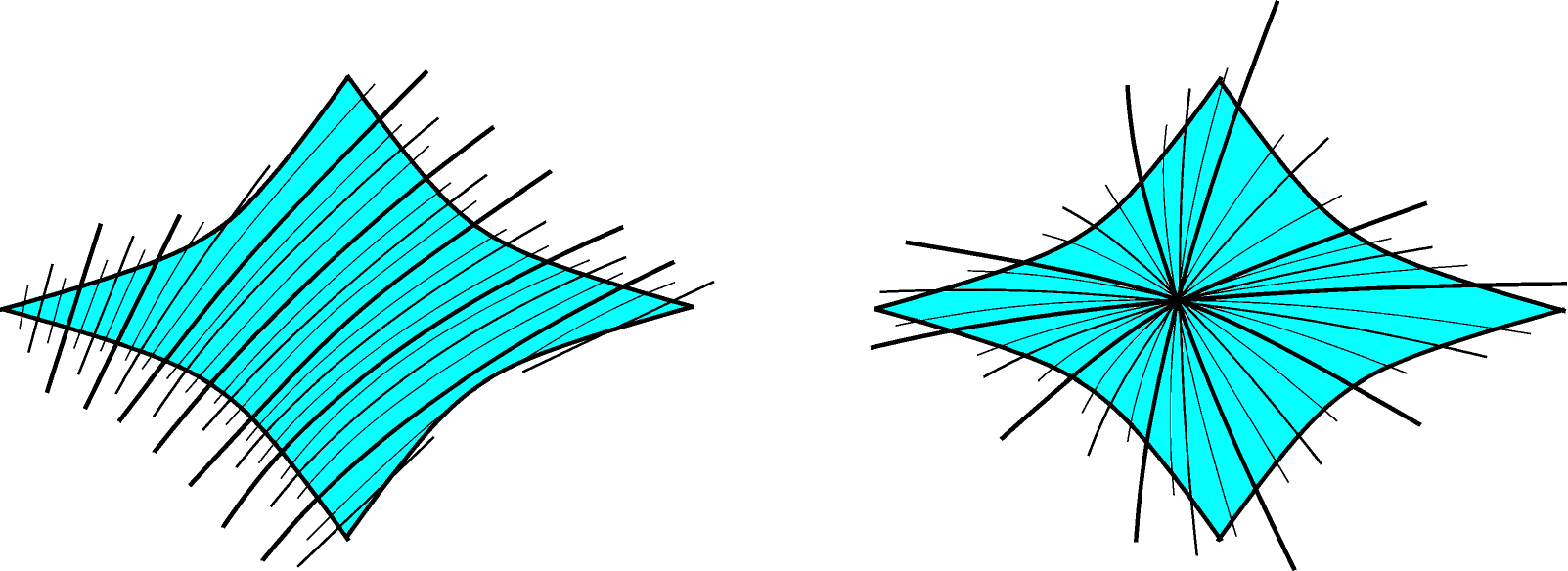} 
\caption{The set of directions of $\, {\bf B} \, $ near Stability Zones, 
corresponding to the appearance of (unstable) periodic trajectories on 
the Fermi surface.}
\label{DirOutStabZone}
\end{figure}

\begin{figure}[t]
\includegraphics[width=\linewidth]{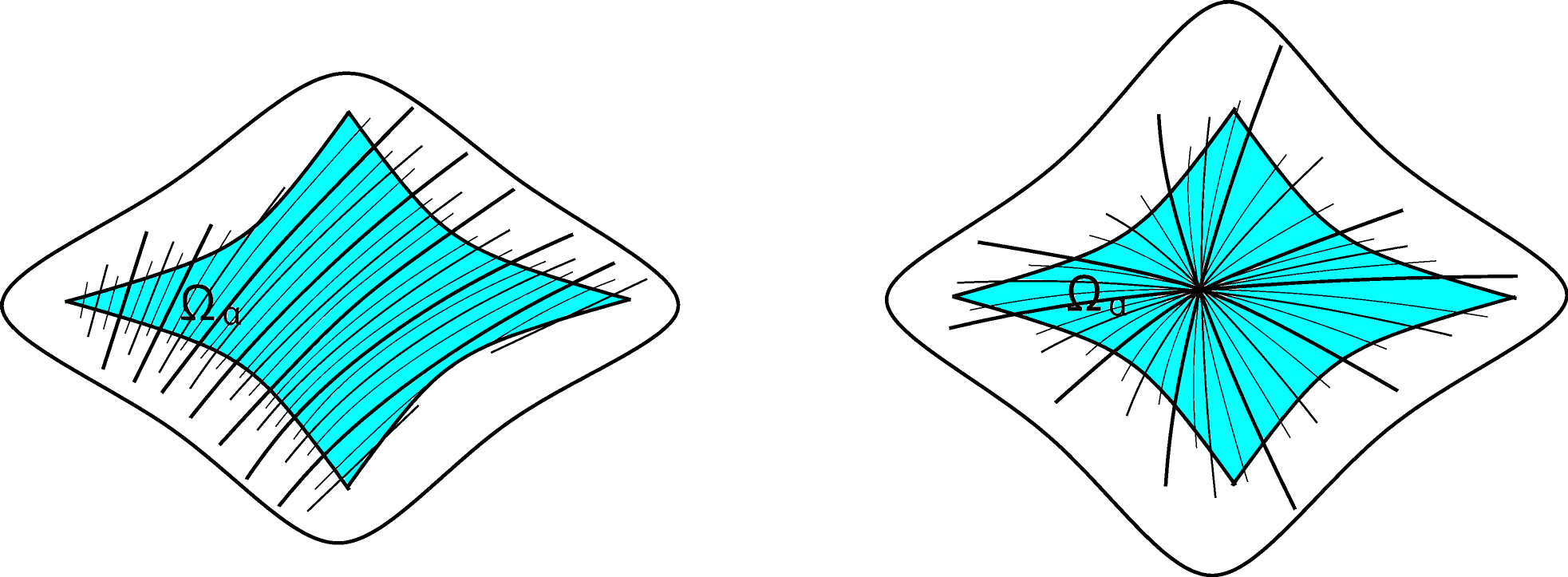} 
\caption{``Experimentally observable'' Stability Zones in experiments on 
direct measurements of conductivity in strong magnetic fields.}
\label{ExtStabZone}
\end{figure}

  We note here at once one feature of the Stability Zones for a fixed Fermi 
surface, which somewhat distinguishes them from the Stability Zones for the 
entire dispersion law. As we have already said, the Stability Zone
$\, \Omega_{\alpha} \, $ defines a set of directions of $\, {\bf B} \, $, 
for which there exist stable open trajectories on the Fermi surface 
corresponding to fixed topological numbers
$\, (M^{1}_{\alpha}, M^{2}_{\alpha}, M^{3}_{\alpha}) \, $.
We can, however, specify additional directions of $\, {\bf B} \, $,
for which there are open trajectories of the system (\ref{MFSyst})
on the Fermi surface that are not stable but are also associated with 
the Zone $\, \Omega_{\alpha} \, $. Such trajectories are periodic and 
appear on the continuations of the segments of large circles described above, 
beyond $\, \Omega_{\alpha} \, $ (Fig. \ref{DirOutStabZone}). It can be shown 
that such trajectories always appear near the boundary of a Zone
$\, \Omega_{\alpha} \, $, which follows from the above representation 
of the Fermi surface (Fig. \ref{FullFermSurf}) for directions
$\, {\bf B}/B \, \in \, \Omega_{\alpha} \, $. The net of the corresponding 
directions of $\, {\bf B} \, $  is everywhere dense on the boundary of
$\, \Omega_{\alpha} \, $, however, the lengths of the additional segments 
of the large circles decrease rapidly with increasing of the integers
$\, (m_{1}, m_{2}, m_{3}) \, $, which determine the direction of the 
corresponding periodic trajectories in the $\, {\bf p}$ - space.
As a consequence, the described nets of additional directions of
$\, {\bf B} \, $ have a finite density at any finite distance from the 
boundary of $\, \Omega_{\alpha} \, $. As was shown in \cite{AnProp}, 
the presence of (unstable) periodic trajectories, as well as very long 
closed trajectories of system (\ref{MFSyst}) near the boundaries of
$\, \Omega_{\alpha} \, $ actually leads to rather complex conductivity 
behavior for the corresponding directions of $\, {\bf B} \, $ and, 
in particular, makes impossible the observation of the exact boundary 
of a Stability Zone in direct measurements of conductivity even 
in rather strong magnetic fields. So, in reality, the 
``experimentally observed'' Stability Zone in such experiments does not 
coincide with the exact mathematical Zone $\, \Omega_{\alpha} \, $ 
(Fig. \ref{ExtStabZone}) and depends on the maximal values of external 
magnetic fields reachable in the experiment.

 At the same time, the exact boundaries of the mathematical Stability Zones
$\, \Omega_{\alpha} \, $ are in fact also observable experimentally with 
a special experimental setup. In particular, as was shown in \cite{CyclRes}, 
they can be determined by observing the picture of oscillation phenomena 
(classical or quantum) in strong magnetic fields. In this paper, we will be 
interested in the distribution of precisely exact mathematical Stability Zones 
on an angular diagram of conductivity. 

 As can be seen from the above picture, which appears on the Fermi surface 
in the case of presence of stable open trajectories of the system (\ref{MFSyst}),
for each of the Stability Zones $\, \Omega_{\alpha} \, $ it is also natural to 
introduce the ``second boundary'' on the angular diagram. The second boundary of 
the Zone $\, \Omega_{\alpha} \, $ is determined by the disappearance of one more 
cylinder of closed trajectories at Fig. \ref{FullFermSurf} and the appearance of 
jumps of trajectories of (\ref{MFSyst}) between pairs of ``merged'' carriers of 
open trajectories defined for the Zone $\, \Omega_{\alpha} \, $. 
It can be seen that in the region $\, \Omega^{\prime}_{\alpha} \, $,
which lies between the first and second boundaries of the Stability Zone,  
there can not be open trajectories of the system (\ref{MFSyst}), other than 
the unstable periodic trajectories described above. The domain
$\, \Omega^{\prime}_{\alpha} \, $ can be called the ``derivative'' of 
the Stability Zone $\, \Omega_{\alpha} \, $, since the behavior of trajectories
of the system (\ref{MFSyst}) in this area is actually determined by the structure 
of this system in the Zone $\, \Omega_{\alpha} \, $. We note here that the  
picture of the behavior of conductivity, as well as the oscillation phenomena, 
in the region $\, \Omega^{\prime}_{\alpha} \, $ is, in general, rather complicated
(see \cite{SecBound}). As well as for the first boundary of a Stability Zone
$\, \Omega_{\alpha} \, $, the position of its second boundary is also most 
convenient to determine with the help of the study of oscillation phenomena 
in strong magnetic fields (\cite{SecBound}).

\begin{figure*}
\begin{center}
\includegraphics[width=175mm]{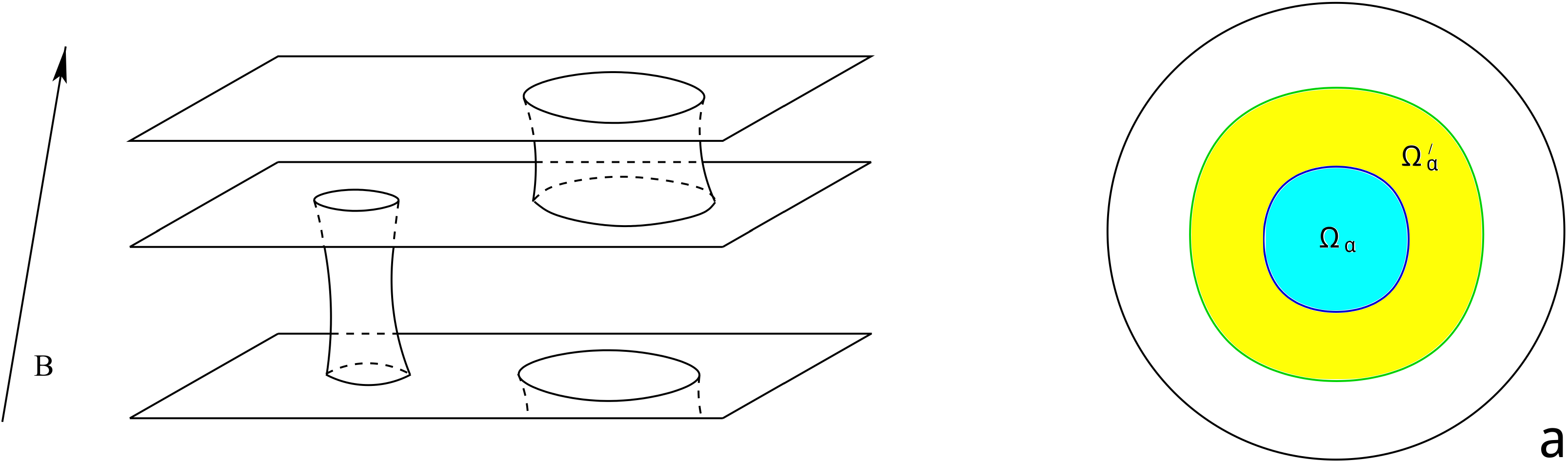} 
\end{center}
\begin{center}
\includegraphics[width=175mm]{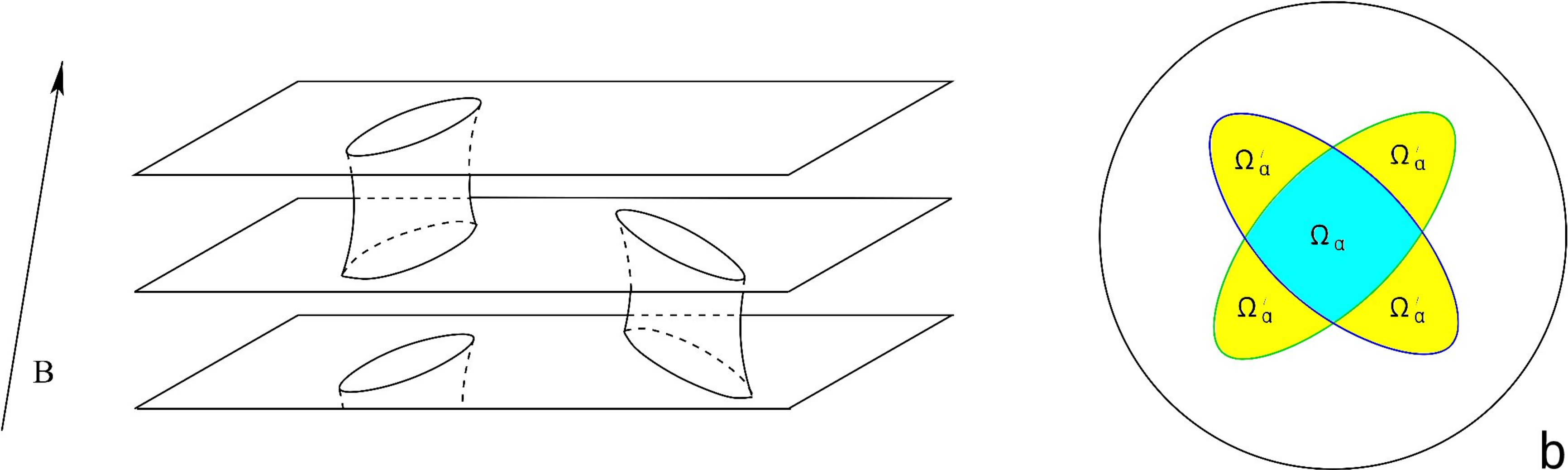} 
\end{center}
\begin{center}
\includegraphics[width=175mm]{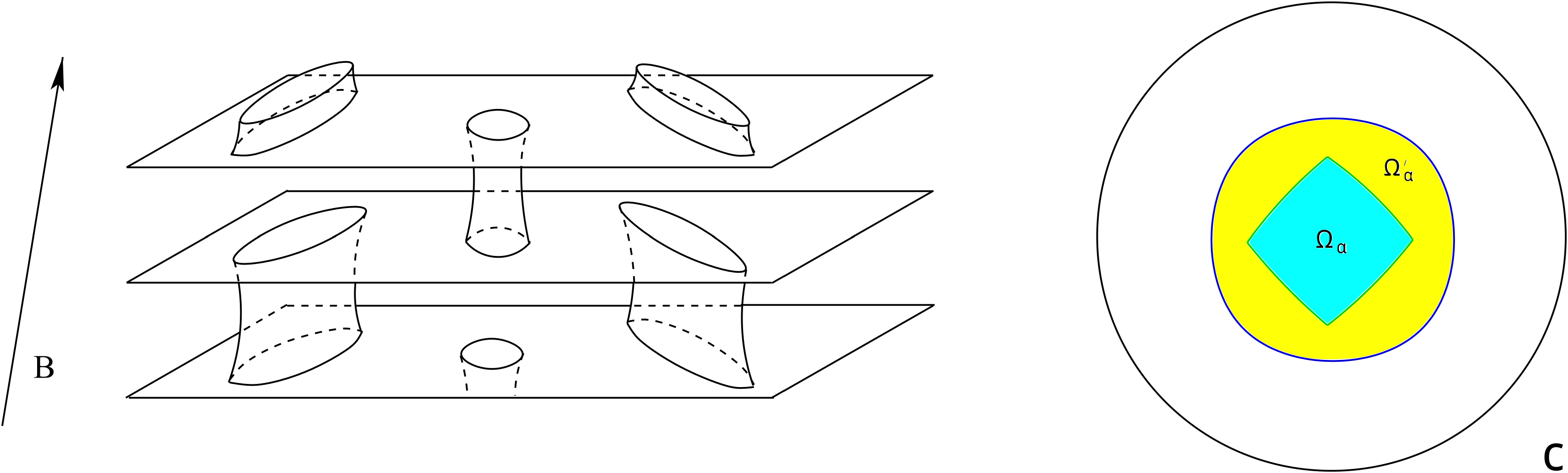}
\end{center}
\caption{Internal and external boundaries bounding the domains
$\, \Omega^{\prime}_{\alpha} \, $.}
\label{SecBounds}
\end{figure*}

\begin{figure*}
\begin{center}
\includegraphics[width=175mm]{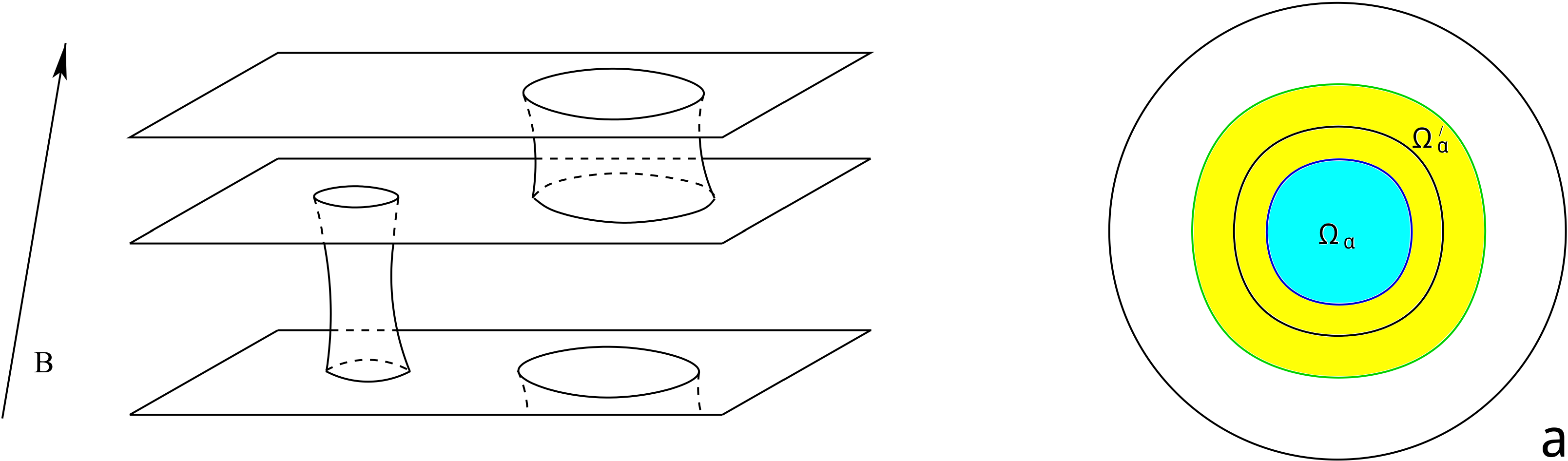} 
\end{center}
\begin{center}
\includegraphics[width=175mm]{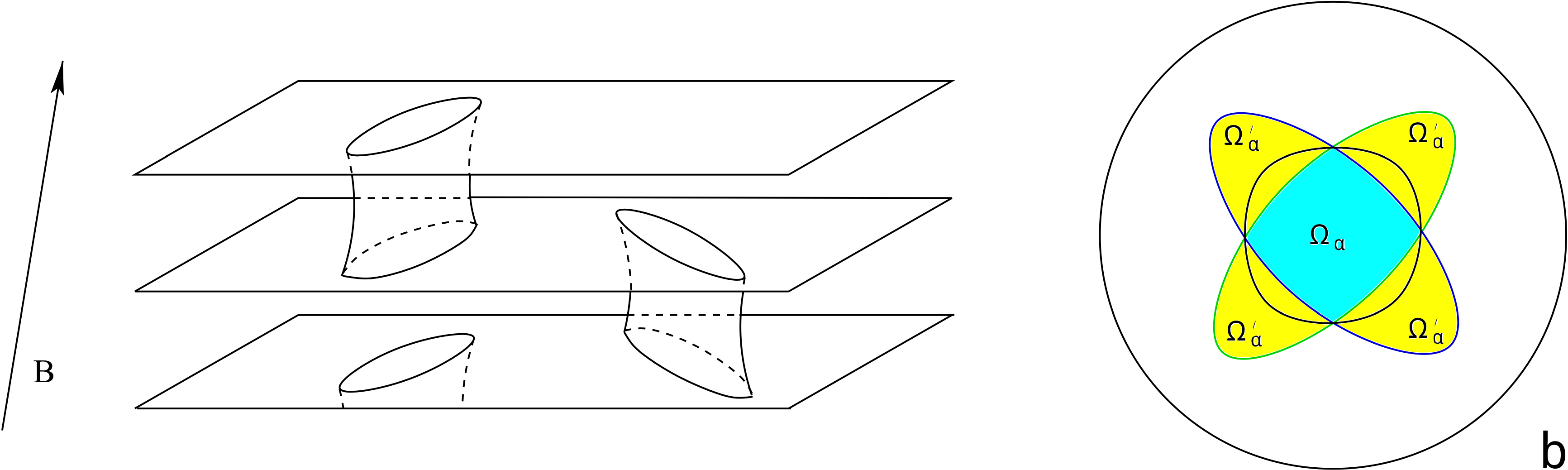} 
\end{center}
\begin{center}
\includegraphics[width=175mm]{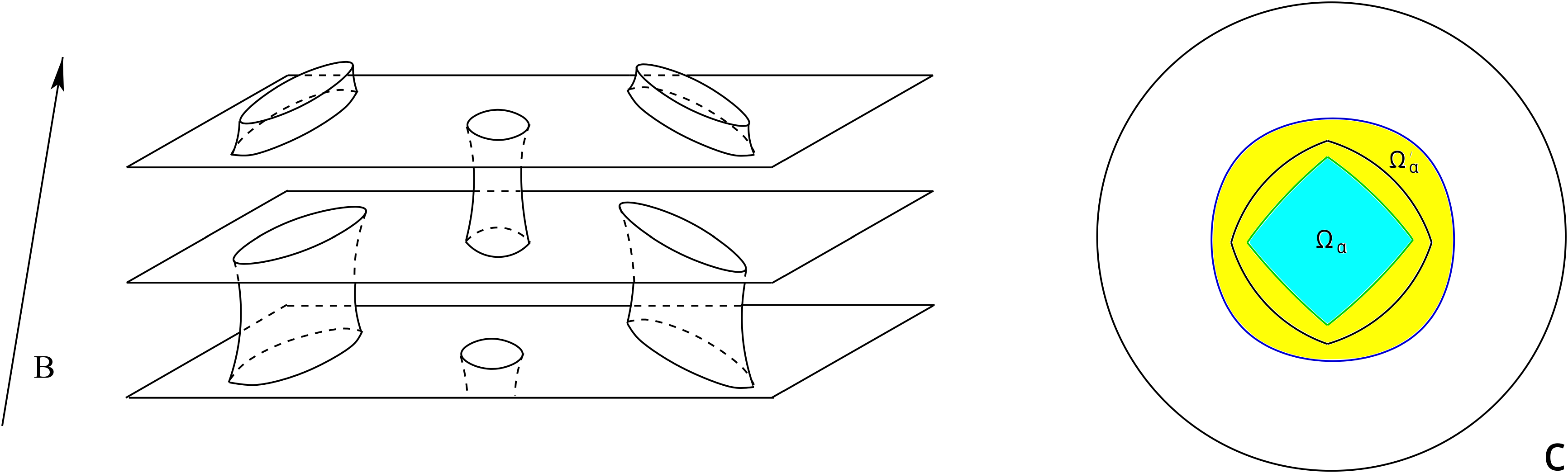}
\end{center}
\caption{The boundaries of the Zones $\, \Omega^{*}_{\alpha} \, $ 
inside the domains $\, \Omega^{\prime}_{\alpha} \, $.}
\label{FullBounds}
\end{figure*}

 As can be seen, in the general case, the second boundary of a
Stability Zone can be located at an arbitrary distance from its first boundary.
In the general case, the domain $\, \Omega^{\prime}_{\alpha} \, $ may consist of 
several connected components bounded by parts of the first boundary of the
Zone $\, \Omega_{\alpha} \, $ from the ``internal'' side and by parts 
of the second boundary from the ``external'' side (Fig. \ref{SecBounds}). 
It can also be noted that for each of the connected components of the domain
$\, \Omega^{\prime}_{\alpha} \, $  its ``internal'' and ``external'' boundaries 
correspond to the disappearance of cylinders of closed trajectories of opposite 
types at Fig. \ref{FullFermSurf}. Within each domain 
$\, \Omega^{\prime}_{\alpha} \, $ there is also the boundary of the Stability
Zone $\, \Omega^{*}_{\alpha} \, $, which has both the ``electron'' and the 
``hole'' type (Fig. \ref{FullBounds}). It is also a common circumstance that 
for a Stability Zone with a compound boundary that has sections corresponding 
to the disappearance of cylinders of closed trajectories of different types, 
the ``corner'' points of its boundaries (separating sections corresponding 
to the disappearance of cylinders of closed trajectories of different types) 
are also ``corner'' points for the second boundary of the Stability Zone, and 
also belong to the boundary of the corresponding Zone 
$\, \Omega^{*}_{\alpha} \, $ (Fig. \ref{FullBounds}, b). At the same time, 
for a Stability Zone, whose boundaries completely correspond to the disappearance 
of cylinders of closed trajectories of the same type, its first boundary does not 
intersect the second boundary in the generic case, and the boundary of the 
corresponding Zone $\, \Omega^{*}_{\alpha} \, $  lies inside the area bounded 
by the first and second boundaries (Fig. \ref{FullBounds}, a, c).

\vspace{1mm}

 It will be convenient for us here to introduce some special notations for 
different types of Zones $\, \Omega_{\alpha} \, $. Namely, let's say that 
the Zone $\, \Omega_{\alpha} \, $  is of type $(+)$  if all points of its 
boundary correspond to the disappearance of cylinders of closed electron-type 
trajectories. Similarly, let's say that $\, \Omega_{\alpha} \, $ is of type
$(-)$ if all points of its boundary correspond to the disappearance of cylinders 
of closed hole-type trajectories. Finally, the Zone $\, \Omega_{\alpha} \, $
is of the type $(\pm)$ if at different parts of its boundary the cylinders 
of closed trajectories of both the electron and hole type disappear.

\vspace{1mm}

 It is not difficult to show that on the sections of the boundaries 
of $\, \Omega_{\alpha} \, $ corresponding to the disappearance of cylinders 
of closed electron-type trajectories we have in generic case the relation
$$\epsilon_{F} \,\,\, = \,\,\, 
\tilde{\epsilon}_{2} ({\bf B}/B) \,\,\, > \,\,\,
\tilde{\epsilon}_{1} ({\bf B}/B) $$

 Similarly, on the sections of the boundaries of $\, \Omega_{\alpha} \, $
corresponding to the disappearance of cylinders of closed hole-type trajectories 
we have in generic case the relation
$$\epsilon_{F} \,\,\, = \,\,\, 
\tilde{\epsilon}_{1} ({\bf B}/B) \,\,\, < \,\,\,
\tilde{\epsilon}_{2} ({\bf B}/B) $$

 A special situation takes place at the ``corner'' (separating parts of the 
boundary of different types) points of the Zones of type $(\pm)$, where we have 
the relation
$$\epsilon_{F} \,\,\, = \,\,\, 
\tilde{\epsilon}_{1} ({\bf B}/B) \,\,\, = \,\,\,
\tilde{\epsilon}_{2} ({\bf B}/B) $$

\vspace{1mm}

 It can be seen here that for a generic Fermi surface all the Stability Zones
$\, \Omega_{\alpha} \, $ of type $(+)$ can be covered with open domains
$\, U^{+}_{\alpha} \, $, and the Stability Zones $\, \Omega_{\beta} \, $ 
of type $(-)$ - with open domains $\, U^{-}_{\beta} \, $, such that:

 1) All domains $\, U^{+}_{\alpha} \, $ and $\, U^{-}_{\beta} \, $ 
do not intersect with each other, as well as with the Stability Zones
$\, \Omega_{\alpha} \, $ of type $(\pm)$;

 2) Everywhere on the sets
$\, U^{+}_{\alpha} \, \backslash \, \overline{\Omega_{\alpha}} \, $ 
takes place the relation
\begin{equation}
\label{U+sootn}
\epsilon_{F} \,\,\, > \,\,\, \tilde{\epsilon}_{2} ({\bf B}/B) \,\,\, , 
\end{equation}
and on the sets
$\, U^{-}_{\beta} \, \backslash \, \overline{\Omega_{\beta}} \, $ 
- the relation
\begin{equation}
\label{U-sootn}
\epsilon_{F} \,\,\, < \,\,\, \tilde{\epsilon}_{1} ({\bf B}/B) 
\end{equation}

\vspace{1mm}

\begin{figure*}
\begin{center}
\vspace{5mm}
\includegraphics[width=175mm]{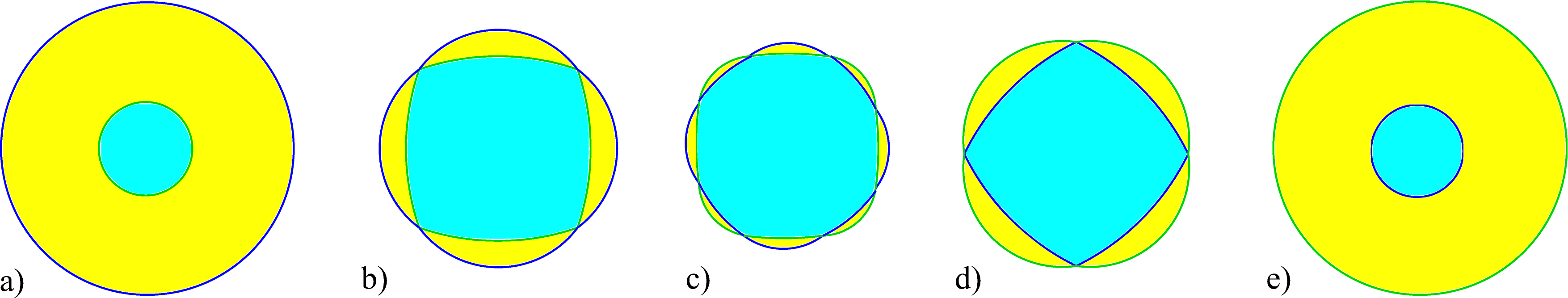} 
\end{center}
\caption{Change of the shape of the Zones $\, \Omega_{\alpha} \, $ 
and $\, \Omega^{\prime}_{\alpha} \, $ with increasing the value of
$\, \epsilon_{F} \, $ for a fixed dispersion relation
$\, \epsilon ({\bf p}) \, $. (a) The appearance and growth of a
Stability Zone $\, \Omega_{\alpha} \, $ having the type $(-)$. 
(b) The connection of the first and second boundaries of the Zone
$\, \Omega_{\alpha} \, $ and the formation of a Zone of type $(\pm)$.
(c) Evolution of the Zone of type $(\pm)$. (d) Change of type to $(+)$.
(e) Decrease and disappearance of a Zone of Type $(+)$.}
\label{ZoneEvolution}
\end{figure*}

 Coming back to the full dispersion relation, it can be noted that each Zone 
$\, \Omega_{\alpha} \, $ passes in generic case all types $\, (-)$ , 
$\, (\pm) \, $ and $\, (+) \, $ when the Fermi level changes from some minimal 
to some maximal value, determined by the existence of this Zone.
Fig. \ref{ZoneEvolution} presents a possible scheme of the evolution of a Zone
$\, \Omega_{\alpha} \, $ when changing $\, \epsilon_{F} \, $ in the interval
$\, [\epsilon_{min} , \epsilon_{max}] \, $.

\vspace{1mm}

\vspace{1mm}

  As we have already said, we will be interested here in the questions 
of the complexity of angular diagrams for conductivity in metals with 
arbitrary (physical) dispersion relations. We also note here that in the work
\cite{SecBound} it was proposed to divide all non-trivial (containing 
Stability Zones) angular diagrams of metals into simpler (type A) and more 
complex (type B). As noted in \cite{SecBound}, these features of angular 
diagrams are related to the behavior of the Hall conductivity outside 
the Stability Zones in the limit
$\, \omega_{B} \tau \, \rightarrow \, \infty \, $.

 More precisely, for generic directions of $\, {\bf B} \, $, lying outside 
any Stability Zone, on the Fermi surface there exist only closed trajectories 
of the system (\ref{MFSyst}). The conductivity tensor in the plane orthogonal 
to $\, {\bf B} \, $ can be represented in this case as a regular series in 
inverse powers of the parameter $\, \omega_{B} \tau \, $ in the limit
$\, \omega_{B} \tau \, \rightarrow \, \infty \, $
and has in the main order the following form
\begin{multline}
\label{ClosedTraj}
\sigma^{\alpha\beta} \,\,\, = \,\,\, \left(
\begin{array}{cc}
a^{11} \, (\omega_{B} \tau )^{-2}  \, &  
\, a^{12} \, ( \omega_{B} \tau )^{-1}  \cr
- a^{12} \, ( \omega_{B} \tau )^{-1}  \, &  
\, a^{22} \, ( \omega_{B} \tau )^{-2}
\end{array}  \right) \,\, ,  \\
\omega_{B} \tau \, \rightarrow \, \infty
\end{multline}

 The value $\, \sigma^{12} \, = \, a^{12} (\omega_{B} \tau)^{-1} \, $
can be in this case represented as
\begin{equation}
\label{SigmaNeNh}
\sigma^{12} \,\,\, = \,\,\, {ec \over B} \, 
\left( n_{e} - n_{h} \right) \,\,\, , 
\end{equation}
where the values $\, n_{e} \, $ and $\, n_{h} \, $ 
represent the ``electrons'' and ``holes'' concentration for a given 
direction of $\, {\bf B} \, $. The values $\, n_{e} \, $ and $\, n_{h} \, $
can be defined by the formulae
$$n_{e} \,\,\, = \,\,\, 2 V_{e} / (2\pi \hbar)^{3} 
\quad , \quad \quad
n_{h} \,\,\, = \,\,\, 2 V_{h} / (2\pi \hbar)^{3} \,\,\, , $$
where $\, V_{e} \, $ and $\, V_{h} \, $ 
represent the volumes bounded by all nonequivalent cylinders of closed 
trajectories (on the Fermi surface) of the electron and hole type in
$\, {\bf p}$ - space.

\begin{figure}[t]
\begin{center}
\vspace{5mm}
\includegraphics[width=\linewidth]{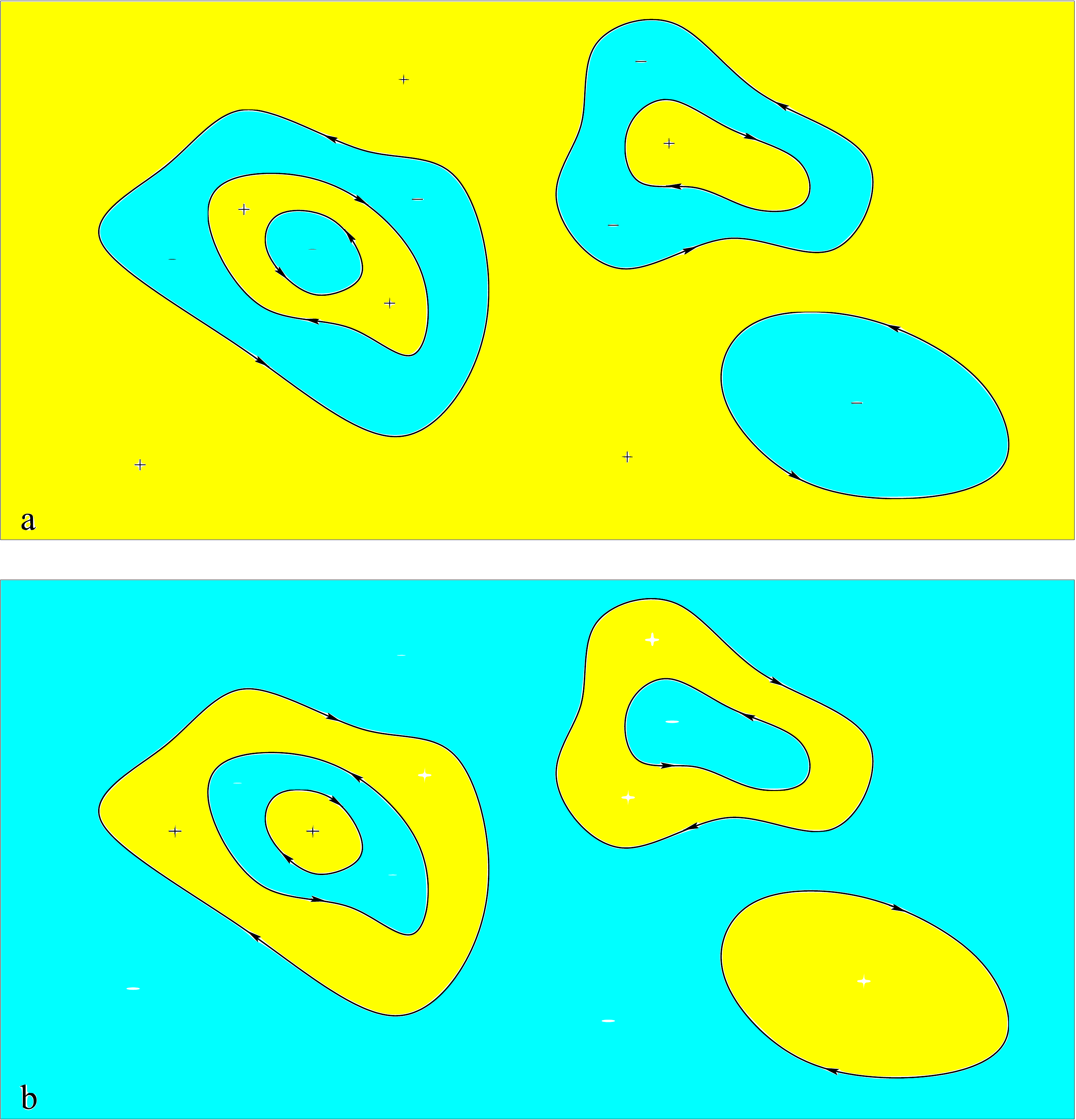}
\end{center}
\caption{Two possible pictures in planes orthogonal to
$\, {\bf B} \, $ in the case when the Fermi surface contains 
only closed trajectories in the $\, {\bf p}$ - space.}
\label{PlaneSea}
\end{figure}

 An important feature of the picture arising for generic directions of
$\, {\bf B} \, $, lying outside any Stability Zone, is that in all planes 
orthogonal to $\, {\bf B} \, $ we have the same type of the behavior of
trajectories of system (\ref{MFSyst}), determined by one of the following 
possibilities:

\vspace{1mm}

I) The region of the higher energy values  
$\, \epsilon ({\bf p}) \, > \, \epsilon_{F} \, $
represents the ``sea'' in the plane orthogonal to
$\, {\bf B} \, $, while the areas of the lower energy values
$\, \epsilon ({\bf p}) \, < \, \epsilon_{F} \, $
represent finite ``islands'' (``islands'' may contain ``lakes'' 
of higher energy values, etc., Fig. \ref{PlaneSea}, a).

\vspace{1mm}

II) The region of the lower values of energy
$\, \epsilon ({\bf p}) \, < \, \epsilon_{F} \, $
represents the ``sea'' in the plane orthogonal to
$\, {\bf B} \, $,  while the areas of the higher energy values
$\, \epsilon ({\bf p}) \, > \, \epsilon_{F} \, $
represent finite ``islands'' (``islands'' may contain ``lakes'' 
of lower energy values, etc., Fig. \ref{PlaneSea}, b).

\vspace{1mm}

 As a consequence of the picture described above, the value of 
$\, \sigma^{12} \, $ in this case is actually determined by one 
of the following relations:
\begin{equation}
\label{Sigma12Elekt}
\sigma^{12} \,\,\, = \,\,\, 
{2 ec \over (2\pi \hbar)^{3} B} \,\,\,  V_{-}
\quad \quad {\rm (\text{case I})}
\end{equation}
\begin{equation}
\label{Sigma12Hole}
\sigma^{12} \,\,\, = \,\,\, 
- \, {2 ec \over (2\pi \hbar)^{3} B} \,\,\,  V_{+}
\quad \quad  {\rm (\text{case II})}
\end{equation}
where $\, V_{-} \, $ and $\, V_{+} \, $ 
represent the volumes, defined by the conditions
$\, \epsilon ({\bf p}) \, < \, \epsilon_{F} \, $
and $\, \epsilon ({\bf p}) \, > \, \epsilon_{F} \, $
in the Brillouin zone.

 As a result, the magnitude of the Hall conductivity is locally constant 
(at a constant value of $\, B $) everywhere outside the Stability Zones 
under the condition $\, \omega_{B} \tau \, \gg \, 1 \, $. 
(In fact, the condition $\, \tau \, \gg \, T \, $,
where $\, T \, $ is typical time of electron motion along a closed trajectory 
for a given direction of $\, {\bf B} \, $. As was noted in \cite{SecBound},
this condition may be stronger.) This circumstance, as is easy to see, gives 
a simple tool of experimental determination of each of the two described 
situations for directions of $\, {\bf B} \, $ lying outside the Stability Zones.
In this case, the areas corresponding to different situations described above 
should be separated by Stability Zones, which in the generic case form 
``chains'' consisting of an infinite number of Zones.

\vspace{1mm}

 Let us define the type (A or B) of the angular diagram of conductivity of 
metal as follows:

\vspace{1mm}

 We say that the angular diagram of the conductivity of a metal is of type A, 
if for all generic directions of $\, {\bf B} \, $  outside the Zones
$\, \overline{\Omega_{\alpha}} \, $ we observe only one type of the 
Hall conductivity in the limit
$\, \omega_{B} \tau \, \rightarrow \, \infty \, $.
We say that the angular diagram of the conductivity of a metal is of type B, 
if for different generic directions of $\, {\bf B} \, $ outside the Zones
$\, \overline{\Omega_{\alpha}} \, $ both types of the Hall conductivity 
can be observed in the limit
$\, \omega_{B} \tau \, \rightarrow \, \infty \, $.

\vspace{1mm}

 Let us make here one more remark concerning the Fermi surfaces of real metals.
As we have said above, we assume here for simplicity that the Fermi surface 
consists of one connected component, on which we consider the trajectories 
of the system (\ref{MFSyst}). In fact, the Fermi surface of a real metal 
usually consists of several components and very often contains components 
of the genus 0 or 1 along with complex components. In calculating the 
Hall conductivity, we must in fact add up the contributions from all 
components of the Fermi surface, which also have the properties described 
above. In this case, it may turn out that the type of total Hall conductivity 
may differ from the type of contribution to this conductivity from the 
component we are considering. Therefore, it would be more strict to say 
that the angular diagram of the conductivity of a metal is of type A, 
if for all generic directions of $\, {\bf B} \, $ outside the Zones
$\, \overline{\Omega_{\alpha}} \, $ there is only one value of the 
Hall conductivity (for a given value of $\, B$) under the condition
$\, \omega_{B} \tau \, \gg \, 1 \, $, and the angular diagram of the 
conductivity of a metal is of type B, if for different generic directions
of $\, {\bf B} \, $ outside the Zones $\, \overline{\Omega_{\alpha}} \, $ 
at least two values of the Hall conductivity can be observed under the same 
condition. We will, however, keep here the terminology introduced above 
for greater transparency of the presentation.

\begin{figure}[t]
\begin{center}
\vspace{5mm}
\includegraphics[width=\linewidth]{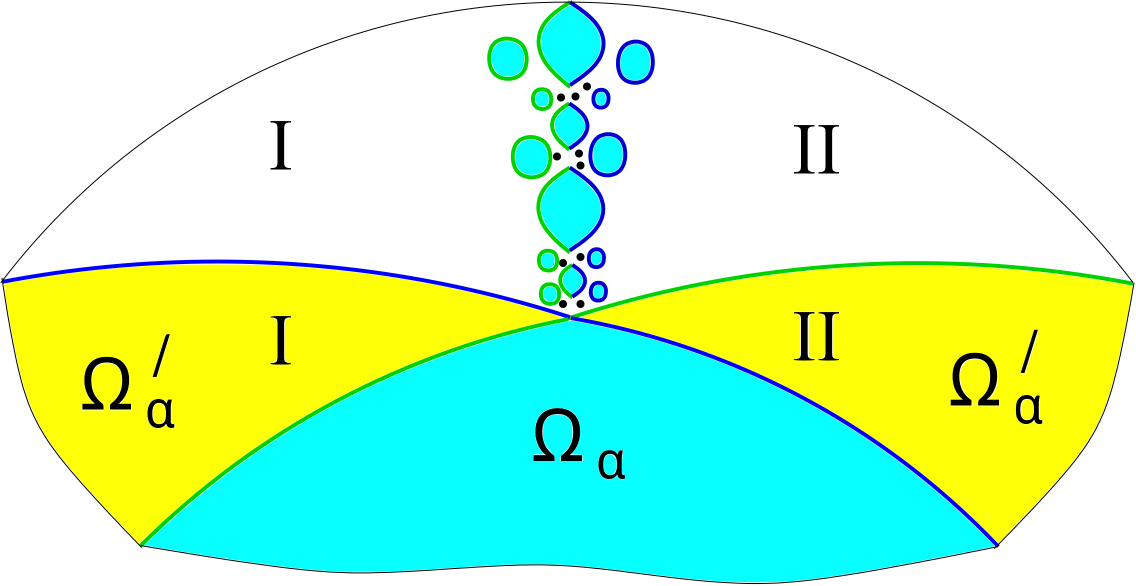}
\end{center}
\caption{Regions with different values of Hall conductivity and a chain of 
decreasing Stability Zones and ``chaotic'' directions of $\, {\bf B} \, $
near the ``corner point'' of the boundary of a Stability Zone of type $(\pm)$.}
\label{CornerTypeB}
\end{figure}

 As was also noted in \cite{SecBound}, an angular diagram has type B, 
in particular, if it contains at least one Stability Zone of type
$(\pm)$. Indeed, as can be shown (see \cite{SecBound}), in different 
connected components of the corresponding domain
$\, \Omega^{\prime}_{\alpha} \, $ there present in this case different types 
of the Hall conductivity for generic directions of $\, {\bf B} \, $
(Fig. \ref{CornerTypeB}). It can be seen that in this situation a chain 
of Stability Zones should adjoin the ``corner point'' of 
$\, \Omega_{\alpha} \, $, separating regions with different Hall conductivity 
that appear near the corresponding sections of the boundary of 
$\, \Omega_{\alpha} \, $. In this case, the Zone $\, \Omega_{\beta} \, $ can 
adjoin directly to the ``corner point'' of $\, \Omega_{\alpha} \, $ only if the 
corresponding direction of $\, {\bf B} \, $ corresponds to the appearance of 
periodic trajectories (with the direction given by the intersection of the 
planes $\, \Gamma_{\alpha} $ and $\, \Gamma_{\beta} \, $) on the Fermi surface, 
which corresponds to a non-generic situation. In the generic case, a chain 
of an infinite number of decreasing Stability Zones $\, \Omega_{\beta} \, $
should adjoin the ``corner point'' of $\, \Omega_{\alpha} \, $ 
(Fig. \ref{CornerTypeB}).

\vspace{1mm}

 Let us note here that accumulation points of decreasing Stability Zones, 
which are not boundary points of Stability Zones, represent generically  
directions of $\, {\bf B} \, $ corresponding to the appearance of chaotic 
trajectories (of Tsarev or Dynnikov type) on the Fermi surface. The set 
of the corresponding directions of $\, {\bf B} \, $ represents in the general 
case a rather complex set on the angular diagram. As can be shown 
(see \cite{dynn3}), the full measure of such directions on the angular diagram 
is zero for a generic Fermi surface. According to the conjecture of S.P. Novikov
(\cite{BullBrazMathSoc,JournStatPhys}), the fractal dimension of the set
$\, {\cal D} \, $ of such directions for a generic Fermi surface is strictly 
less than 1 (but may be larger for special Fermi surfaces). We also note here 
that, unlike the angular diagrams for the entire dispersion relation, on the 
angular diagrams for a generic Fermi surface the accumulation of decreasing 
Stability Zones cannot occur at the ``regular'' points of the boundaries
of $\, \Omega_{\alpha} \, $, since each of the Zones $\, \Omega_{\alpha} \, $ 
is surrounded here with the additional area $\, \Omega^{\prime}_{\alpha} \, $. 
The only exceptions are the ``corner'' points of Zones of type $(\pm)$, 
where the second boundary of the Zone $\, \Omega_{\alpha} \, $ is adjacent 
to the first one.

 The set 
$${\cal M} \quad = \quad
{\cal D} \,\, \bigcup \,\, \left( \cup \, \overline{\Omega_{\alpha}}\right) $$
represents a closed set on $\, \mathbb{S}^{2} \, $.

 At the points of the set $\, \mathbb{S}^{2} \backslash {\cal M} \, $
we always have one of the relations
$\, \epsilon_{F} \, < \, \tilde{\epsilon}_{1} ({\bf B}/B) \, $ or
$\, \epsilon_{F} \, > \, \tilde{\epsilon}_{2} ({\bf B}/B) \, $,
and the corresponding subsets are separated by points of the set 
$\, {\cal M} \, $ on $\, \mathbb{S}^{2} \, $. It can be seen (see \cite{dynn3}) 
that the first of the relations corresponds to the picture (I) described above 
for generic directions of $\, {\bf B} \, $,  while the second relation corresponds 
to the picture (II) for generic directions of $\, {\bf B} \, $. It should be noted 
that the points of the set $\, {\cal D} \, $, as well as the Stability Zones
$\, \Omega_{\alpha} \, $ play an important role in the separation of the set
$\, \mathbb{S}^{2} \backslash {\cal M} \, $ into components with different 
behavior of the Hall conductivity in the limit
$\, \omega_{B} \tau \, \rightarrow \, \infty \, $. Note also that 
in the entire set of Stability Zones $\, \Omega_{\alpha} \, $ only Zones of 
the type $\, (\pm) \, $ are important for the separation of such components, 
since Zones of the types $\, (+) \, $ and $\, (-) \, $ do not play a role in the 
separation of the mentioned components due to the above properties (1) and (2) 
and the relations (\ref{U+sootn}) - (\ref{U-sootn}).

\vspace{1mm}

 In our situation it is natural to introduce the following values
$$\epsilon^{\cal B}_{1} \,\,\, = \,\,\, \min \,\, 
\tilde{\epsilon}_{2} ({\bf B}/B) \,\,\, , \quad
\epsilon^{\cal B}_{2} \,\,\, = \,\,\, \max \,\,
\tilde{\epsilon}_{1} ({\bf B}/B) $$
for each dispersion relation. We note right away that for the complex angular 
diagrams we are considering (an infinite number of Stability Zones for the entire 
dispersion relation) we always have the relation
$$\epsilon^{\cal B}_{2} \,\,\, \geq \,\,\, \epsilon^{\cal B}_{1} \,\,\, ,$$
since on the boundaries of the Stability Zones $\, \Omega^{*}_{\alpha} \, $
(as well as for ``chaotic'' directions of $\, {\bf B} $) we have the relation
$\, \tilde{\epsilon}_{1} ({\bf B}/B) \, = \, 
\tilde{\epsilon}_{2} ({\bf B}/B) \, $. For generic dispersion relations we will 
in fact have 
$$\epsilon^{\cal B}_{2} \,\,\, > \,\,\, \epsilon^{\cal B}_{1} \,\,\, ,$$
since for such directions of $\, {\bf B} \, $ the corresponding values
$$\tilde{\epsilon}_{1} ({\bf B}/B) \,\,\, = \,\,\,
\tilde{\epsilon}_{2} ({\bf B}/B) \,\,\, = \,\,\,
\epsilon_{0} ({\bf B}/B) $$
do not belong to a single energy level in the general case. As follows from 
our reasoning, the angular diagram of the conductivity of a metal has type B, 
if the Fermi level $\, \epsilon_{F} \, $ belongs to the interval
$\, (\epsilon^{\cal B}_{1}, \epsilon^{\cal B}_{2}) \, $.

  At the boundary points of the interval
$\, \epsilon_{F} \, = \, \epsilon^{\cal B}_{1} \, $ and
$\, \epsilon_{F} \, = \, \epsilon^{\cal B}_{2} \, $ 
on all set $\, \mathbb{S}^{2} \backslash {\cal M} \, $ 
there is only one type of the Hall conductivity (respectively, electron and hole) 
for generic directions of $\, {\bf B} \, $, at the same time, the angular 
diagram of the conductivity of a metal contains in generic case an infinite 
number of Stability Zones $\, \Omega_{\alpha} \, $. It can also be seen, that as 
the value of $\, \epsilon_{F} \, $ increases in the interval
$\, [\epsilon^{\cal B}_{1}, \epsilon^{\cal B}_{2}] \, $, the full measure 
of the components $\, \mathbb{S}^{2} \backslash {\cal M} \, $ corresponding to 
the electron Hall conductivity decreases, while the full measure of the components
corresponding to the hole Hall conductivity grows.

 It is natural to introduce also the values
$$\epsilon^{\cal A}_{1} \,\,\, = \,\,\, \min \,\, 
\tilde{\epsilon}_{1} ({\bf B}/B) \,\,\, , \quad
\epsilon^{\cal A}_{2} \,\,\, = \,\,\, \max \,\,
\tilde{\epsilon}_{2} ({\bf B}/B) $$

 As follows from our reasoning, the angular diagram of the conductivity of 
a metal belongs to type A, if the Fermi level $\, \epsilon_{F} \, $
belongs to one of the intervals
$\, ( \epsilon^{\cal A}_{1}, \epsilon^{\cal B}_{1}] \, $ or
$\, [\epsilon^{\cal B}_{2}, \epsilon^{\cal A}_{2}) \, $. 
It is not difficult to see also that the first case corresponds to 
the Hall conductivity of the electron type at
$\, {\bf B}/B \, \in \, \mathbb{S}^{2} \backslash {\cal M} \, $,
while the second case corresponds to the hole-type Hall conductivity on 
the same set of (generic) directions of $\, {\bf B} \, $.

\vspace{1mm}

\begin{figure*}
\begin{tabular}{lll}
\includegraphics[width=56mm]{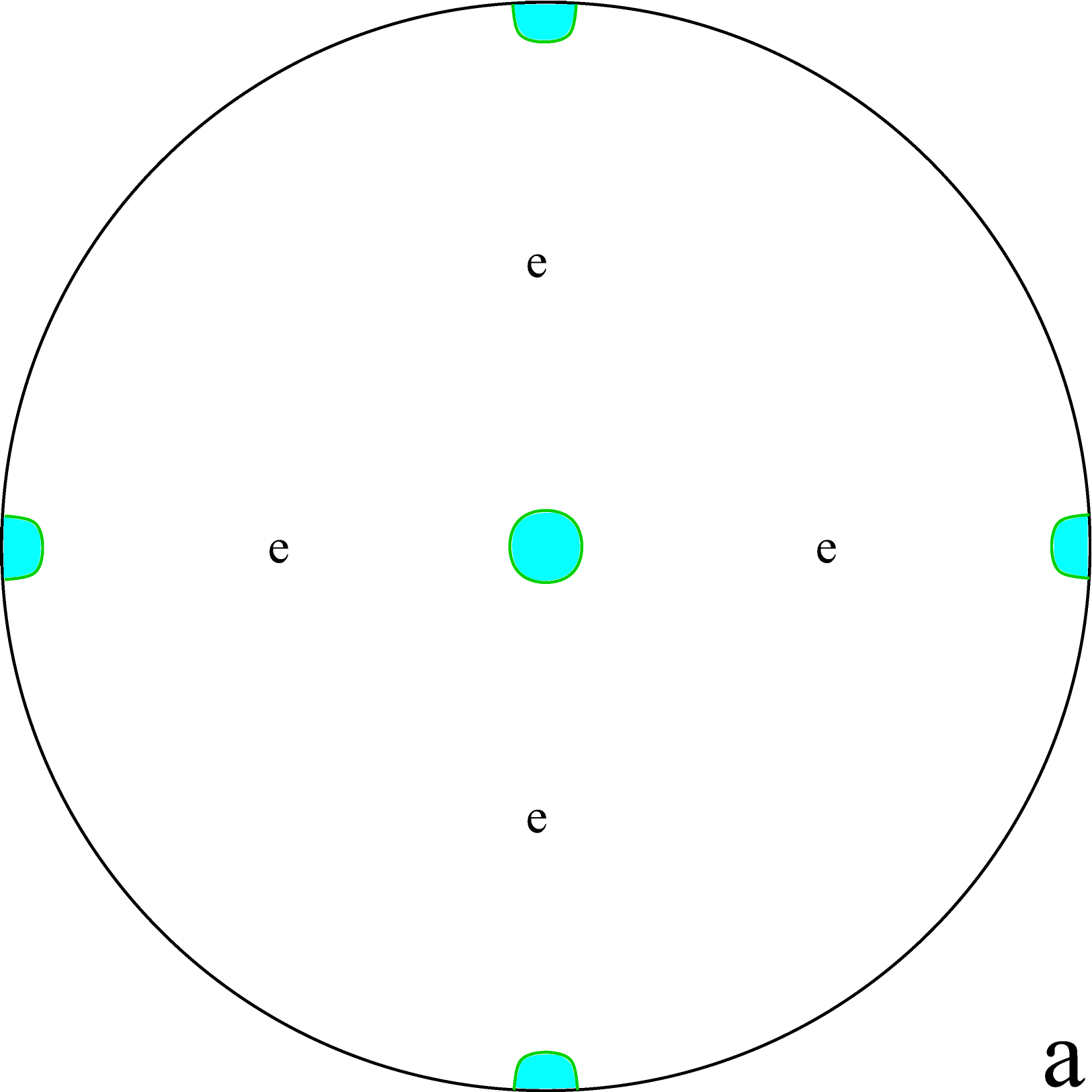}  &
\hspace{2mm}
\includegraphics[width=56mm]{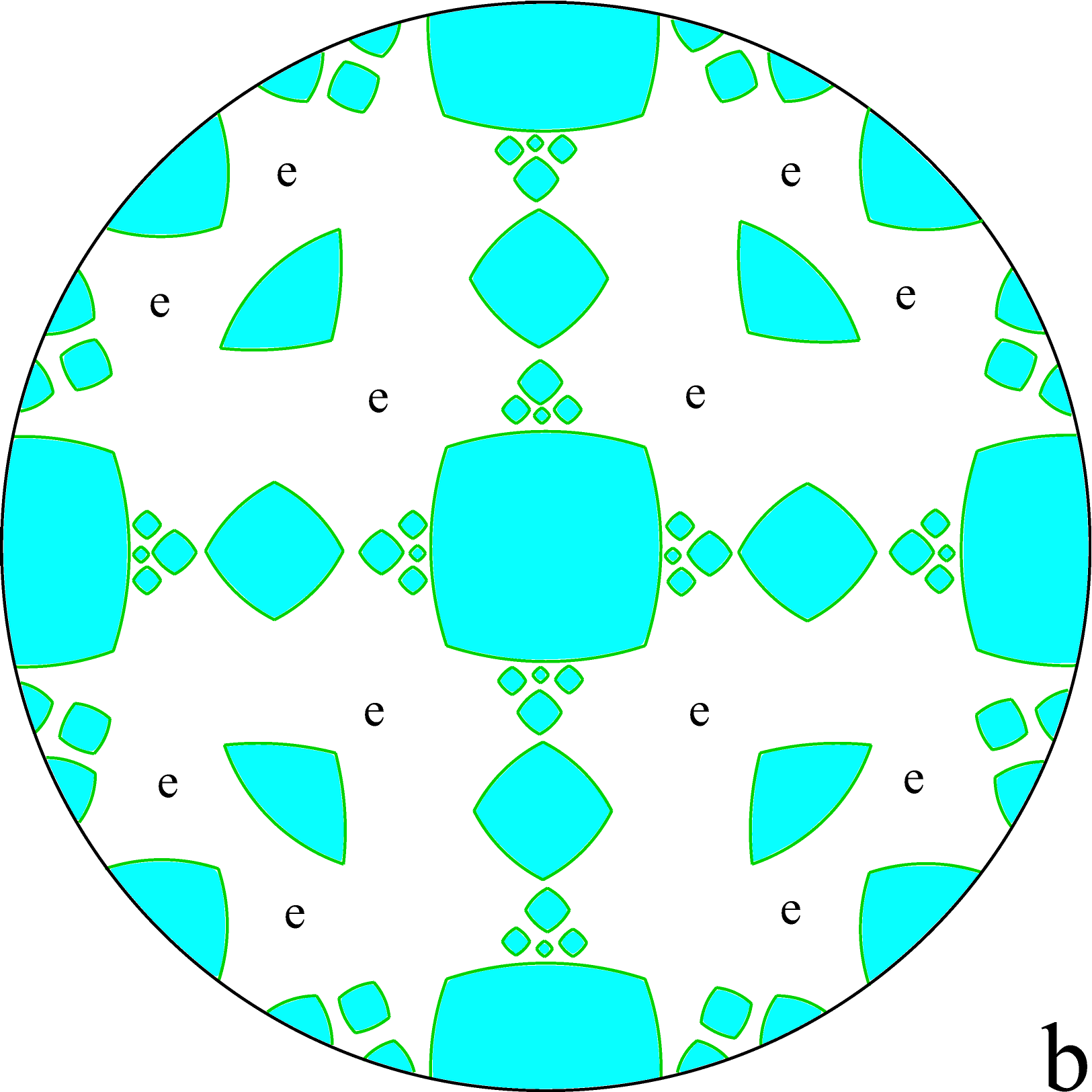}  &
\hspace{2mm}
\includegraphics[width=56mm]{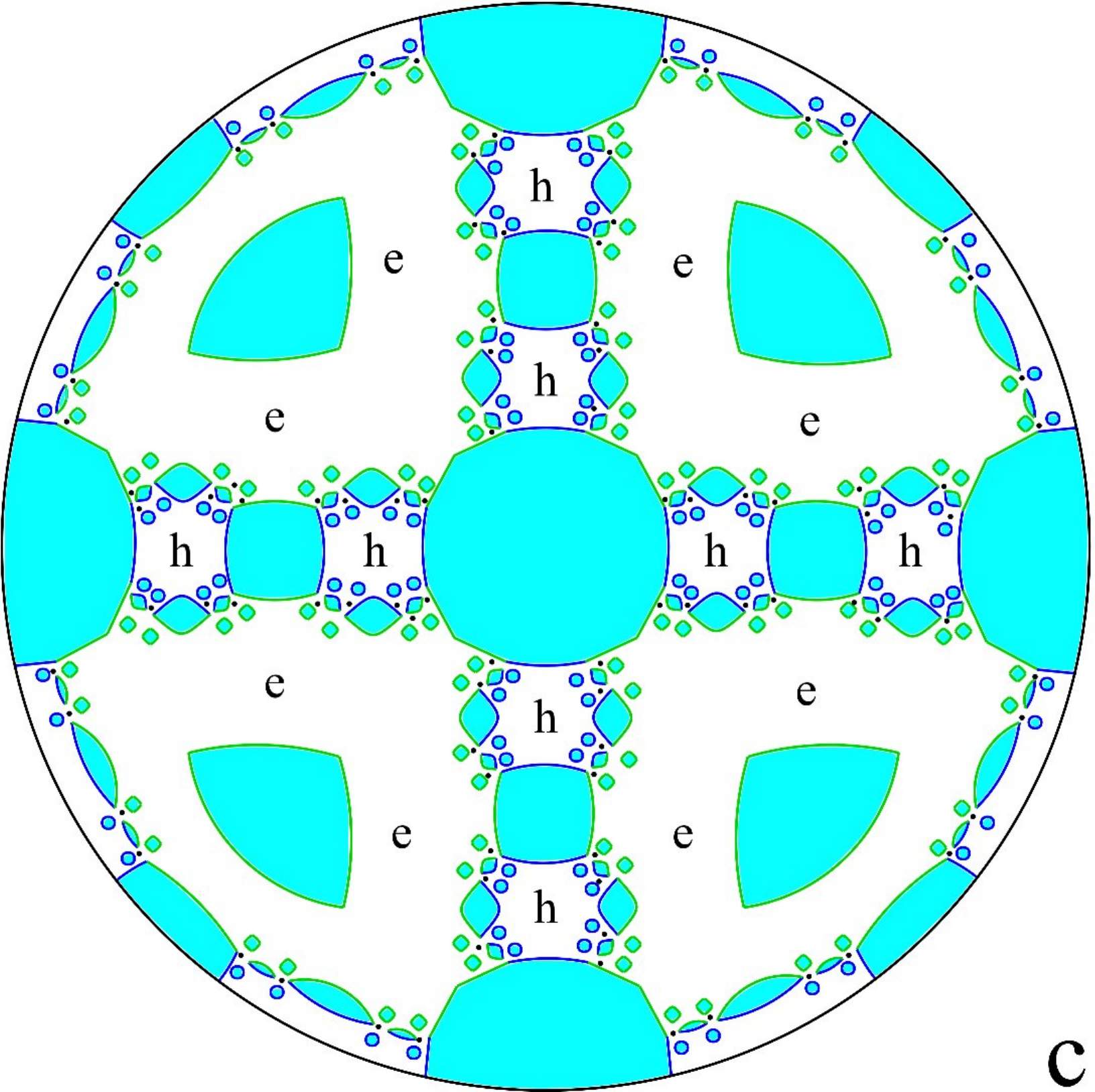}  
\end{tabular}

\vspace{5mm}

\begin{tabular}{lll}
\includegraphics[width=56mm]{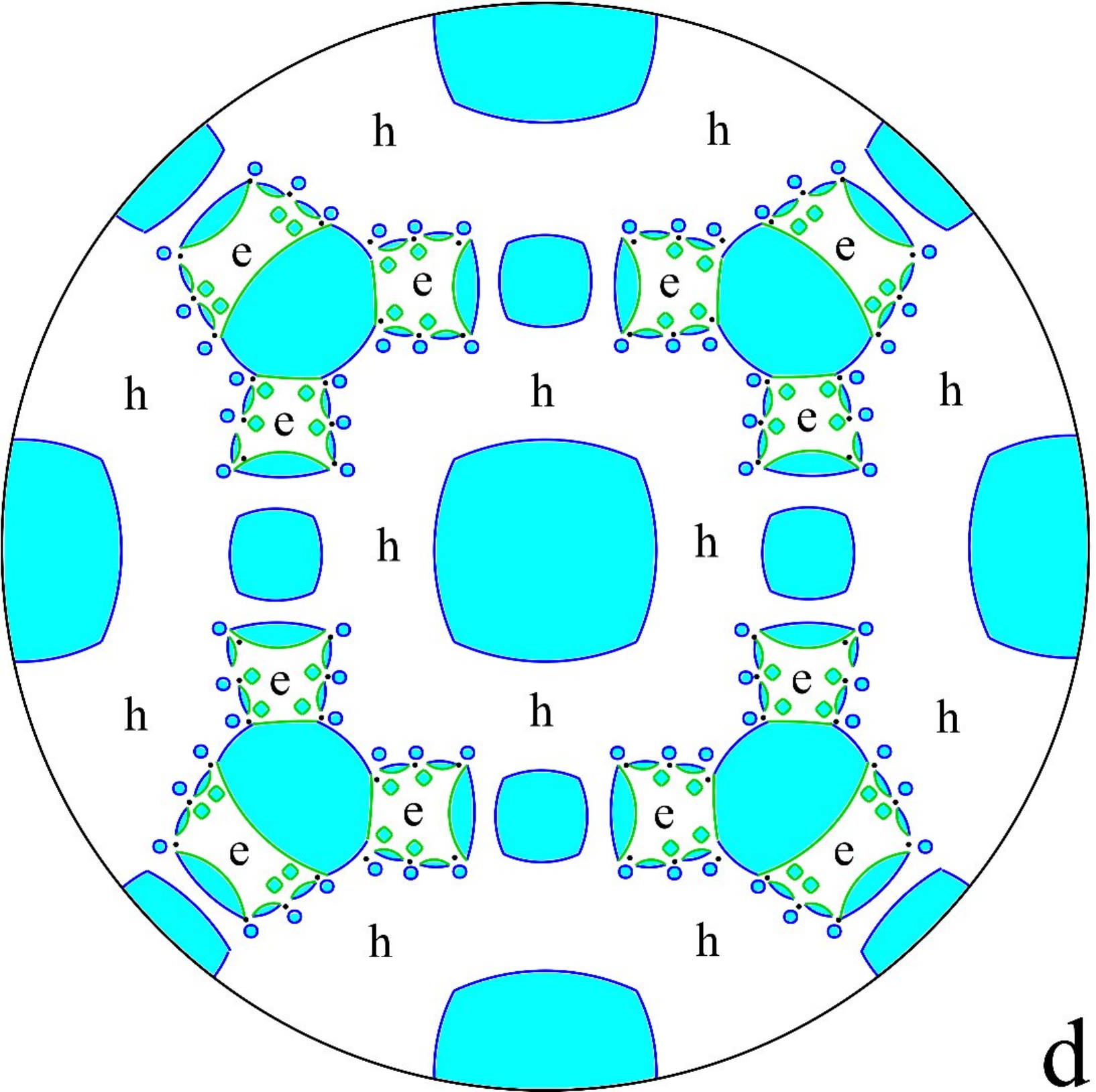}  &
\hspace{2mm}
\includegraphics[width=56mm]{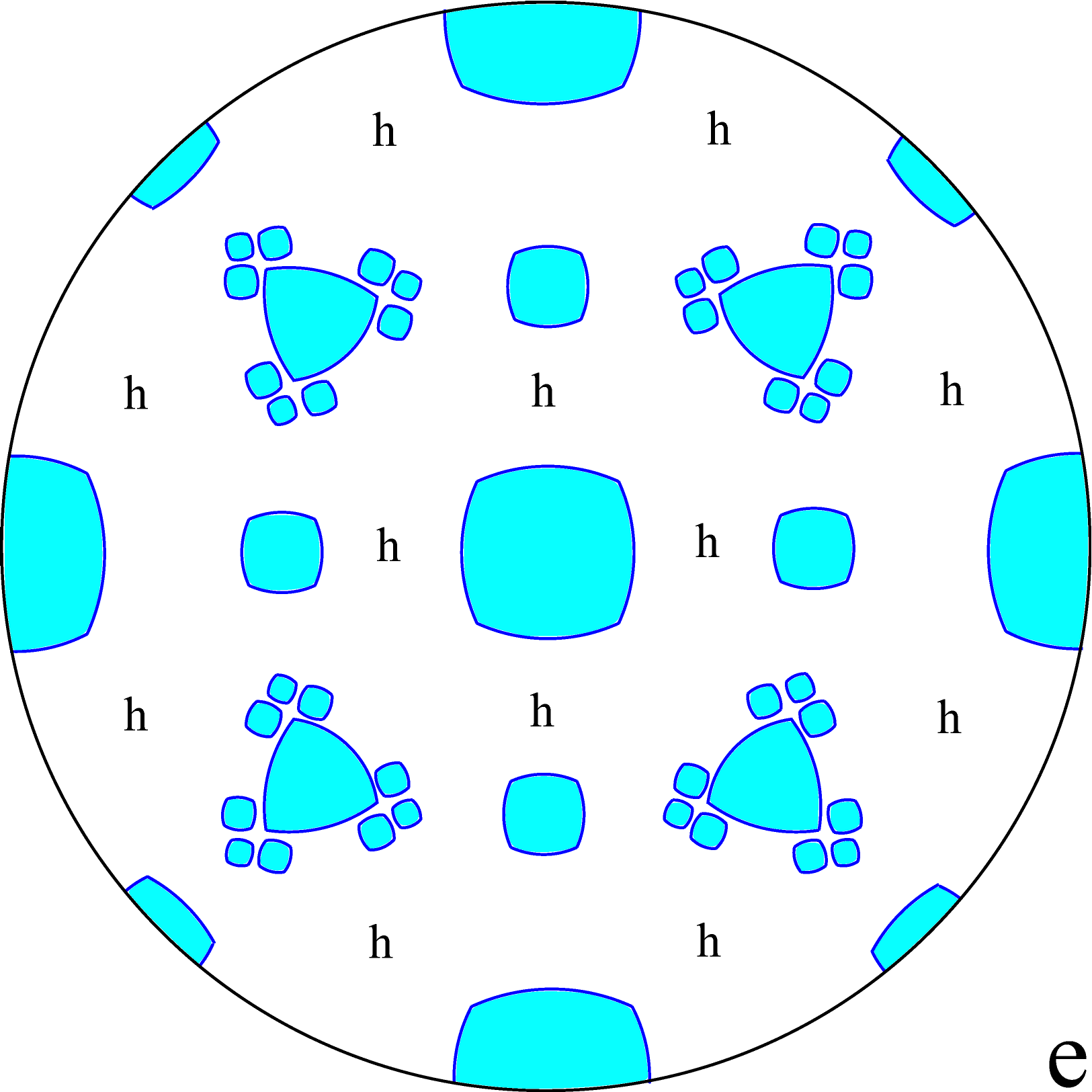}  &
\hspace{2mm}
\includegraphics[width=56mm]{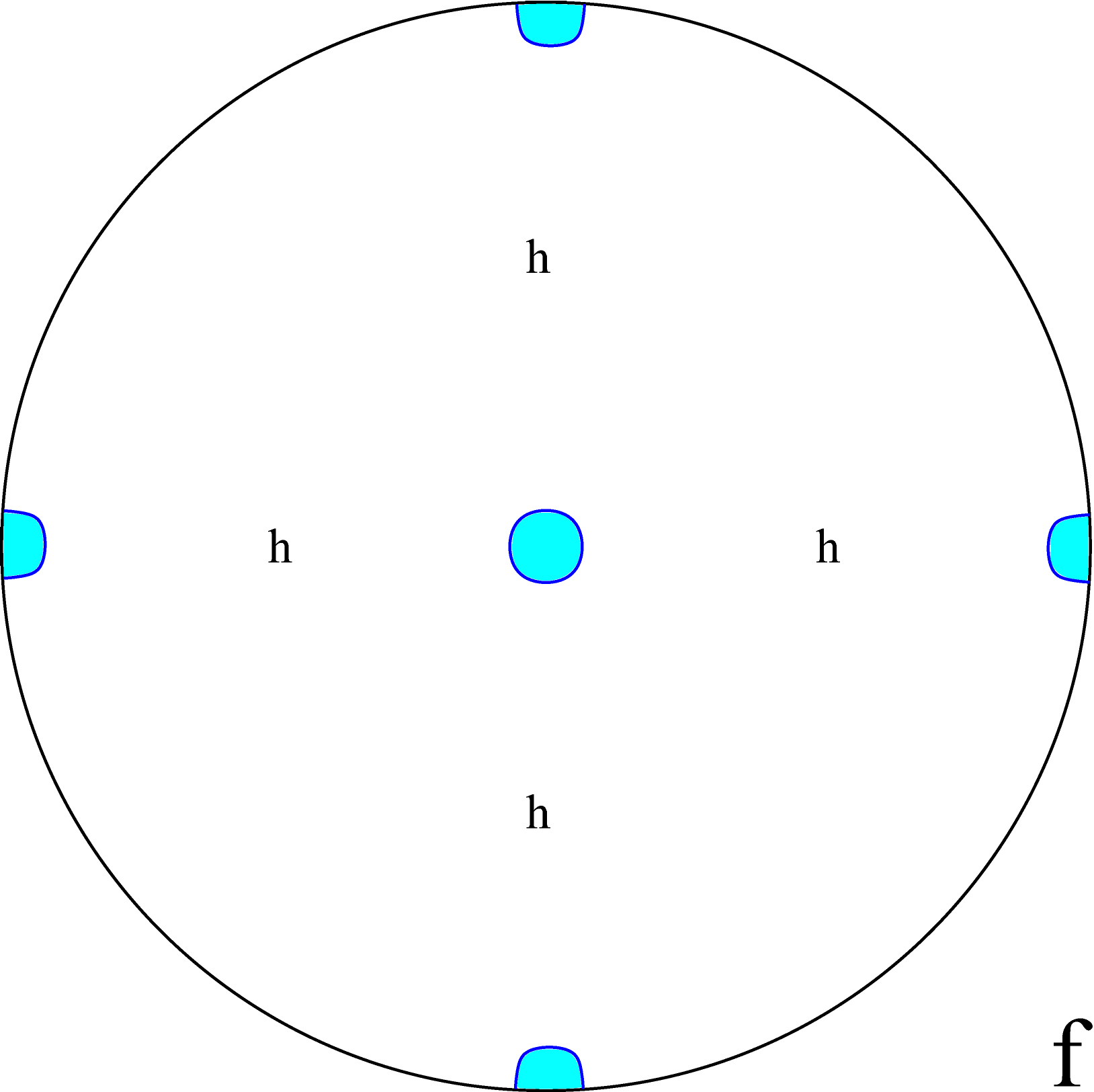} 
\end{tabular}
\caption{(a) The appearance of the first Stability Zones of the type $(-)$ 
after passing the value $\, \epsilon^{\cal A}_{1} \, $ by the Fermi level
$\, \epsilon_{F} \, $. (b) An increase in the number of Zones of type
$(-)$ and the appearance of points of concentration of small Zones at
$\, \epsilon_{F} \rightarrow \epsilon^{\cal B}_{1} \, $.
(c) Formation of the hole Hall conductivity regions and Zones of types
$(\pm)$ and $(+)$ and the appearance of ``chaotic'' directions of 
$\, {\bf B} \, $ after passing the value $\, \epsilon^{\cal B}_{1} \, $ 
by the Fermi level $\, \epsilon_{F} \, $. (d) The disappearance of the 
electron Hall conductivity regions and Zones of the types $(-)$ and $(\pm)$ 
at $\, \epsilon_{F} \rightarrow \epsilon^{\cal B}_{2} \, $. 
(e) The disappearance of the points of concentration of small Zones and 
the decrease in the number and size of Zones of type $(+)$
after passing the value $\, \epsilon^{\cal B}_{2} \, $ by the Fermi level
$\, \epsilon_{F} \, $. (f) Disappearance of the Zones of type $(+)$ 
in the limit $\, \epsilon_{F} \rightarrow \epsilon^{\cal A}_{2} \, $.}
\label{Diagrams}
\end{figure*}

 Fig. \ref{Diagrams} presents (very schematically) a typical change 
of the angular diagram of conductivity with a change of the value of
$\, \epsilon_{F} \, $ in the interval
$\, (\epsilon^{\cal A}_{1}, \epsilon^{\cal A}_{2}) \, $.
Thus, after passing the level
$\,  \epsilon_{F} \, = \, \epsilon^{\cal A}_{1} \, $ 
one can observe the appearance of the first Stability Zones and then 
further increase of their number and size, while everywhere outside 
the Stability Zones we have an electron type of the Hall conductivity 
for the generic directions of $\, {\bf B} \, $. The number of the 
Stability Zones tends to infinity at
$\, \epsilon_{F} \rightarrow \epsilon^{\cal B}_{1} \, $, and at the point
$\, \epsilon^{\cal B}_{1} \, $ the first points of concentration of 
the Stability Zones are observed on the angular diagram. After passing the 
value of $\, \epsilon^{\cal B}_{1} \, $ one can observe the presence of 
regions outside the Stability Zones, corresponding to both the electron 
and hole Hall conductivity and separated by an infinite set of Stability 
Zones and ``chaotic'' directions of $\, {\bf B} \, $ (in the generic case). 
As the value of $\, \epsilon_{F} \, $ increases, the measure of the regions 
corresponding to the electron Hall conductivity decreases, while the measure 
of the regions corresponding to the hole Hall conductivity increases.
In the interval $\, ( \epsilon^{\cal B}_{1}, \epsilon^{\cal B}_{2}) \, $
one can also observe the appearance of new Zones of type $\, (-) \, $,
the formation of Zones of type $\, (\pm) \, $, and then Zones of type
$\, (+) \, $, and disappearance of Zones of type $\, (+) \, $. At the point
$\,  \epsilon_{F} \, = \, \epsilon^{\cal B}_{2} \, $ the areas corresponding 
to the electron Hall conductivity completely disappear, and after passing 
through the value $\, \epsilon^{\cal B}_{2} \, $ we again have an angular 
diagram of type A, corresponding to the hole Hall conductivity for the
generic directions of $\, {\bf B} \, $ outside the Zones
$\, \Omega_{\alpha} \, $. At the same time, the number of Stability Zones 
and their sizes decrease with the increasing value of $\, \epsilon_{F} \, $ 
and in the limit $\, \epsilon_{F} \rightarrow \epsilon^{\cal A}_{2} \, $ 
the Zones disappear completely. As can be seen, the most complex structure 
of angular diagrams should be observed in the general case along 
``quasi-one-dimensional'' sets separating regions of different values 
of the Hall conductivity.

\vspace{1mm}

 As we noted above, a generic angular diagram of type B contains an infinite 
number of Stability Zones with arbitrarily large values of topological numbers
$\, (M^{1}_{\alpha}, M^{2}_{\alpha}, M^{3}_{\alpha})$. As it is not difficult 
to show, a generic angular diagram of type A, on the contrary, contains only 
a finite number of Zones $\, \Omega_{\alpha} \, $. More precisely, one can see 
that if
$\, \epsilon_{F} \, \in \, (\epsilon^{\cal A}_{1}, \epsilon^{\cal B}_{1}) \, $
or
$\, \epsilon_{F} \, \in \, (\epsilon^{\cal B}_{2}, \epsilon^{\cal A}_{2}) \, $,
then the corresponding angular diagram of the conductivity of the metal contains 
only a finite number of Stability Zones. Indeed, the presence of an infinite number 
of Stability Zones means the presence of a point of accumulation of such Zones on
$\, \mathbb{S}^{2} \, $, in which we must have the relation
$\, {\tilde \epsilon}_{1} ({\bf B}/B) \, = \,
{\tilde \epsilon}_{2} ({\bf B}/B) \, $, which is impossible in the indicated 
energy intervals.

\vspace{1mm}

 It can be seen, therefore, that an angular diagram of type A can contain 
an infinite number of Stability Zones only in the case
$\, \epsilon_{F} \, = \, \epsilon^{\cal B}_{1} \, $ or
$\, \epsilon_{F} \, = \, \epsilon^{\cal B}_{2} \, $.
It can be also seen that, for the same reason, an angular diagram of 
conductivity cannot contain ``chaotic'' directions of 
$\, {\bf B} \, $ at $\, \epsilon_{F} \, < \, \epsilon^{\cal B}_{1} \, $
or $\, \epsilon_{F} \, > \, \epsilon^{\cal B}_{2} \, $.
Recall also that we are considering here angular diagrams of a complex type, 
i.e. diagrams containing more than one Stability Zone when 
$\, \epsilon_{F} \, $ changes in the interval
$\, (\epsilon^{\cal A}_{1}, \epsilon^{\cal A}_{2}) \, $.

 As can be also seen, both types of complex diagrams (A and B) represent generic 
diagrams and should be observed experimentally. It should be noted, however, that 
the diagrams of type B should probably be rather rare due to a small size of the 
segment $\, (\epsilon^{\cal B}_{1}, \epsilon^{\cal B}_{2}) \, $
for real dispersion relations. It can also be noted that, although for dispersion 
laws with close values of $\, \epsilon^{\cal B}_{1} \, $ and
$\, \epsilon^{\cal B}_{2} \, $ the probability of observing a type B diagram is
rather small, the corresponding diagrams most closely approach the diagrams 
corresponding to the full dispersion relation in level of complexity.

\vspace{1mm}

 For the most complete description of all types of angular diagrams arising 
in the theory of normal metals, it is natural to introduce also the values
$$\epsilon^{{\cal A}\prime}_{1} \,\,\, = \,\,\, \inf \,\, 
\epsilon_{1} ({\bf B}/B) \,\,\, , \quad
\epsilon^{{\cal A}\prime}_{2} \,\,\, = \,\,\, \sup \,\,
\epsilon_{2} ({\bf B}/B) $$

 Note, that here we do not concentrate just on the ``complicated'' 
Fermi surfaces and include all their types in the consideration.

 For values of $\, \epsilon_{F} \, $, lying in the intervals
$\, (\epsilon^{{\cal A}\prime}_{1} , \epsilon^{\cal A}_{1} ] \, $ and
$\, [ \epsilon^{\cal A}_{2} , \epsilon^{{\cal A}\prime}_{2}) \, $,
the angular diagram of conductivity does not contain finite Stability Zones 
and can only contain Zones that degenerate into points at
$\, \epsilon_{F} = \epsilon^{\cal A}_{1} \, $ or
$\, \epsilon_{F} = \epsilon^{\cal A}_{2} \, $. For values of
$\, \epsilon_{F} \, $, lying in the intervals
$\, (\epsilon^{{\cal A}\prime}_{1} , \epsilon^{\cal A}_{1} ) \, $ and
$\, (\epsilon^{\cal A}_{2} , \epsilon^{{\cal A}\prime}_{2} ) \, $,
only (unstable) periodic trajectories of the system (\ref{MFSyst}) appear 
on the Fermi surface for special directions of $\, {\bf B} \, $. 
The corresponding directions of $\, {\bf B} \, $ are represented here 
by segments of big circles, each of which corresponds to a certain rational 
direction of open trajectories of the system (\ref{MFSyst}). 
It is not difficult to see also that in the intervals
$\, (\epsilon^{{\cal A}\prime}_{1} , \epsilon^{\cal A}_{1} ) \, $ and
$\, (\epsilon^{\cal A}_{2} , \epsilon^{{\cal A}\prime}_{2} ) \, $ 
the number of these segments and the corresponding rational directions 
of the trajectories is always finite. Indeed, an infinite set of segments 
of the indicated type must necessarily contain accumulation points on 
$\, \mathbb{S}^{2} \, $, in each neighborhood of which there are periodic 
trajectories with arbitrarily large denominators of the corresponding rational 
directions. For the corresponding directions of $\, {\bf B} \, $ we have 
the relations
$\, \epsilon_{1} ({\bf B}/B) \, < \, \tilde{\epsilon}_{1} ({\bf B}/B) \, $
(or
$\, \epsilon_{2} ({\bf B}/B) \, > \, \tilde{\epsilon}_{2} ({\bf B}/B) \, $),
however, the values
$\, |\epsilon_{1} ({\bf B}/B) - \tilde{\epsilon}_{1} ({\bf B}/B)| \, $
(or
$\, |\epsilon_{2} ({\bf B}/B) - \tilde{\epsilon}_{2} ({\bf B}/B)| \, $)
rapidly decrease with increasing denominators of rational directions of 
periodic trajectories. As a consequence of this, at the corresponding 
accumulation points we must have the relations
$\, \epsilon_{1} ({\bf B}/B) \, = \, \tilde{\epsilon}_{1} ({\bf B}/B) \, $
or
$\, \epsilon_{2} ({\bf B}/B) \, = \, \tilde{\epsilon}_{2} ({\bf B}/B) \, $,
which is impossible in the specified energy intervals. One can also see that 
when $\, \epsilon_{F} = \epsilon^{\cal A}_{1} \, $ or
$\, \epsilon_{F} = \epsilon^{\cal A}_{2} \, $ the set of the directions of
$\, {\bf B} \, $, corresponding to the appearance of periodic trajectories
on the Fermi surface, may contain an infinite number of segments of big 
circles corresponding to an infinite number of rational directions 
of unstable open trajectories.

\vspace{1mm}

 Finally, in the intervals
$\, [\epsilon_{min} , \epsilon^{{\cal A}\prime}_{1} ] \, $ and
$\, [ \epsilon^{{\cal A}\prime}_{2} , \epsilon_{max} ] \, $
system (\ref{MFSyst}) cannot contain non-singular open trajectories.

\vspace{2mm}

 As a result of our reasoning, we can suggest the division of all angular 
diagrams into ``complexity classes'' according to the position of the 
Fermi level with respect to the set of ``reference'' points defined for 
a given (generic) dispersion relation. Namely:

\vspace{1mm}

(T1) The most simple diagrams:
$$\, \epsilon_{F} \,\,\, \in \,\,\, 
[\epsilon_{min} , \epsilon^{{\cal A}\prime}_{1} ] \, $$

\noindent
The Fermi surface can contain only closed or singular trajectories of 
the system (\ref{MFSyst}) at any direction of $\, {\bf B} \, $. 
The contribution to the Hall conductivity from the corresponding dispersion 
relation has the electron type and a constant value (for a given value of
$\, B \, $, $\, \omega_{B} \tau  \gg 1 $) at all directions of 
$\, {\bf B} \, $.

\vspace{1mm}

(T2) Simple enough diagrams:
$$\, \epsilon_{F} \,\,\, \in \,\,\, 
( \epsilon^{{\cal A}\prime}_{1} , \epsilon^{\cal A}_{1}] \, $$

\noindent
The Fermi surface can contain only closed, singular or unstable periodic 
open trajectories of the system (\ref{MFSyst}). The set of directions of
$\, {\bf B} \, $ corresponding to the appearance of periodic open trajectories 
is represented by a set of segments of big circles, each of which corresponds 
to a certain rational direction of periodic trajectories. For
$\, \epsilon_{F} \, \in \, ( \epsilon^{{\cal A}\prime}_{1} , 
\epsilon^{\cal A}_{1} ) \, $ the set of such segments is finite. For
$\, \epsilon_{F} = \epsilon^{\cal A}_{1} \, $ the number of segments 
and the corresponding rational directions of periodic open trajectories 
can be infinite. For generic directions of $\, {\bf B} \, $ the contribution 
to the Hall conductivity of the corresponding dispersion relation has the 
electron type and a constant value for a given value of $\, B \, $
($\, \omega_{B} \tau  \gg 1 $).

\vspace{1mm}

(T3) Complex type A diagrams:
$$\, \epsilon_{F} \,\,\, \in \,\,\, 
( \epsilon^{\cal A}_{1} , \epsilon^{\cal B}_{1}] \, $$
 
\noindent
In addition to closed and singular trajectories, a generic Fermi surface 
contains stable open trajectories and unstable periodic trajectories for 
certain directions of $\, {\bf B} \, $. Stable open trajectories possess 
the regularity properties described above and arise in special Stability Zones  
$\, \Omega_{\alpha} \, $ on the angular diagram, corresponding to certain values 
of topological quantum numbers
$\, (M^{1}_{\alpha}, M^{2}_{\alpha}, M^{3}_{\alpha}) \, $. For 
$\, \epsilon_{F} \, \in \, ( \epsilon^{\cal A}_{1} , \epsilon^{\cal B}_{1}) \, $
the number of Stability Zones at the angular diagram is finite. The set of 
directions of $\, {\bf B} \, $ corresponding to the appearance of periodic open 
trajectories is represented by an infinite set of segments of big circles, each 
of which corresponds to a certain rational direction of periodic trajectories.
For $\, \epsilon_{F} =  \epsilon^{\cal B}_{1} \, $ the set of Stability Zones
$\, \Omega_{\alpha} \, $ becomes infinite in generic case (i.e., for a
generic dispersion law); moreover, special directions of $\, {\bf B} \, $,
corresponding to the appearance of chaotic open trajectories on the 
Fermi surface, arise on the angular diagram. For generic directions of 
$\, {\bf B} \, $ outside the Zones $\, \Omega_{\alpha} \, $ the contribution 
to the Hall conductivity from the corresponding dispersion relation has 
the electron type and a constant value for a given value of 
$\, B \, $ ($\, \omega_{B} \tau  \gg 1 $).

\vspace{1mm}

(T4) Complex type B diagrams:
$$\, \epsilon_{F} \,\,\, \in \,\,\, 
( \epsilon^{\cal B}_{1} , \epsilon^{\cal B}_{2}) \, $$

\noindent
In addition to closed and singular trajectories, a generic Fermi surface contains 
stable open trajectories, unstable periodic trajectories, and chaotic trajectories 
for certain directions of $\, {\bf B} \, $. The set of Stability Zones
$\, \Omega_{\alpha} \, $, the set of additional segments of big circles, 
corresponding to the appearance of periodic open trajectories, and the set 
of special directions of $\, {\bf B} \, $, corresponding to the appearance 
of chaotic open trajectories on the Fermi surface, are in generic case infinite.
For generic directions of $\, {\bf B} \, $ outside the Zones 
$\, \Omega_{\alpha} \, $ the contribution to the Hall conductivity
from the corresponding dispersion relation has different (electron and hole) 
types in different areas of the angular diagram, separated in generic  case 
by infinite ``chains'' of Stability Zones and ``chaotic'' directions of 
$\, {\bf B} \, $ ($\, \omega_{B} \tau  \gg 1 $).

\vspace{1mm}

(T5) Complex type A diagrams:
$$\, \epsilon_{F} \,\,\, \in \,\,\, 
[ \epsilon^{\cal B}_{2} , \epsilon^{\cal A}_{2} ) \, $$

\noindent
In addition to closed and singular trajectories, a generic Fermi surface 
contains stable open trajectories and unstable periodic trajectories for certain 
directions of $\, {\bf B} \, $. For
$\, \epsilon_{F} \, \in \, ( \epsilon^{\cal B}_{2} , \epsilon^{\cal A}_{2} ) \, $
the number of Stability Zones at the angular diagram is finite. The set of 
directions of $\, {\bf B} \, $ corresponding to the appearance of periodic open 
trajectories is represented by an infinite set of segments of big circles, each 
of which corresponds to a certain rational direction of periodic trajectories. 
For $\, \epsilon_{F} =  \epsilon^{\cal B}_{2} \, $ the set of Stability Zones
$\, \Omega_{\alpha} \, $ is infinite in generic case, besides that, 
special directions of $\, {\bf B} \, $, corresponding to the appearance of 
chaotic open trajectories on the Fermi surface, present on the angular diagram.
For generic directions of $\, {\bf B} \, $ outside the Zones 
$\, \Omega_{\alpha} \, $ the contribution to the Hall conductivity from the 
corresponding dispersion relation has the hole type and a constant value for 
a given value of $\, B \, $ ($\, \omega_{B} \tau  \gg 1 $).

\vspace{1mm}

(T6) Simple enough diagrams:
$$\, \epsilon_{F} \,\,\, \in \,\,\, 
[ \epsilon^{\cal A}_{2} , \epsilon^{{\cal A}\prime}_{2} ) \, $$

\noindent
The Fermi surface can contain only closed, singular or unstable periodic 
open trajectories of the system (\ref{MFSyst}). The set of directions of 
$\, {\bf B} \, $, corresponding to the appearance of periodic open trajectories, 
is represented by a set of segments of big circles, each of which corresponds 
to a certain rational direction of periodic trajectories. For 
$\, \epsilon_{F} \, \in \, ( \epsilon^{\cal A}_{2} , 
\epsilon^{{\cal A}\prime}_{2} ) \, $
the set of such segments is finite. For
$\, \epsilon_{F} = \epsilon^{\cal A}_{2} \, $ the number of segments and 
the corresponding rational directions of periodic open trajectories can be 
infinite. For generic directions of $\, {\bf B} \, $ the contribution to 
the Hall conductivity of the corresponding dispersion relation has the hole 
type and a constant value for a given value of $\, B \, $
($\, \omega_{B} \tau  \gg 1 $).

\vspace{1mm}

(T7) The most simple diagrams:
$$\, \epsilon_{F} \,\,\, \in \,\,\, 
[ \epsilon^{{\cal A}\prime}_{2} , \epsilon_{max} ] \, $$

\noindent
The Fermi surface can contain only closed or singular trajectories of the system
(\ref{MFSyst}) at any direction of $\, {\bf B} \, $. The contribution to the Hall 
conductivity from the corresponding dispersion relation has the hole type and 
a constant value (for a given value of  $\, B \, $,
$\, \omega_{B} \tau  \gg 1 $) for all directions of $\, {\bf B} \, $.

\vspace{2mm}

 Thus, all angular diagrams of conductivity in metals in strong magnetic fields 
can be assigned to one of the above classes (T1) - (T7), determined by the position 
of the Fermi level and the dispersion law $\, \epsilon ({\bf p}) \, $. Note again 
that the full picture of the angle diagrams for all values of $\, \epsilon_{F} \, $ 
is somewhat abstract in nature, since only the diagram for the real value of
$\, \epsilon_{F} \, $ is experimentally observable. In particular, in experimental 
change of the position of $\, \epsilon_{F} \, $ (for example, using external pressure) 
the change in the corresponding conductivity diagram is not required to be determined 
by the initial relation $\, \epsilon ({\bf p}) \, $, since the corresponding external 
influence changes also the parameters of the dispersion relation. The above remark, 
however, does not cancel the fact that the type of a specific angular diagram is 
determined unambiguously on the basis of the theoretical consideration given above.
Let us note again that our considerations here can be considered as quite general
for arbitrary ``physically realistic'' dispersion relation $\, \epsilon ({\bf p}) \, $.
At the same time, we omit here some theoretically possible additional features
(f.e. non-simply-connectedness of Stability Zones) which are extremely unlikely
in a real situation.

 Another remark that can be made here relates to the fact that, in the general case, 
the full Fermi surface represents the union of several components related to different
dispersion relations. It is easy to see that the angular diagram of conductivity is 
given in this case by the ``overlapping'' of the diagrams defined by all the components.
An important circumstance here is that if different components do not intersect 
with each other, then the intersection of the corresponding Stability Zones is possible 
only if these Zones correspond to the same topological numbers
$\, (M^{1}_{\alpha}, M^{2}_{\alpha}, M^{3}_{\alpha}) \, $.
As a consequence, such intersecting Zones can be viewed as complex ``composite'' 
Zones with a more complex border structure.

\vspace{1mm}

 Here, of course, we should also say a few words about angular diagrams 
(for fixed values of $\, \epsilon_{F} $) arising for dispersion relations 
with only one Stability Zone $\, \Omega \, $ (occupying the entire sphere
$\, \mathbb{S}^{2} $). It must be said that the dispersion relations of this 
type also represent generic relations and may well occur in real crystals.
(In particular, such relations include relations for which in a certain energy 
interval the surfaces $\, \epsilon ({\bf p}) \, = \, const \, $ represent a set 
of periodically deformed (``corrugated'') integral planes that are not connected 
to each other.) We omit here a detailed analysis of the situation arising in this 
case and give only a general description of the complexity classes of angular 
diagrams arising here for different values of $\, \epsilon_{F} \, $.

 Namely, as in the case of dispersion relations with ``complex'' angular diagrams, 
in the situation we are considering it is also possible in generic case to introduce 
some ``reference'' values
$$\epsilon_{\min} \, < \, \hat{\epsilon}^{{\cal A}\prime}_{1} \, < \,
\hat{\epsilon}^{\cal A}_{1} \, < \, \hat{\epsilon}^{\cal B}_{1} \, < \, 
\hat{\epsilon}^{\cal B}_{2} \, < \, \hat{\epsilon}^{\cal A}_{2} \, < \, 
\hat{\epsilon}^{{\cal A}\prime}_{2} \, < \, \epsilon_{\max} $$
and classes of diagrams (T1) - (T7), similar to those considered above.
Here the classes (T1), (T2), (T6), (T7) completely coincide with the corresponding 
classes described above, and for the diagrams of the classes (T3) and (T5), only 
the words ``the number of Stability Zones at the angular diagram is finite'' need 
to be replaced to  ``we have one Stability Zone $\, \Omega \, $, occupying part of 
the sphere $\, \mathbb{S}^{2} \, $'' .  (We do not require here the connectedness of 
the Zone $\, \Omega \, $ and, in particular, we consider as the same Zone 
possibly separated opposite parts on $\, \mathbb{S}^{2} $.) As for the class (T4), 
the diagrams of this class contain here just one Stability Zone $\, \Omega \, $,
which occupies the whole sphere $\, \mathbb{S}^{2} \, $. Note here, that for the 
class (T4) theoretically we could also have a Stability Zone $\, \Omega \, $,
occupying a part of $\, \mathbb{S}^{2} \, $, with different signs of the Hall
conductivity in different regions outside $\, \Omega \, $. This situation,
however, corresponds to non-simply-connected Zone $\, \Omega \, $ and physically
is extremely unlikely, so, we do not bring it here.

 Note also that the above mentioned ``reference'' energy values are quite simply 
related to the ``reference'' values defined for dispersion relations, which have 
``complex'' angular diagrams. Thus, we can actually write the relations
$$\hat{\epsilon}^{{\cal A}\prime}_{1} \, = \, 
\epsilon^{{\cal A}\prime}_{1} \,\, , \quad
\hat{\epsilon}^{\cal A}_{1} \, = \, 
\epsilon^{\cal A}_{1} \,\, , \quad
\hat{\epsilon}^{\cal A}_{2} \, = \, 
\epsilon^{\cal A}_{2} \,\, , \quad
\hat{\epsilon}^{{\cal A}\prime}_{2} \, = \, 
\epsilon^{{\cal A}\prime}_{2} $$
according to the definition of the values
$\, \epsilon^{{\cal A}\prime}_{1} \, $,
$\, \epsilon^{\cal A}_{1} \, $, $\, \epsilon^{\cal A}_{2} \, $,
$\, \epsilon^{{\cal A}\prime}_{2} \, $, given above.

 As for the values $\, \hat{\epsilon}^{\cal B}_{1} \, $ and 
$\, \hat{\epsilon}^{\cal B}_{2} \, $, for them we have the relations
$$\hat{\epsilon}^{\cal B}_{1} \,\,\, = \,\,\,
\epsilon^{\cal B}_{2} \,\,\, = \,\,\, \max \,\,
\tilde{\epsilon}_{1} ({\bf B}/B) \,\,\, , $$
$$\hat{\epsilon}^{\cal B}_{2} \,\,\, = \,\,\,
\epsilon^{\cal B}_{1} \,\,\, = \,\,\, \min \,\, 
\tilde{\epsilon}_{2} ({\bf B}/B) \,\,\, ,$$
since for the dispersion laws of the type under consideration we have in
our case the relation 
$\, \epsilon^{\cal B}_{1} \, > \, \epsilon^{\cal B}_{2} \, $.
(Note again, that theoretically also the situation
$\, \epsilon^{\cal B}_{1} < \epsilon^{\cal B}_{2} \, $ and 
$\, \hat{\epsilon}^{\cal B}_{1} = \epsilon^{\cal B}_{1} \, $,
$\,\, \hat{\epsilon}^{\cal B}_{2} = \epsilon^{\cal B}_{2} \, $
is possible here but is extremely unlikely according to our remarks above). 

 It is not difficult to see also that in this case no angular diagrams 
can contain directions of $\, {\bf B} \, $ corresponding to the appearance 
of ``chaotic'' trajectories on the Fermi surface.

\vspace{2mm}

 In conclusion, we make here a comparison of the complexity of the location 
of the Stability Zones on angular diagrams for fixed Fermi surfaces with the 
location of the Zones $\, \Omega^{*}_{\alpha} \, $ on the diagrams for the 
entire dispersion law. More specifically, we will consider here an analogue 
of the structure of Zones, presented at Fig. \ref{Adjacent}, for diagrams
defined for a fixed value of $\, \epsilon_{F} \, $.

 As we have already seen, angular diagrams of conductivity for fixed values 
of $\, \epsilon_{F} \, $ may, in particular, be quite simple and not have a 
level of complexity comparable to the level of complexity of angular diagrams  
for the entire relation  $\, \epsilon ({\bf p}) \, $. We, therefore, will be 
interested in diagrams (of type A or B) containing the Stability Zones
$\, \Omega_{\alpha} \, $, which are close in shape to the corresponding 
Stability Zones $\, \Omega^{*}_{\alpha} \, $, defined for the entire dispersion 
relation. As it is not difficult to see, such Zones are Stability Zones, the 
first and second borders of which are located quite close to each other 
(Fig. \ref{ZoneEvolution}, b, c, d). As we have indicated earlier, in the 
domain $\, \Omega^{\prime}_{\alpha} \, $, bounded by the first and second 
boundary of the Zone $\, \Omega_{\alpha} \, $, there can not appear stable
open trajectories of system (\ref{MFSyst}), and, correspondingly, there can 
be no other Stability Zones for a given value of $\, \epsilon_{F} \, $. 
It can be seen, therefore, that structures similar to those shown at 
Fig. \ref{Adjacent}, if they arise, must be connected in this case with 
the second boundary of the Zone $\, \Omega_{\alpha} \, $.

 To describe the picture that appears in our case, it is necessary to turn 
again to the structure of the Fermi surface at 
$\, {\bf B}/B \, \in \, \Omega_{\alpha} \, $ (Fig. \ref{FullFermSurf}),
as well as to the mechanism of destruction of this structure on the first 
and second boundaries of $\, \Omega_{\alpha} \, $ by the ``jumps'' of 
trajectories from one carrier of open trajectories to another represented 
at Fig. \ref{BoundCylind}. To simplify the description, it is convenient 
to consider the projection of the `carriers of open trajectories' on the 
plane $\, \Pi ({\bf B}/B) \, $  generated by the direction of 
$\, {\bf B} \, $, and also the direction of intersection of the plane 
orthogonal to $\, {\bf B} \, $ and the plane $\, \Gamma_{\alpha} \, $.
All trajectories of system (\ref{MFSyst}) will be represented in this plane 
by parallel straight lines orthogonal to $\, {\bf B} \, $. After crossing 
the first boundary of $\, \Omega_{\alpha} \, $ and arising of trajectories 
jumps from one carrier to another (Fig. \ref{BoundCylind}), for each carrier 
on the projection we can mark ``places'', where trajectories jump between 
pairs of ``merged'' carriers. In this case, both former carriers from each 
resulting pair can be represented by the same projection on the plane
$\, \Pi ({\bf B}/B) \, $, and the ``places'' where the jumps occur can be 
denoted by circles of a small radius (although it is not very precise).
It can be said that such circles represent ``traps'', falling into which 
a trajectory jumps from one former carrier of open trajectories 
(for a fixed pair) to another and back. 

 Each of the carriers of open trajectories represents a periodically deformed 
integral plane in $\, {\bf p}$ - space and, thus, a picture arising in the 
plane $\, \Pi ({\bf B}/B) \, $ is always doubly periodic. (Note also that 
the holes (flat disks) in the carriers of open trajectories disappear on the 
projection, being orthogonal to $\, {\bf B}$). In particular, the arrangement 
of the ``traps'' described by us for each pair of ``merged'' carriers of open 
trajectories is doubly periodic.

\begin{figure}[t]
\begin{center}
\includegraphics[width=\linewidth]{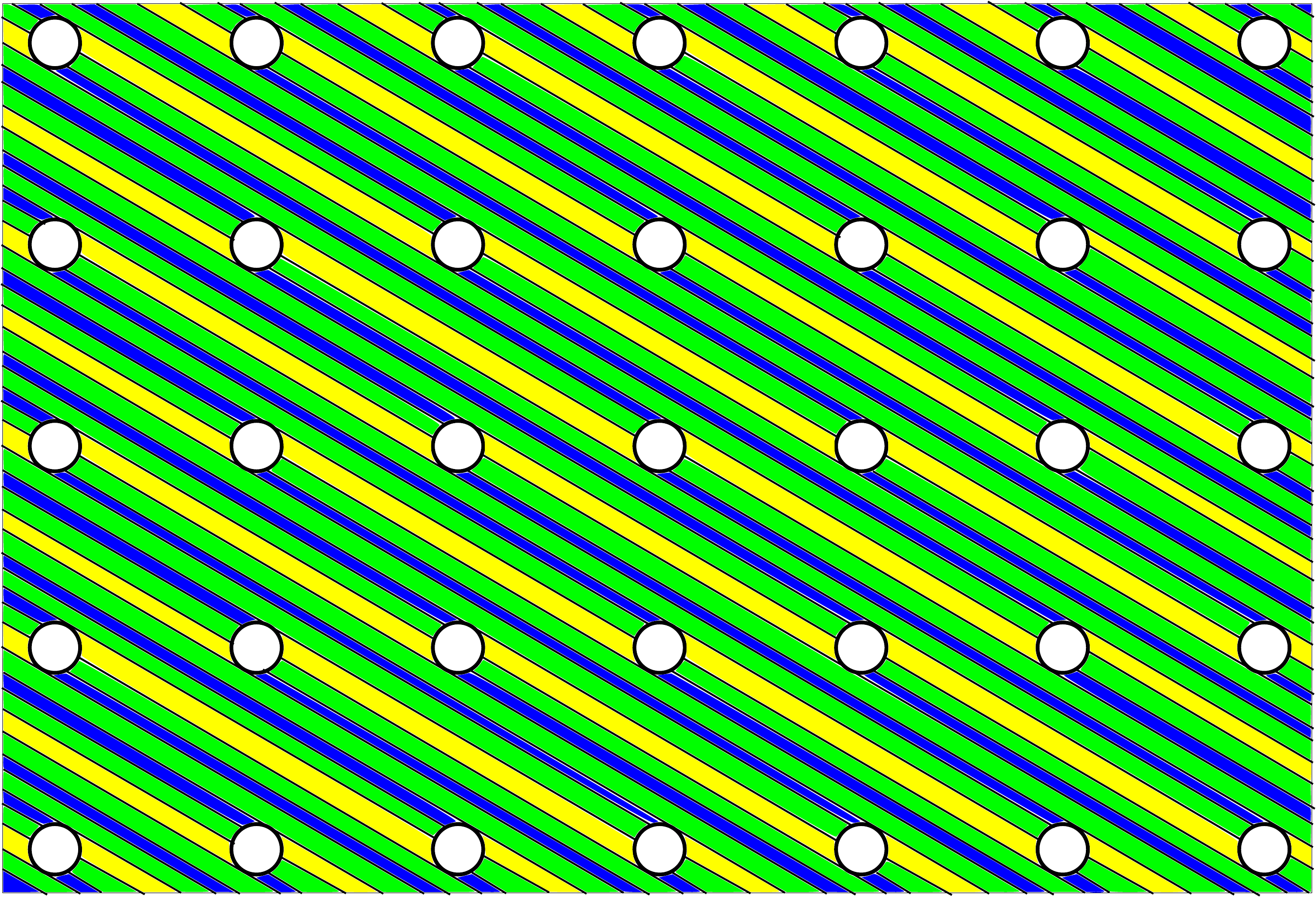}
\end{center}
\begin{center}
\includegraphics[width=\linewidth]{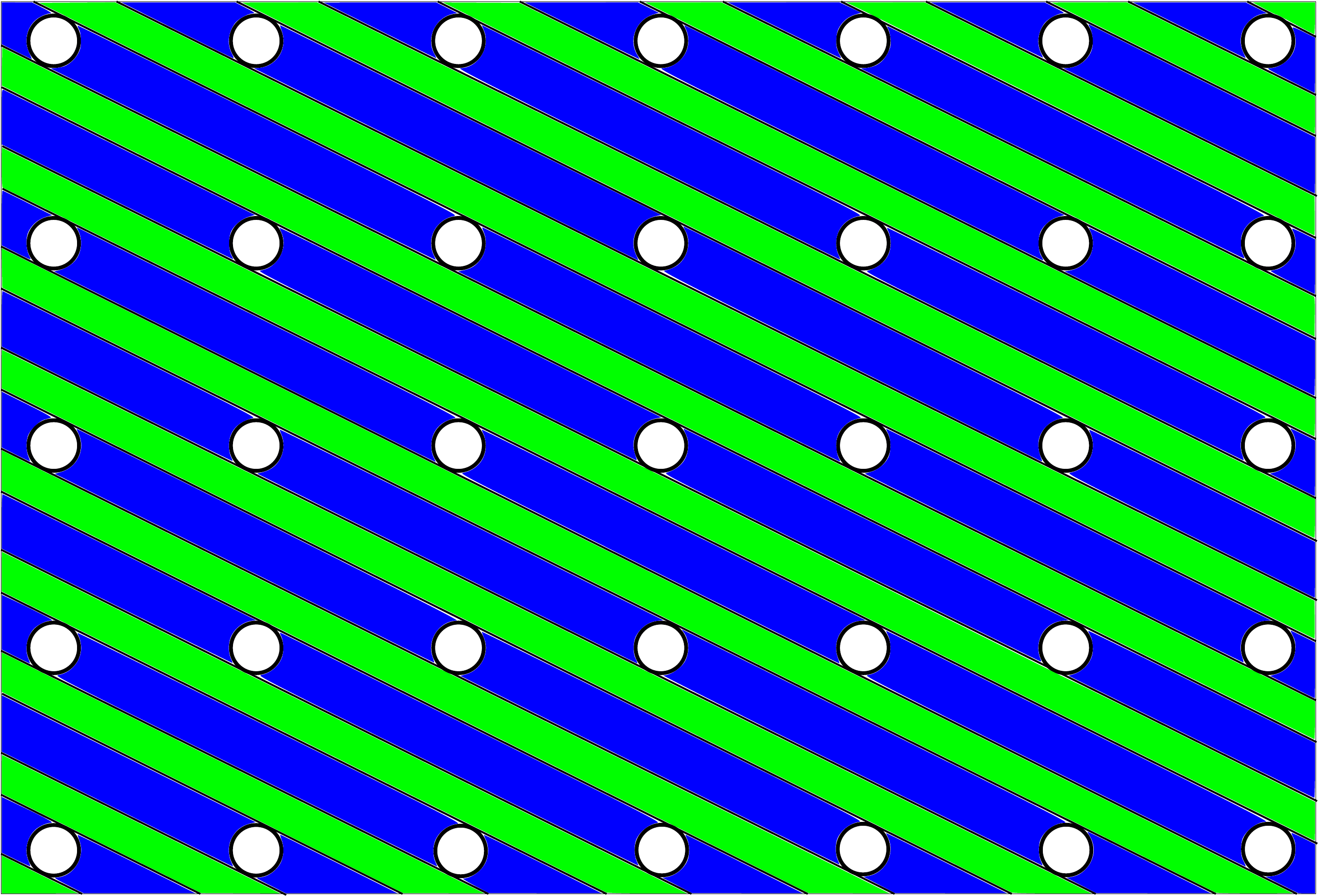}
\end{center}
\caption{(a) Projections of long closed trajectories of system (\ref{MFSyst})
arising after crossing the boundary of a Stability Zone for an irrational mean 
direction of open trajectories. (b) Projections of long closed and open periodic 
trajectories of system (\ref{MFSyst}) present after crossing the boundary of 
a Stability Zone for a rational mean direction of open trajectories.}
\label{PerNet}
\end{figure}

 Straight lines that represent the trajectories of the system (\ref{MFSyst})
after crossing the boundary of $\, \Omega_{\alpha} \, $ become in general
``cut'' into segments of finite length. It is not difficult to see that each 
of these segments, bounded by ``traps'' at both its ends, represents a (long) 
closed trajectory of the system (\ref{MFSyst}) arising on a pair of former 
carriers of open trajectories after crossing the boundary of a Stability Zone.
It can also be seen that if the intersection of a plane, orthogonal to
$\, {\bf B} \, $, and $\, \Gamma_{\alpha} \, $ has an irrational direction, 
then all former open trajectories turn into closed ones after crossing the 
boundary of $\, \Omega_{\alpha} \, $ (Fig. \ref{PerNet}, a). At the same time, 
if the mean direction of open trajectories is rational, then immediately after 
crossing the first boundary of $\, \Omega_{\alpha} \, $ on a pair of former 
carriers of open trajectories both long closed trajectories and ``intact''
periodic open trajectories will be present (Fig. \ref{PerNet}, b).

 The diameters of the ``traps'' tend to zero when approaching the boundary of 
the Zone $\, \Omega_{\alpha} \, $ from the outside and increase when moving away 
from it (near $\, \Omega_{\alpha}$). It can be seen here that for each rational 
direction of open trajectories there actually exists a critical radius of 
``traps'', for which the layers of open trajectories disappear due to their 
complete overlap by ``traps''. In addition, it can also be seen that the critical 
radius of ``traps'' decreases rather quickly with increasing of ``denominators'' 
of the corresponding rational directions of the trajectories. As a consequence, 
the segments adjacent to the boundary of a Stability Zone and corresponding to 
the appearance of unstable periodic trajectories on the Fermi surface
(Fig. \ref{DirOutStabZone}) form an everywhere dense set on the boundary of
$\, \Omega_{\alpha} \, $, however, at any finite distance from the boundary 
the number of such segments is finite. For our consideration here, the segments 
that reach the second boundary of a Stability Zone will be important.

\begin{figure}[t]
\begin{center}
\includegraphics[width=\linewidth]{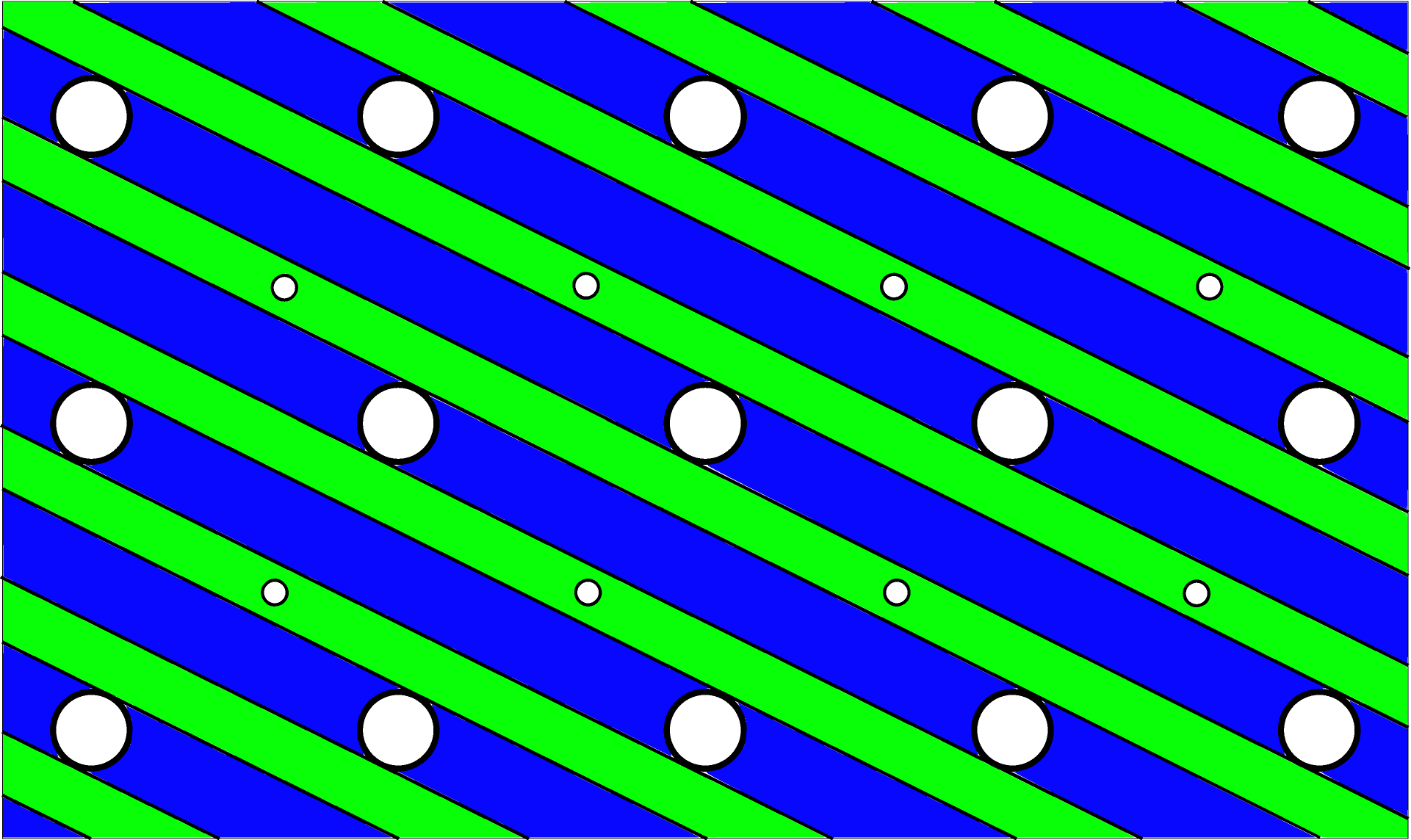}
\end{center}
\caption{Traps of the second type, arising after crossing the second boundary 
of the Zone $\, \Omega_{\alpha} \, $.}
\label{NewTraps}
\end{figure}

 As was noted above, in each connected component of the domain
$\, \Omega^{\prime}_{\alpha} \, $, adjacent to a certain part of the first 
boundary of the Zone $\, \Omega_{\alpha} \, $, for all generic directions of 
$\, {\bf B} \, $ the same type of the Hall conductivity appear, which is opposite 
to the type of the cylinder of closed trajectories vanishing at the corresponding 
part of the first boundary of $\, \Omega_{\alpha} \, $. At the corresponding 
section of the second boundary of $\, \Omega_{\alpha} \, $ the cylinder of 
closed trajectories of the opposite type (relative to the type of the cylinder 
disappearing at the first boundary) disappears and trajectories start to jump 
between pairs of ``merged'' former carriers of open trajectories 
(Fig. \ref{FullFermSurf}). Thus, on the above projection of the former 
carriers of open trajectories on the plane $\, \Pi ({\bf B}/B) \, $ 
``traps of the second type'' appear, getting into which a trajectory jumps 
between pairs of ``merged'' carriers. Like the ``traps of the first type'', 
the ``traps of the second type'' form a periodic structure in the plane 
$\, \Pi ({\bf B}/B) \, $ (Fig. \ref{NewTraps}).

 At Fig. \ref{NewTraps} we have actually represented the projections of only 
half of the ``traps of the second type'' arising on a pair of ``merged'' carriers 
of open trajectories, namely, the ``traps'' lying on the ``upper'' carrier and 
corresponding to jumping to one of the neighboring pairs of similar carriers.
The projection of the neighboring pair of ``merged'' carriers onto the plane 
$\, \Pi ({\bf B}/B) \, $ coincides with the projection of the pair under 
consideration shifted by some vector in this plane. It can be said, therefore, 
that each of the ``traps of the second type'' actually has an upper and a lower 
base, lying in different parts of the picture, represented at
Fig. \ref{PerNet}a or Fig. \ref{PerNet}b.

 The diameters of the ``traps of the second type'' tend to zero when 
approaching the second boundary of the Zone $\, \Omega_{\alpha} \, $ from 
the outside and increase when moving away from it (near $\, \Omega_{\alpha}$). 
After crossing the second boundary of $\, \Omega_{\alpha} \, $ 
``traps of the second type'' arise in the form of points at Fig. \ref{PerNet}a 
or Fig. \ref{PerNet}b, having ``upper and lower bases'' on cylinders of closed 
trajectories or on layers of periodic trajectories (if any) of the system 
(\ref{MFSyst}).

 Here we will consider only ``physical'' dispersion laws that satisfy 
the requirement $\, \epsilon ({\bf p}) \, = \, \epsilon (- {\bf p}) \, $, 
which also implies invariance of the Fermi surface under the transformation
$\, {\bf p} \, \rightarrow \, - {\bf p} \, $. It is not difficult to check 
then, that the invariance of the structure of the system (\ref{MFSyst}) 
with respect to the same transformation also implies in our situation 
the symmetry of the upper and lower bases of the ``traps of the second type''.
In particular, each newly appeared after crossing the second boundary of the 
Zone $\, \Omega_{\alpha} \, $ (small) ``trap of the second type'' interconnects 
(in the generic case) either long closed trajectories of the system (\ref{MFSyst})
or periodic open trajectories. As can be seen, these two cases correspond 
to substantially different structures of trajectories of system (\ref{MFSyst}) 
 on the Fermi surface after crossing the second boundary of the Zone
$\, \Omega_{\alpha} \, $.

\begin{figure}[t]
\begin{center}
\includegraphics[width=\linewidth]{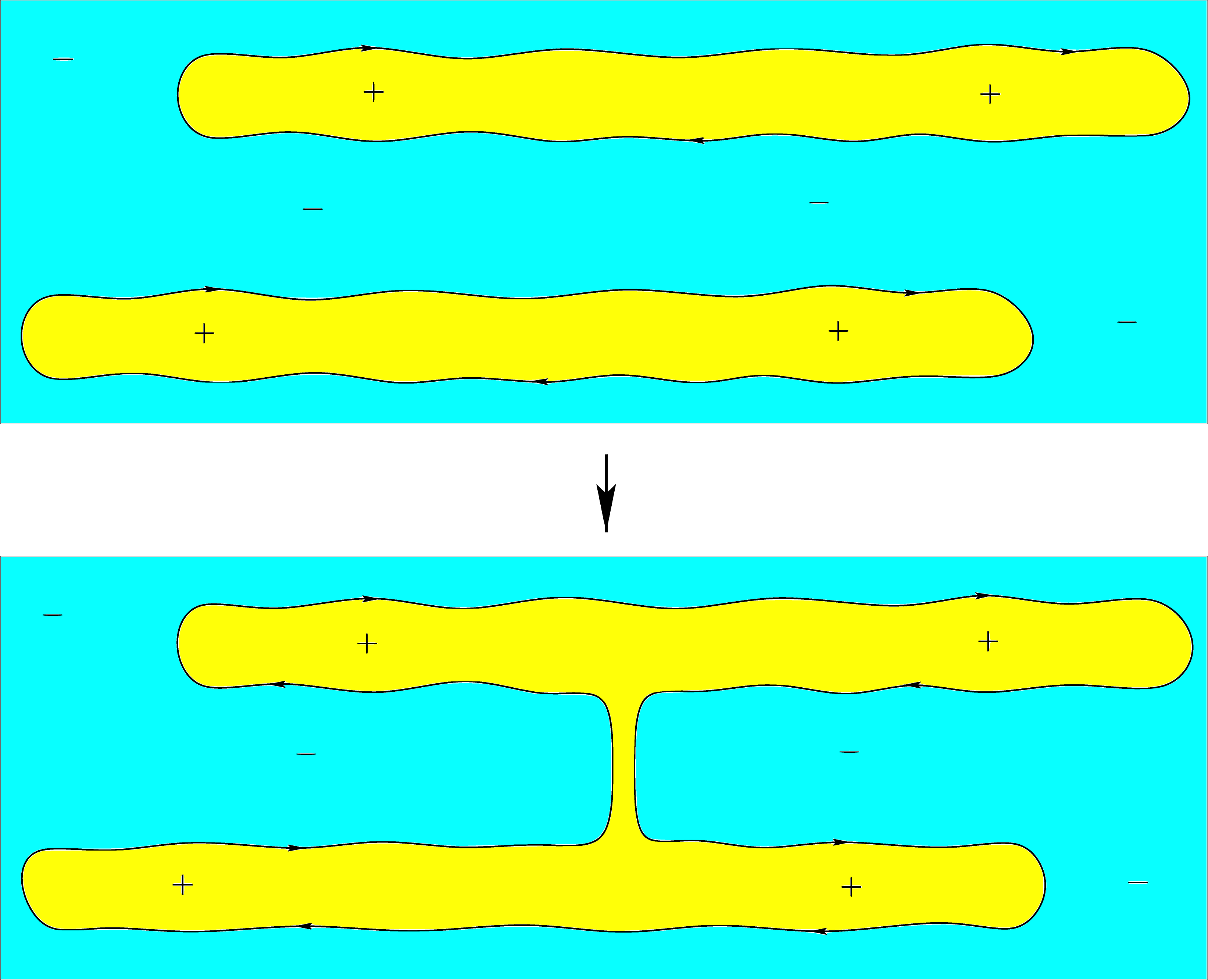}
\end{center}
\caption{The reconstruction of trajectories of system (\ref{MFSyst}) after
crossing of the second boundary of $\, \Omega_{\alpha} \, $  in the case when 
the ``traps of the second type'' appear on the cylinders of closed trajectories.}
\label{Perestr1}
\end{figure}

 Indeed, as is not difficult to see, the reconstruction of the second type 
corresponds to the picture presented at Fig. \ref{Perestr1}. It can also be seen 
that, both before the restructuring and after the restructuring of the 
trajectories, the Fermi surface contains only closed trajectories of the system
(\ref{MFSyst}) and pictures in all planes orthogonal to  $\, {\bf B} \, $
have the same type (I or II) in both cases. Thus, after the reconstruction of
the trajectories described above, the corresponding directions of
$\, {\bf B} \, $  lie outside any of the Stability Zones $\, \Omega_{\beta} \, $ 
and the total contribution of the Fermi surface to the Hall conductivity does not 
change. This situation takes place, in particular, when crossing the second boundary 
of the Zone $\, \Omega_{\alpha} \, $  at generic points, as well as at points 
corresponding to rational directions of intersection of $\, \Gamma_{\alpha} \, $ 
and the plane orthogonal to $\, {\bf B} \, $ if the diameter of the 
``traps of the first type'' exceeds the critical one (the corresponding 
``additional segments'' do not reach the second boundary).

\begin{figure}[t]
\begin{center}
\includegraphics[width=\linewidth]{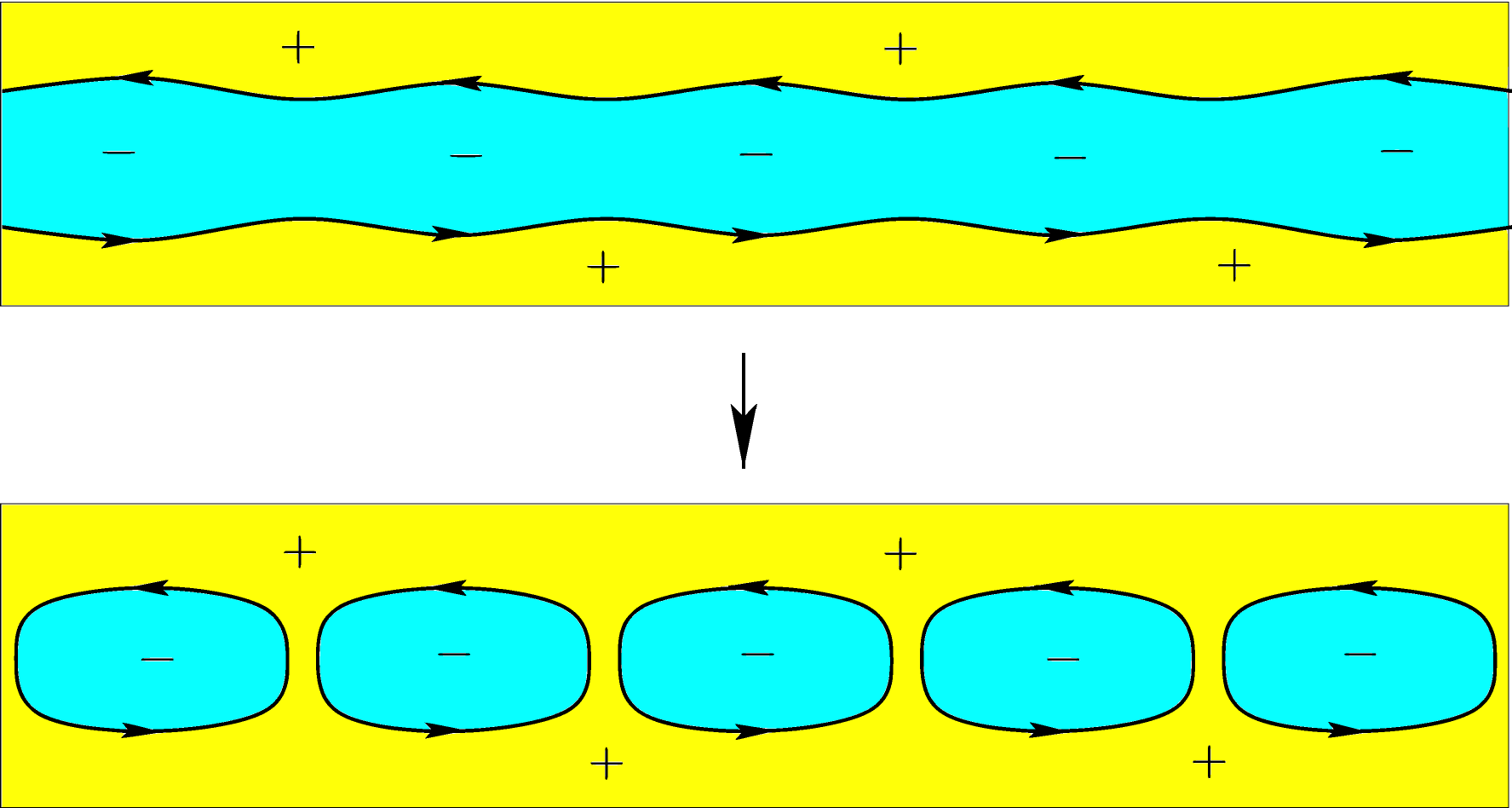}
\end{center}
\caption{The reconstruction of trajectories of system (\ref{MFSyst}) after
crossing of the second boundary of $\, \Omega_{\alpha} \, $  in the case when 
the ``traps of the second type'' appear on layers of periodic open trajectories.}
\label{Perestr2}
\end{figure}

\begin{figure}[t]
\begin{center}
\includegraphics[width=\linewidth]{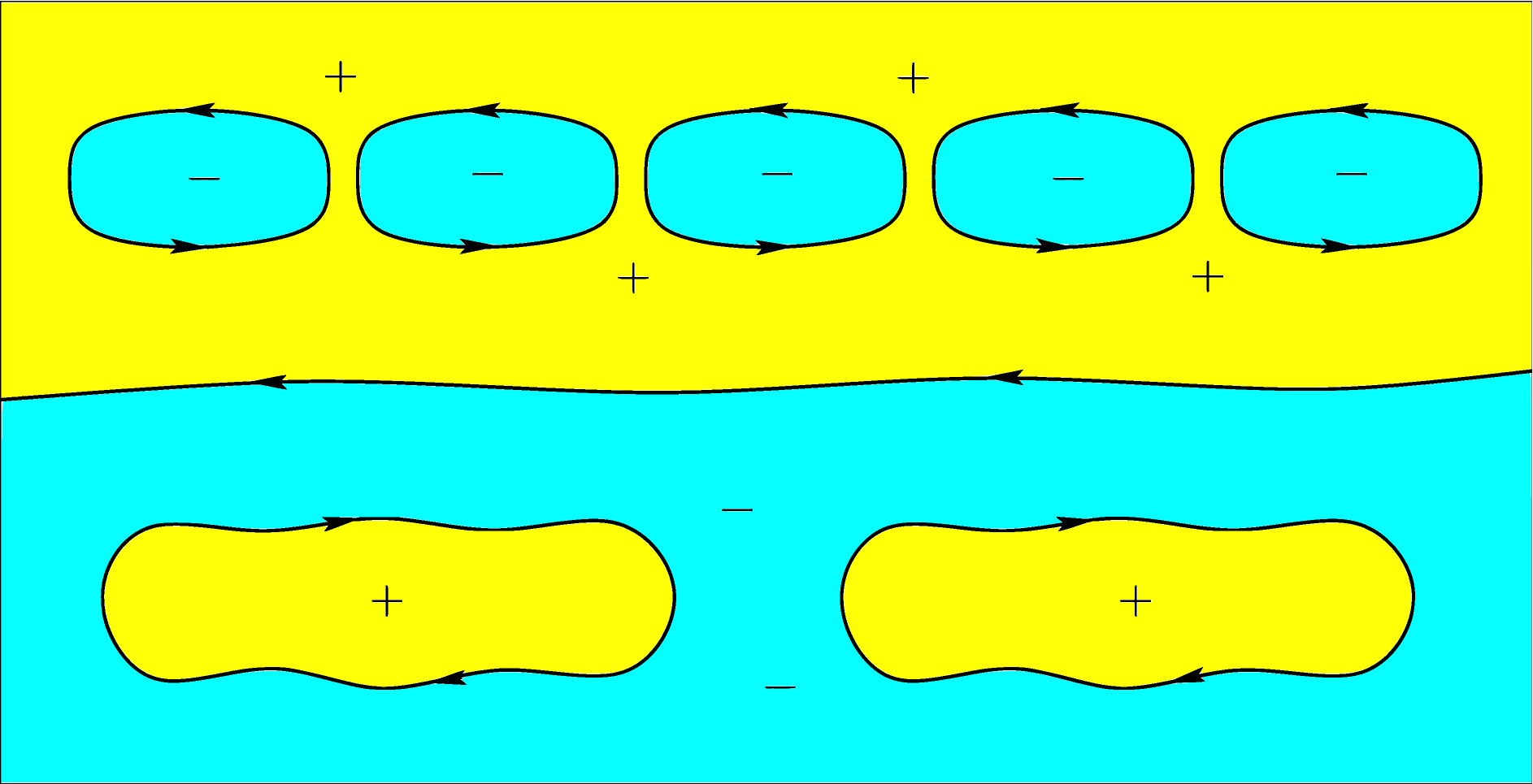}
\end{center}
\caption{Situations I and II in a plane orthogonal to  $\, {\bf B} \, $,
separated by open trajectories of (\ref{MFSyst}).}
\label{TwoPictures}
\end{figure}

 In another case, when the ``traps of the second type'' appear on the layers 
of periodic open trajectories of the system (\ref{MFSyst}), the corresponding 
reconstruction of the trajectories can be represented by the picture shown at 
Fig. \ref{Perestr2}. In this case both pictures (I and II) appear in the 
planes orthogonal to $\, {\bf B} \, $ being separated by open trajectories of 
the system (\ref{MFSyst}) (Fig. \ref{TwoPictures}). Such a situation can occur 
only when the second boundary of a Stability Zone is crossed at a point 
corresponding to the presence of periodic open trajectories of (\ref{MFSyst})
on the Fermi surface, i.e. at a point of intersection of the second boundary 
with an additional segment adjacent to the boundary of the Stability Zone 
and corresponding to the appearance of unstable periodic trajectories on the 
Fermi surface. We note at the same time, that even in this case the situation 
can also be described by the picture presented at Fig. \ref{Perestr1}
if the ``traps of the second type'' appear on cylinders of closed trajectories.

 The closed trajectories shown at Fig. \ref{TwoPictures} cut the Fermi surface
into new carriers of open trajectories, corresponding to the appearance of a new
Stability Zone  $\, \Omega_{\beta} \, $. The Zone $\, \Omega_{\beta} \, $, 
defined for a fixed Fermi surface, belongs to one of the Zones 
$\, \Omega^{*}_{\beta} \, $ shown at Fig. \ref{Adjacent}. 

\begin{figure}[t]
\begin{center}
\includegraphics[width=\linewidth]{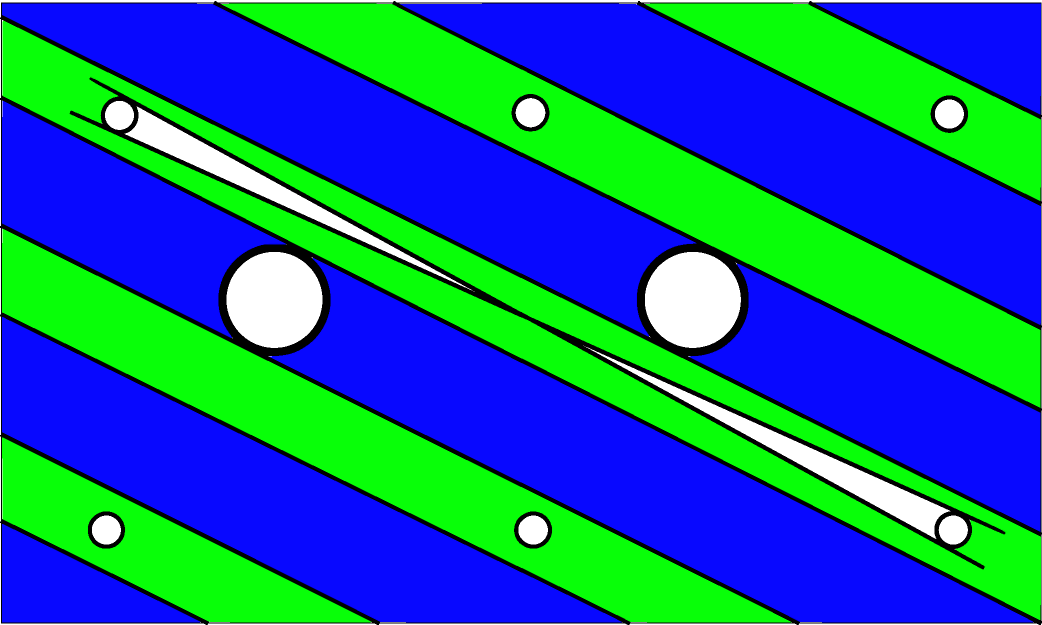}
\end{center}
\caption{The allowable angle of deviation of the intersection of
$\, \Gamma_{\alpha} \, $ with the plane orthogonal to $\, {\bf B} \, $
for a fixed diameter of the ``traps of the second type''.}
\label{AdmAngle}
\end{figure}

\begin{figure}[t]
\begin{center}
\includegraphics[width=\linewidth]{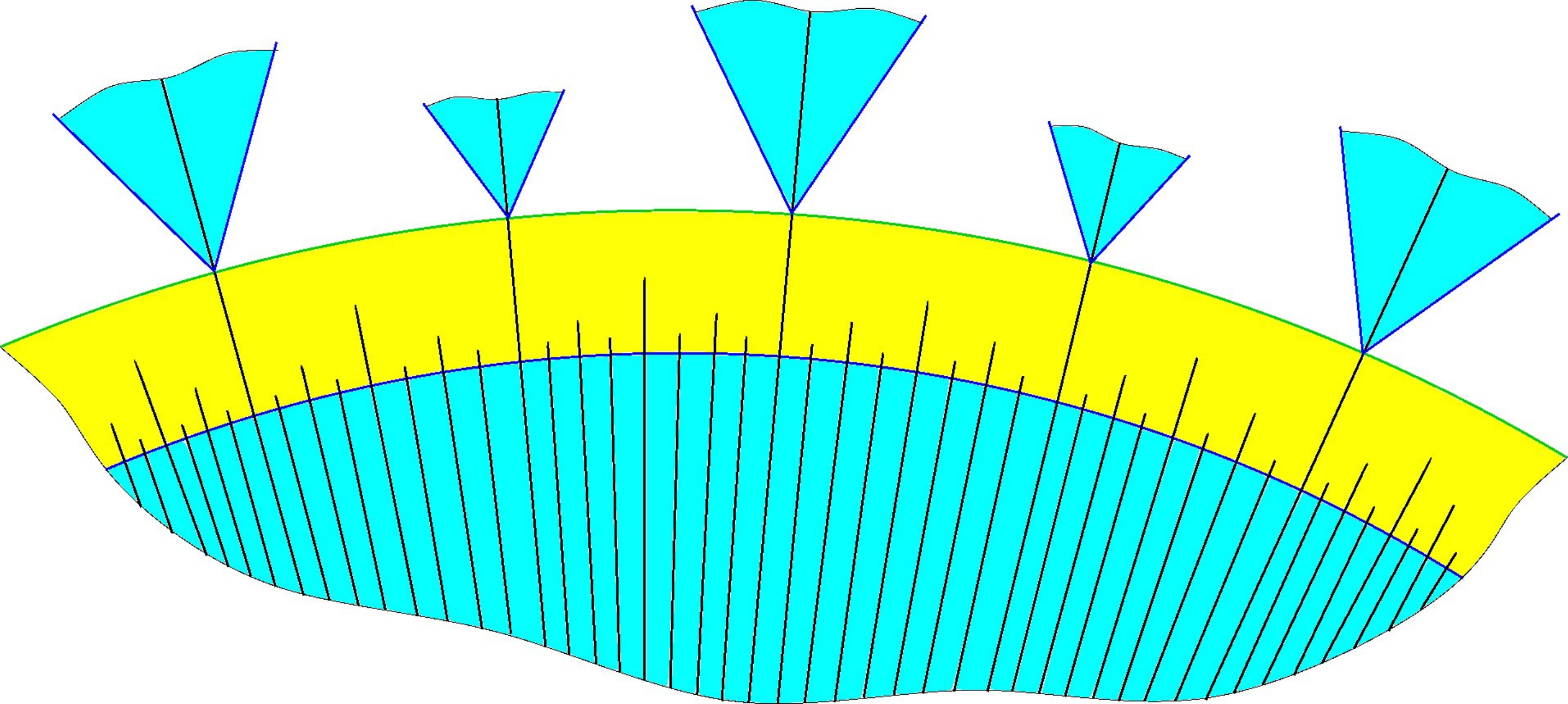}
\end{center}
\caption{The Stability Zones $\, \Omega_{\beta} \, $ adjacent to the second 
boundary of the Zone $\, \Omega_{\alpha} \, $ in part of its ``rational'' 
points.}
\label{CornerOmegaBeta}
\end{figure}

 The picture presented at Fig. \ref{TwoPictures} is locally stable 
(with respect to small rotations of the direction of $\, {\bf B}$)
and is preserved until a reconstruction of the closed trajectories shown
on it occurs. It can be seen that near the second boundary of the Zone
$\, \Omega_{\alpha} \, $ the region of stability of the picture shown at 
Fig. \ref{TwoPictures} is determined by the diameter of the 
``traps of the second type'', which in this situation is small compared 
to the diameter of the ``traps of the first type''. Thus, the allowable angle 
of deviation of the direction of intersection of the plane 
$\, \Gamma_{\alpha} \, $ with the plane orthogonal to $\, {\bf B} \, $ 
for a fixed small diameter of the ``traps of the second type'' is proportional 
to this diameter, which determines the allowable angle of deviation of the 
direction of $\, {\bf B} \, $ in parallel to the second boundary
(Fig. \ref{AdmAngle}). Since the diameter of the ``traps of the second type''
is proportional to the distance to the second boundary of the Zone
$\, \Omega_{\alpha} \, $, it can be seen also that the boundary of the
Zone $\, \Omega_{\beta} \, $ must have a ``corner'' on the second boundary 
of $\, \Omega_{\alpha} \, $ (Fig. \ref{CornerOmegaBeta}).

\begin{figure}[t]
\begin{center}
\includegraphics[width=\linewidth]{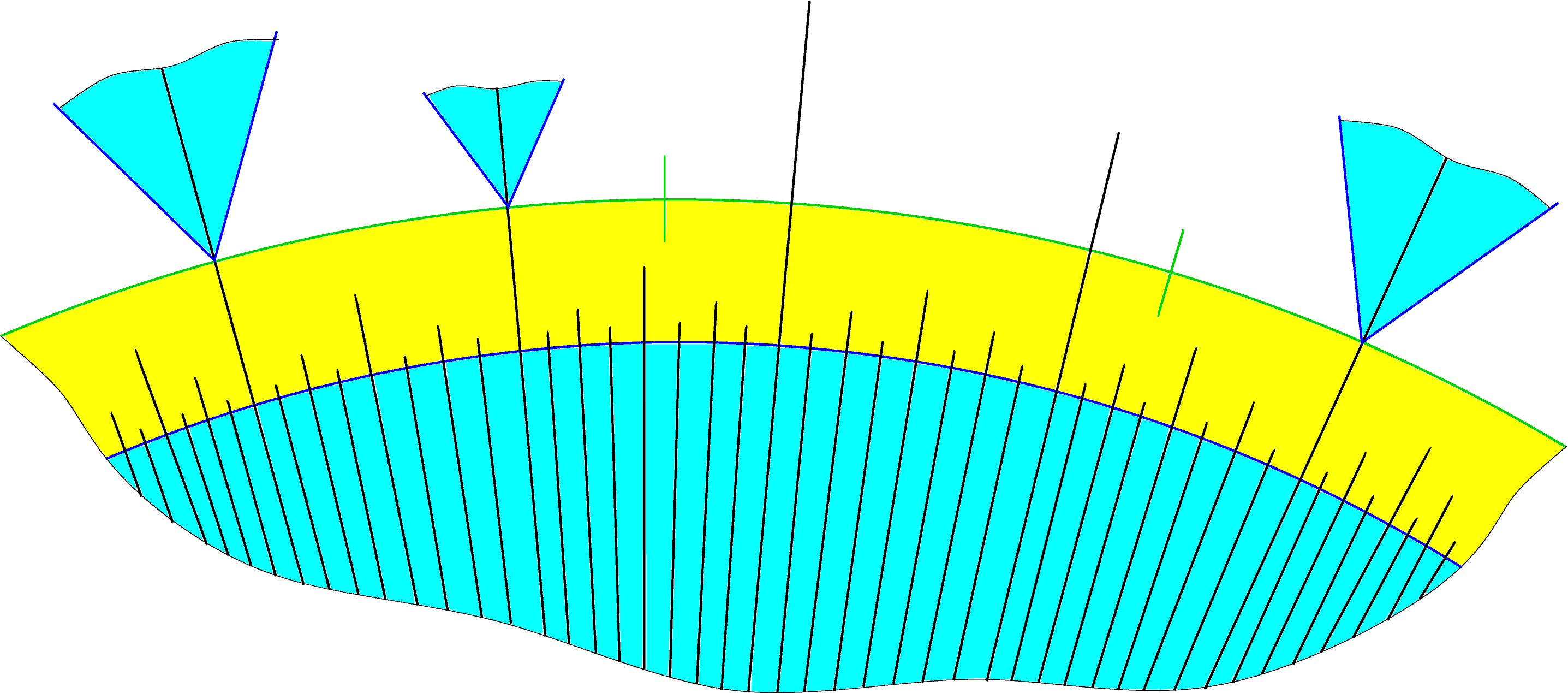}
\end{center}
\caption{Segments of the boundary of a Zone $\, \Omega_{\alpha} \, $ 
corresponding to the appearance and absence of new Stability Zones at 
the points of intersection of the second boundary with segments of special 
directions of $\, {\bf B} \, $.}
\label{Segments}
\end{figure}

 As we have already said above, the appearance of a new Stability Zone on the 
second boundary of the Zone $\, \Omega_{\alpha} \, $ may not occur even at the 
intersection points of the second boundary with ``additional segments'' adjacent 
to $\, \Omega_{\alpha} \, $ and representing the directions of $\, {\bf B} \, $
corresponding to the appearance of unstable periodic trajectories on the Fermi 
surface. This situation occurs if, despite the presence of layers of periodic 
trajectories on the Fermi surface, the appearance of the ``traps of the second type''
occurs on cylinders of closed trajectories of the system (\ref{MFSyst}). 
Thus, in the general case, the boundary of $\, \Omega_{\alpha} \, $ can be divided 
into segments in which new Zones are born at the described intersection points, and 
into segments in which new Zones are not born (Fig. \ref{Segments}). It can be noted 
that, since the height of the cylinders of closed trajectories is determined in this 
case by the diameter of the ``traps of the first type'', the general measure of the 
segments of the second type will be the smaller, the closer the second boundary of 
the Zone $\, \Omega_{\alpha} \, $ is to its first boundary. In addition, one can also 
notice that for segments of the second type, the appearance of new Zones
$\, \Omega_{\beta} \, $ on the segments of special directions of $\, {\bf B} \, $ 
theoretically can occur at some distance from the second boundary of 
$\, \Omega_{\alpha} \, $ as a result of mutual displacement of ``traps'' of the 
first and second types.

 In general, it can be noted that the structure of the angular diagrams for a fixed 
Fermi surface can be quite close to the structure of the diagrams for the entire 
dispersion law near the special Stability Zones discussed above.

\section{Conclusion}
\setcounter{equation}{0}

 We considered angular diagrams for conductivity in metals with arbitrary 
(physical) Fermi surfaces. As it turns out, it is natural to divide all angular 
diagrams into a finite number of ``complexity classes'' that have a quite 
effective geometric description. The most interesting, from our point of view, 
are the most complex diagrams (of type B), containing an infinite number of 
Stability Zones, as well as directions of $\, {\bf B} \, $, corresponding 
to the appearance of ``chaotic'' trajectories on the Fermi surface. 
The paper also provides some comparison of the complexity of such diagrams 
with diagrams defined for the entire dispersion law.

\vspace{2mm}

The study was carried out at the expense of a grant from the 
Russian Science Foundation (project \textnumero $\, $ 18-11-00316).

\end{document}